\documentclass[lettersize,journal]{IEEEtran}
\usepackage{amsmath,amsfonts,bm}
\usepackage{algorithm}
\usepackage{array}
\usepackage{textcomp}
\usepackage{stfloats}
\usepackage{url}
\usepackage{verbatim}
\usepackage{graphicx}
\usepackage{cite}
\usepackage[capitalise]{cleveref}
\crefname{assumption}{Assumption}{Assumptions}
\usepackage{selinput}\SelectInputMappings{adieresis={ä},germandbls={ß}}
\usepackage{epsfig}
\usepackage{times}
\usepackage{float}
\usepackage{tikz,pgfplots}
\usepackage[english]{babel}
\usepackage{algpseudocode}
\usepackage{verbatim}
\usepackage{makecell}
\usepackage{multirow}

\usepackage{xcolor}
\usetikzlibrary{arrows,shapes,backgrounds,patterns,fadings,matrix,arrows,calc,
	intersections,decorations.markings,
	positioning,arrows.meta}
\usepgfplotslibrary{fillbetween}
\usepgfplotslibrary{statistics}
\usepackage{caption}
\usepackage{subcaption}
\pgfplotsset{width=5\columnwidth /5, compat = 1.13,
	height = 60\columnwidth /100, grid= major,
	legend cell align = left, ticklabel style = {font=\scriptsize},
	every axis label/.append style={font=\small},
	legend style = {font={\scriptsize}},title style={yshift=-7pt, font = \small} }

\newtheorem{assumption}{Assumption}
\newtheorem{theorem}{Theorem}
\newtheorem{corollary}{Corollary}
\newtheorem{lemma}{Lemma}
\newtheorem{remark}{Remark}

\newtheorem{proposition}{Proposition}
\newtheorem{property}{Property}

\newcommand{\todo}[1]{{\color{black} #1}}

\xdefinecolor{blueZ}{RGB}{0,77,163}
\xdefinecolor{redZ}{RGB}{232,61,117}

\definecolor{pariedLight_blue}{RGB}{166,206,227}
\definecolor{pariedLight_green}{RGB}{178,223,138}
\definecolor{pariedLight_red}{RGB}{251,154,153}
\definecolor{pariedLight_orange}{RGB}{253,191,111}

\definecolor{paried_blue}{RGB}{31,120,180}
\definecolor{paried_green}{RGB}{51,160,44}
\definecolor{paried_red}{RGB}{227,26,28}
\definecolor{paried_orange}{RGB}{255,127,0}
\definecolor{paried_purple}{RGB}{106,61,154}

\newif\ifarxiv
\arxivtrue

\begin{document}
	
	\title{
		Cooperative Online Learning for Multi-Agent System Control via Gaussian Processes with Event-Triggered Mechanism
		\ifarxiv \\ \todo{(Extended Version)} \fi
	}
	
	\author{
		Xiaobing Dai$^{1}$, Zewen Yang$^{1*}$,~\IEEEmembership{Member,~IEEE}, Sihua Zhang$^{1,2}$, Di-Hua Zhai$^{2}$, \\
		Yuanqing Xia$^{2}$,~\IEEEmembership{Fellow,~IEEE}, Sandra Hirche$^{1}$,~\IEEEmembership{Fellow,~IEEE}
		\thanks{
			$^{*}$Corresponding author.
		}
		\thanks{
			$^{1}$Xiaobing Dai, Zewen Yang, Sihua Zhang and Sandra Hirche are with the Chair of Information-oriented Control (ITR), School of Computation, Information and Technology (CIT), Technical University of Munich (TUM), 80333 Munich, Germany (email: xiaobing.dai, zewen.yang, sihua.zhang, hirche@tum.de).
			$^{2}$Sihua Zhang, Di-Hua Zhai and Yuanqing Xia are with the School of Automation, Beijing Institute of Technology, Beijing, People’s Republic of China (email: sihua.zhang, zhaidih, xia\_yuanqing@bit.edu.cn).
		}
		\ifarxiv
		\thanks{
			This work is supported by the Federal Ministry of Education and Research of Germany in the programme of “Souverän. Digital. Vernetzt.” under the joint project 6G-life with identification number: 16KISK002, by the European Research Council (ERC)  Consolidator  Grant  ”Safe  data-driven  control  for human-centric systems (CO-MAN)” under grant agreement number 864686, and National Natural Science Foundation of China under Grant 62173035.
		}
		\fi
	}
	
	\markboth{Journal of \LaTeX\ Class Files,~Vol.~14, No.~8, August~2021}%
	{Shell \MakeLowercase{\textit{et al.}}: A Sample Article Using IEEEtran.cls for IEEE Journals}
	
	
	\maketitle
	
	\begin{abstract}
		In the realm of the cooperative control of multi-agent systems (MASs) with unknown dynamics, Gaussian process (GP) regression is widely used to infer the uncertainties due to its modeling flexibility of nonlinear functions and the existence of a theoretical prediction error bound.
		Online learning, which involves incorporating newly acquired training data into Gaussian process models, promises to improve control performance by enhancing predictions during the operation. 
		Therefore, this paper investigates the online cooperative learning algorithm for MAS control.
		Moreover, an event-triggered data selection mechanism, inspired by the analysis of a centralized event-trigger, is introduced to reduce the model update frequency and enhance the data efficiency.
		With the proposed learning-based control, the practical convergence of the MAS is validated with guaranteed tracking performance via the Lynaponve theory. 
		Furthermore, the exclusion of the Zeno behavior for individual agents is shown. 
		Finally, the effectiveness of the proposed event-triggered online learning method is demonstrated in simulations.
		\looseness=-1
	\end{abstract}
	
	
	\section{Introduction}
	
	Cooperative control for the multi-agent system (MAS) with unknown models or environmental uncertainties has drawn large attention over the past two decades, particularly in fields such as aerial drones \cite{cui2019multi}, underwater vehicles \cite{yan2019discrete} and  networked sensors\cite{gu2012spatial}.
	To compensate for the uncertainties, machine learning methods are employed on each agent to learn the unknown components from collected data, and then subsequently integrate into the model-based controller design. 
	Specifically, among machine learning techniques, Gaussian process (GP) regression \cite{williams2006gaussian} is popular for model estimation in safe control \cite{chowdhary2014bayesian, he2022adaptive} due to its capability of inferring unknown dynamics with a probabilistically guaranteed prediction performance \cite{srinivas2012information}.  \looseness=-1
	
	The efficacy of learning-based control systems is usually contingent upon the accuracy of the predictions from GP models, which can be enhanced by leveraging more training data \cite{liu2020gaussian}.
	Therefore, cooperative learning is utilized to augment the inference accuracy while managing large datasets within MAS by dividing the overall data set into several sets of individual agents, where the predictions are aggregated via the communication networks. 
	For instance, incorporating the communication graph among agents, a topology-aware aggregation method is first introduced in \cite{yang2021distributed}, inspired by the conventional aggregation framework of distributed GP like the product of experts (PoE) \cite{cao2014generalized}. 
	Additionally, applying consensus theory, a dynamics average consensus algorithm is incorporated into GP aggregation \cite{lederer2022cooperative} to synchronize the individual predictions from each agent.   
	However, the performance of cooperative learning with aggregation methods remains constrained by the precision of the local GP regression models.
	
	To surmount this limitation and further improve prediction accuracy, the integration of online learning emerges as a promising strategy during system operation.
	Among online learning methods for MAS, collective online learning of GP is proposed to bolster the precision of the local GP model in each agent, by real-time optimization of hyper-parameters through the streaming data \cite{hoang2019collective}.
	However, considering the design of this algorithm is strongly based on empirical approximations, the extension of the prediction error bound from the exact GP regression is blocked and thus its practical application in safety-critical scenarios is constrained.
	In addition to the above methods, online data collection yields more accurate prediction models by generating larger data sets \cite{wang2022gaussian}, while maintaining the existence of the prediction error bound \cite{umlauft2019feedback}.
	The online data collection for multi-agent system is introduced in \cite{beckers2021online} to achieve formation control, and in \cite{beckers2022learning} for rigidity-based flocking control.
	Note that the prediction in \cite{beckers2021online, beckers2022learning} uses merely local GP models on each agent, i.e., without cooperative learning.
	The combination of online and cooperative learning for control performance improvement of MAS, to the best knowledge of the authors, has not been addressed yet.
	Moreover, these works, while intuitive and practical for improving the local GP models, raise concerns about data storage and computational resource requirements due to the accumulated data. 
	Furthermore, while the adopted time-trigger serves as an intuitive and practical strategy to reduce computational demands, it overlook the varying significance of data with respect to control performance, often resulting in the collection of unnecessary data and diminished data efficiency. 
	
	In response to the concern of data efficiency, a smart data selection strategy becomes imperative, ensuring the exclusive collection of necessary data.
	Event-triggered data selection methods, recognized for their efficiency in control scenarios, offer benefits in terms of data storage and computational resources while maintaining desired control performance~\cite{Solowjow2018,castaneda2022probabilistic}. 
	The concept of event-triggered learning has been extensively explored in the control of MASs based on neural networks (NNs).
	The trigger mechanisms are designed to determine the instances for updating control inputs \cite{liang2020neural} or broadcasting the weights in the fully connected layer of NN for cooperative learning \cite{gao2019neural}, aiming to alleviate communication burdens. 
	While these methods are shown effective for NN-based control, they strongly depend on the unique regression form of NNs, i.e., linear regression with nonlinear features, which hampers a straightforward extension to other learning-based control in MAS, including GP-based methods \cite{beckers2021online}.
	In the realm of event-triggered online learning with GPs via data collection, studies have been conducted on single-system control using feedback linearization \cite{dai2023can} and back-stepping \cite{jiao2022backstepping}. 
	However, these studies presume complete knowledge of system states, rendering them impractical for deployment in networked MASs. 
	Therefore, for the robustness and scalability of the proposed event-trigger mechanism, the capability of distributed computation becomes pivotal.
	Although event-triggered learning mechanisms have been investigated in model-based and NN-based control, as far as we are aware, the study of effective GP-based online learning control using event-triggered data collection especially in distributed ways has not been explored in the existing literature.

	\subsection{Contribution and Structure}
	In this work, we develop online cooperative learning strategies for GP-based MAS control with event-triggered mechanisms.
	To begin, a learning-based leader-follower time-varying formation control framework for high-order MASs in directed topology with unknown dynamics is proposed, where the derived methodology extends naturally to other control tasks, such as consensus control.
	To improve the learning performance during the operation, a general cooperative online learning strategy based on aggregation and online data collection is proposed, and its prediction performance is analyzed.
	Furthermore, to enhance streaming data collection efficiency and alleviate computational burdens from prediction model updates, a distributed event-triggered online learning strategy is designed, which is inspired by the analysis of centralized approach. 
	To obviate the requirement for a central node with access to all agents, we further propose a fully distributed event-triggered approach, which not only exhibits enhanced scalability but also entails fewer model updates, all while maintaining guaranteed prediction performance.
	Moreover, the achievement of desired control performance, i.e., ensuring an overall tracking error bound around the equilibrium point, is substantiated by using the proposed event-triggered online learning mechanisms. 
	Additionally, a rigorous analysis is provided to show the exclusion of Zeno behavior for each agent within the MAS.
	Finally, the effectiveness of the proposed event-triggered online learning algorithm is demonstrated through simulations, which shows the enhancement of the control performance with less frequent model updates compared to time-triggered learning and offline learning. \looseness=-1
	
	The remainder of the paper is structured as follows:
	\cref{section_problem_setting} outlines the problem setting.
	In \cref{section_cooperative_online_learning}, cooperative online learning with Gaussian process regression is discussed for MAS control with their performance analysis.
	The centralized and distributed event-triggered online learning mechanisms are proposed in \cref{section_event_triggered_learning} with the discussion of the Zeno behavior.
	Numerical simulations are presented in \cref{section_simulation} to demonstrate the effectiveness of the proposed method.
	Finally, \cref{section_conclusion} concludes the paper.	\looseness=-1
	
	\subsection{Notation and Graph Theory}
	The natural numbers with/without zero are denoted by $\mathbb{N}$/$\mathbb{N}_+$, real positive numbers with and without zero by $\mathbb{R}_{0,+}$ and $\mathbb{R}_+$, respectively. 
	Minimum/maximum eigenvalues of a square matrix $\bm{A}$ are denoted by $\underline{\lambda}(\bm{A})$/$\bar{\lambda}(\bm{A})$. 
	Unless explicitly specified, $| \cdot |$ refers to the element-wise absolute operator, and $\| \!\cdot\! \|$ represents the Euclidean norm.
	The $i$-th entry of a vector $\boldsymbol{a}$ is represented as $a_i$, whereas $a_{ij}$ signifies the element at the intersection of the $i$-th row and the $j$-th column within matrix $\boldsymbol{A}$. 
	The symbol $\otimes$ denotes the Kronecker product.
	The diagonal operator for scalar inputs denotes $\mathrm{diag} (\cdot)$, and the block diagonal operator for vector/matrix inputs denotes $\mathrm{blkdiag} (\cdot)$. 
	The identity matrix with the dimension of $m\times m$ is denoted by $\boldsymbol{I}_m$, and the column vector  $m\times 1$ vector with all components equal to one is denoted as $\mathbf{1}_m$.
	\looseness=-1
	
	The communication network for information exchange among $N\in\mathbb{N}_+$ agents is defined by a directed graph $\mathcal{G} = \{ \mathcal{V}, \mathcal{E}\}$, where $\mathcal{V} = \{ 1, \cdots, N \}$ is the vertex set representing the indices of agents and $\mathcal{E} \in \mathcal{V} \times \mathcal{V}$ is the edge set. The agent $i$ can receive the information from agent $j$ when $(j,i) \in \mathcal{E}, \forall i,j \in \mathcal{V}$. 
	The topology of the graph is characterized by a weighted adjacency matrix denoted as $\bm{\mathcal{A}} \in \mathbb{R}^{N \times N}$. 
	The entries $a_{ij}>0$ when there exists a communication channel from agent $j$ to agent $i$, i.e., $(j,i) \in \mathcal{E}$, otherwise $a_{ij} = 0, \forall i,j \in \mathcal{V}$. 
	Furthermore, the out-degree Laplacian matrix is defined as $\bm{\mathcal{L}} = \{ l_{ij} \}_{i,j \in \mathcal{V}} \in \mathbb{R}^{N \times N}$ with $l_{ii} = \sum_{j=1}^N a_{ij}$ and $l_{ij} = - a_{ij}$ for $j \ne i$. 
	The set $\mathcal{N}_i$ contains all the neighbor agents of agent $i$, i.e., $\mathcal{N}_i = \{ j \in \mathcal{V} | a_{ij} > 0 \}$, and the set $\bar{\mathcal{N}}_i$ is defined as $\bar{\mathcal{N}}_i = \{ j \in \mathcal{V} | j \in \mathcal{N}_i ~\wedge~ i \in \mathcal{N}_j \}$ such that agent $j$ can get $\bm{x}_i$ from agent $i$ and vice versa. 	\looseness=-1
	\section{Problem Setting and Preliminaries} \label{section_problem_setting}
	
	\subsection{Multi-Agent System} \label{subsection_system_description}
	In this paper, we consider a MAS with $N \in \mathbb{N}_+$ homogeneous agents.
	The $i$-th agent follows a $n$-order continuous dynamical system with $n \in \mathbb{N}_+$ described as
	\begin{align} \label{eqn_agent_dynamics}
		&\dot{\bm{x}}_{i,k} = \bm{x}_{i,k+1}, ~ \forall k = 1, \cdots, n - 1, \nonumber \\ 
		&\dot{\bm{x}}_{i,n} = \bm{h}(\bm{x}_i) + \bm{g}(\bm{x}_i) \bm{u}_{i} + \bm{f}(\bm{x}_i), ~ i \in \mathcal{V},
	\end{align}
	where the $p \in \mathbb{N}_+$ dimension states denote $\bm{x}_{i,j} \in \mathbb{R}^p, \forall j = 1, \cdots, n$.
	The concatenated system states and control input for agent $i$ denote $\bm{x}_{i} = [\bm{x}_{i,1}^T, \cdots, \bm{x}_{i,n}^T]^T \in \mathbb{X} \subset \mathbb{R}^{n p}$ and $\bm{u}_i \in \mathbb{R}^q$ with $q \in \mathbb{N}_+$, respectively. 
	The function $\bm{h}(\cdot): \mathbb{X} \to \mathbb{R}^n$ and the non-singular function $\bm{g}(\cdot): \mathbb{X} \to \mathbb{R}^{np \times q}$ represent the known parts of the system dynamics, which are usually obtained by using the first principle.
	Note that the non-singularity of $\bm{g}(\cdot)$ is a prerequisite for feedback controller design \cite{khalil2015nonlinear}, which ensures each agent is controllable at any state in $\mathbb{X}$ \cite{lederer2022cooperative}.
	The unmodeled part and external environmental uncertainties are encoded into function $\bm{f}(\cdot): \mathbb{X} \!\rightarrow\! \mathbb{R}^p$, so it is considered as unknown but identical for each agent. 
	
	\ifarxiv
	\todo{
		The high-order system \eqref{eqn_agent_dynamics} is a general form for MAS aligned with many applications, such as robotics \cite{gao2023quasi} with $n \!=\! 2$ and electrohydraulic systems \cite{kaddissi2007identification} with $n \!=\! 4$. 
	}
	\fi
	In this paper, we set the dimension of the state in every order, i.e., $\bm{x}_{i,k}$, and control input $\bm{u}_i$ to be scalar respectively, indicating $p = q = 1$ for notational simplicity.
	The derived results can be directly extended to high-dimensional systems by using Kronecker product and multi-output machine learning methods.
	
	The control objective is to achieve a leader-follower time-varying formation, where the dynamics of the leader follows 
	\begin{align} \label{eqn_leader_dynamics}
		&\dot{x}_{l,k} = x_{l,k + 1}, ~ \forall k = 1, \cdots, n-1, \nonumber \\
		&\dot{x}_{l,n} = x_{l,r}(t),
	\end{align}
	where the leader states denote $\bm{x}_l = [x_{l,1}, \cdots, x_{l,n}]^T \in \mathbb{R}^n$ with $x_{l,k} \in \mathbb{R}, \forall k = 1, \cdots, n$ and the known continuous function $x_{l,r}(\cdot): \mathbb{R}_{0,+} \rightarrow \mathbb{R}$.
	Moreover, each agent tracks the leader but keeps a predefined relative time-varying distance as
	\begin{align} \label{eqn_reference}
		&\dot{s}_{i,k} = s_{i,k+1},  ~ \forall k = 1, \cdots, n-1, \nonumber \\
		&\dot{s}_{i,n} = s_{i,r}(t), ~i \in \mathcal{V},
	\end{align}
	where its concatenated states $\bm{s}_{i} = [s_{i,1}, \cdots, s_{i,n}]^T \in \mathbb{R}^{n}$ with $s_{i,k} \in \mathbb{R}, \forall k = 1, \cdots, n$ and the known continuous function $s_{i,r}(\cdot): \mathbb{R}_{0,+} \rightarrow \mathbb{R}$. 
	The structures of the leader and relative trajectory in \eqref{eqn_leader_dynamics} and \eqref{eqn_reference} guarantee that all agents $i \in \mathcal{V}$ are capable of following its own reference $\bm{x}_l + \bm{s}_i$  \cite{dong2014time}.
	Furthermore, the desired reference $\bm{x}_l + \bm{s}_i$ for each agent $i$ and the derivative of the highest order of leader dynamics $x_{l,r}(\cdot)$ satisfy the following assumption.
	\begin{assumption} \label{assumption_desired_trajectory}
		There exist well-defined positive constants $F_l, F_{r,i} \in \mathbb{R}_{0,+}$, such that $x_{l,r}(\cdot)$ and the derivative of the references $\dot{\bm{x}}_l + \dot{\bm{s}}_i$ are bounded as $| x_{l,r}(t) | \!\le\! F_l$ and $\| \dot{\bm{x}}_l(t) + \dot{\bm{s}}_i(t) \| \le F_{r,i}$ respectively for $\forall t \!\in\! \mathbb{R}_{0,+}$ and $\forall i \in \mathcal{V}$.
	\end{assumption}
	In practice, the dynamics of the leader $\bm{x}_l$ in \eqref{eqn_leader_dynamics} and relative distances $\bm{s}_i$ in \eqref{eqn_reference} for $\forall i \in \mathcal{V}$ are designed such that $\bm{x}_l + \bm{s}_i \in \mathbb{X}$, indicating each entry in $\bm{x}_l + \bm{s}_i$ is bounded.
	Then, the boundness of its derivative, i.e., $\| \dot{\bm{x}}_l(t) + \dot{\bm{s}}_i(t) \|$, only requires bounded $x_{l,r}(\cdot)$ and $s_{i,r}(\cdot)$ for $\forall t \!\in\! \mathbb{R}_{0,+}$ considering the structures in \eqref{eqn_leader_dynamics} and \eqref{eqn_reference}. Since the dynamics of the leader and relative reference are design choices, \cref{assumption_desired_trajectory} is not restrictive.
	It can be easily satisfied by choosing $x_{l,1}(t)$ and $s_{i,1}(t)$ as at least $(n+1)$-th smooth, i.e., $x_{l,1}(\cdot), s_{i,1}(\cdot) \in \mathcal{C}^{n+1}$.
	
	Note that, while the the states $\bm{s}_{i}$ and $s_{i,r}(\cdot)$ are available for each agent $i$, only the agents connected to the leader are able to obtain the information of leader states $\bm{x}_l$ and $x_{l,r}(\cdot)$. 
	In order to describe the connectivity between the leader and followers, a diagonal matrix $\bm{\mathcal{B}} \!=\! \mathrm{diag}(b_{11}, \!\cdots\!, b_{NN}) \!\in\! \mathbb{R}^{N \!\times\! N}$ is adopted, where $b_{ii} \!=\! 1$ indicates the agent $i$ receives the information from the leader and $b_{ii} \!=\! 0$ otherwise.
	Due to the existence of leader disconnection, the controlled system cannot achieve asymptotically stability with formation error $\bm{\vartheta} \!=\! \bm{0}$, where 
	\looseness=-1
	\begin{align} \label{eqn_vartheta}
		\bm{\vartheta} = \bm{x} - \bm{s} - \bm{1}_N \otimes \bm{x}_l
	\end{align}
	with $\bm{x} = [\bm{x}_1^T, \cdots, \bm{x}_N^T]^T$ and $\bm{s} = [\bm{s}_1^T, \cdots, \bm{s}_N^T]^T$.
	Instead, the formation error $\bm{\vartheta}$ is bounded by a positive constant $\bar{\vartheta} \in \mathbb{R}_{0,+}$, i.e., $\|\bm{\vartheta}\| \le \bar{\vartheta}$.
	To achieve the cooperative control for the leader-follower formation task, the following assumption on the communication topology is required.
	\begin{assumption}[$\!\!$\cite{weirenConsensusSeekingMultiagent2005}] \label{assumption_topology}
		The augmented graph characterized by $\bm{\mathcal{L}}$ and $\bm{\mathcal{B}}$ contains a spanning tree.
		Moreover, the leader is the root without incoming edge from the agents.
	\end{assumption}
	\cref{assumption_topology} is commonly found in MAS control to ensure the information of the leader is directly available by some agents, and propagates through the network~\cite{lederer2022cooperative, yang2021distributed,dong2014time}.
	Moreover, \cref{assumption_topology} is essential to guarantee the achievement of leader-follower task and leads to the following lemma.
	\begin{lemma} [$\!\!$\cite{zhang2012adaptive}] \label{lemma_MAS_Lyapunov}
		Suppose \cref{assumption_topology} holds and choose $\bm{q} \in \mathbb{R}^N$ and $\bm{P} \in \mathbb{R}^{N \times N}$ defined as
		\begin{align*}
			&\bm{q} = [q_1, \cdots, q_N]^T = (\bm{\mathcal{L}} + \bm{\mathcal{B}})^{-1} \bm{1}_N, \\
			&\bm{P} = \mathrm{diag} (q_1^{-1}, \cdots, q_N^{-1}). 
		\end{align*}
		Let $\bm{Q} = \bm{P} (\bm{\mathcal{L}} + \bm{\mathcal{B}}) + (\bm{\mathcal{L}} + \bm{\mathcal{B}})^T \bm{P}  \in \mathbb{R}^{N \times N}$, then the matrices $\bm{P}$ and $\bm{Q}$ are symmetric positive definite.
	\end{lemma}
	\cref{lemma_MAS_Lyapunov} provides a possible choice of the Lyapunov function for high-order MAS \cite{zuo2017fixed}, which is used for stability analysis in \cref{section_cooperative_online_learning}.
	\looseness=-1
	
	\subsection{Distributed Control Law}
	To achieve the leader-follower time-varying formation control task within the multi-agent framework, a distributed control structure for the individual agent $i$ is proposed as
	\begin{align} \label{eqn_control_law}
		u_i \!=\! \frac{1}{g(\bm{x}_i)} \big( \phi(\mathcal{J}_i, \mathcal{I}_i, \mathcal{J}_{r,i}) \!-\! h(\bm{x}_i) \!-\! \hat{f}_i(\bm{x}_i) \big), ~ \forall i \!\in\! \mathcal{V},
	\end{align}
	where the linear consensus law $\phi(\cdot, \cdot)$ is designed using the local information $\mathcal{J}_i = \bm{x}_i - \bm{s}_i$ of the agent $i$, and the received neighboring information set $\mathcal{I}_i = \bigcup_{j \in \mathcal{N}_i} \mathcal{J}_j$ is shared from its neighbors.
	The reference information $\mathcal{J}_{r,i}$ for each agent $i$ is defined according to the connectivity with the leader as
	\begin{align}
		\mathcal{J}_{r,i} = \begin{cases}
			\{ \bm{x}_l, x_{l,r}, \bm{s}_i, s_{i,r} \}, & \text{if} ~ b_{ii} = 1 \\
			\{ \bm{s}_i, s_{i,r} \}, & \text{if} ~ b_{ii} = 0
		\end{cases}.
	\end{align}
	Given the $\mathcal{J}_i$, $\mathcal{I}_i$ and $\mathcal{J}_{r,i}$, the linear consensus law $\phi(\cdot)$ can be evaluated only with the local and neighbor information, allowing distributed computation, which is designed as
	\begin{align} \label{eqn_consensus_control_law}
		\phi(\mathcal{J}_i, \mathcal{I}_i, \mathcal{J}_{r,i}) = -c r_i + s_{i,r} + b_{ii} x_{l,r},
	\end{align}
	where control gain denotes $c \in \mathbb{R}_+$ and the filtered error is $r_i = \sum_{k=1}^{n}  \lambda_k e_{i,k}$.
	The synchronization errors $e_{i,k}$ for the $k$-th dimension are calculated as
	\begin{align} \label{eqn_synchronization_error}
		e_{i,k} = \sum\nolimits_{j \in \mathcal{N}_i} a_{ij} (\tilde{x}_{i,k} - \tilde{x}_{j,k}) + b_{ii} \vartheta_{i,k}
	\end{align}
	for $\forall k = 1, \cdots, n$ with $\vartheta_{i,k} = x_{i,k} - s_{i,k} - x_{l,k}$.
	Meanwhile, the coefficients $\lambda_1, \cdots, \lambda_{n-1} \in \mathbb{R}$ and $\lambda_n \in \mathbb{R}_+$ are chosen such that the matrix $\bm{\Lambda} \in \mathbb{R}^{(n-1) \times (n-1)}$ is Hurwitz with
	\begin{align} \label{eqn_Lambda}
		\bm{\Lambda} = \begin{bmatrix}
			\bm{0}_{(n-2) \times 1} & \bm{I}_{n-2} \\
			- \lambda_{n}^{-1} \lambda_1 & - \lambda_{n}^{-1} [\lambda_2, \cdots, \lambda_{n-1}]
		\end{bmatrix}.
	\end{align}
	The compensation functions $\hat{f}_i(\cdot)$ in \eqref{eqn_control_law} predict the unknown function $f(\cdot)$ through the data-driven method using the data set collected individually by each agent $i \in \mathcal{V}$, which satisfies the following assumption.
	\begin{assumption} \label{assumption_dataset}
		The data pair $\{ \bm{x}_i^{(\varsigma)}, y_i^{(\varsigma)} \}$ is available for each agent $i \in \mathcal{V}$ at any time $t_i^{(\varsigma)} \in \mathbb{R}_{0,+}, \varsigma \in \mathbb{N}_+$, where $y_i^{(\varsigma)} = \dot{x}_{i,n}^{(\varsigma)} + w_i^{(\varsigma)}$.
		The measurement noise $w_i^{(\varsigma)} \in \mathbb{R}$ of $y_i^{(\varsigma)}$ follows a zero-mean, independent and identical Gaussian distribution, i.e., $w_i^{(\varsigma)} \sim \mathcal{N}(0, \sigma_{o,i}^2)$ with $\sigma_{o,i} > 0$.
	\end{assumption}
	\cref{assumption_dataset} indicates each agent collects the data set on their own, without sharing among them.
	\ifarxiv
	\todo{
		While the requirement of full state measurement, i.e., $\bm{x}_i$, is usually found in controller design for nonlinear system \cite{khalil2015nonlinear} and MAS \cite{yang2021distributed, lederer2022cooperative}, it excludes cases where only partial system states are obtainable.
		However, by using proper observers \cite{hong2008distributed, yu2016observer} or filters \cite{talebi2019distributed}, the full system states can be estimated, converting the problems back to full state feedback control.
		Moreover, full state measurement is necessary for machine learning methods, indicating the input of the data driven model is available.
		In practice, the measurement noise on $\bm{x}_i$ is non-neglectable, but its effects can effectively be handled on the output $y$ into $w$ by using Taylor expansion as shown in \cite{mchutchon2011gaussian, kim2023model}, or on the kernel function \cite{wang2022gaussian}.
		Using the above methods, the inputs for the GP model can still be regarded as noise-free.
		Furthermore, it allows noisy observations of $f(\cdot)$ due to known $g(\cdot)$ and $h(\cdot)$ and through numerical approximation of $\dot{x}_{i,n}^{(\varsigma)}$, e.g., via finite difference inducing Gaussian error \cite{dai2023can}.
		The relaxation for noise distribution exists as in \cite{chowdhury2017kernelized, maddalena2021deterministic, hashimoto2022learning, lederer2023gaussian}.
		Note that, this work focus on the event-trigger design for cooperative online learning, therefore the extension to different system classes and noise distributions is left in the future work.
	}
	\fi
	
	Additionally, \cref{assumption_dataset} also facilitates the online data collection for the prediction model update. 
	For GP model update with high data efficiency, an event-triggered mechanism is required for online learning in MAS, that employs a smart data selection strategy to store only the necessary data and ensures a desired control performance. 
	To this end, the event-triggered mechanism is designed such that $\{ \bm{x}_i^{(\varsigma)}, y_i^{(\varsigma)} \}$ is added into the data set of agent $i$ at time $t_i^{(\varsigma)}$ satisfying
	\begin{align} \label{eqn_fake_trigger}
		t_i^{(\varsigma)} = \inf \left\{ t_i: t_i > t_i^{(\varsigma - 1)} ~\wedge~ \rho_i(t_i) > \bar{\rho}_i(t_i) \right\},
	\end{align}
	where $\rho_i(t_i)$ and $\bar{\rho}_i(t_i)$ are the simplified versions of the trigger function and its threshold function, respectively.
	For centralized event-trigger, it has $\rho_i(t_i) \!=\! \rho(\cup_{j \in \mathcal{V}} \{ \mathcal{J}_j(t_i), \mathcal{J}_{r,j}(t_i)\} )$ and $\bar{\rho}_i(t_i) = \bar{\rho}(\cup_{j \in \mathcal{V}} \{ \mathcal{J}_j(t_i), \mathcal{J}_{r,j}(t_i)\} )$ requiring global information for evaluation.
	In the distributed scenario, $\rho_i(t_i)$ and $\bar{\rho}_i(t_i)$ only employ local and neighboring information, i.e., $\rho_i(t_i) = \rho(\mathcal{J}_i(t_i), \mathcal{I}_i(t_i), \mathcal{J}_{r,i}(t_i))$ and $\bar{\rho}_i(t_i) = \bar{\rho}(\mathcal{J}_i(t_i), \mathcal{I}_i(t_i), \mathcal{J}_{r,i}(t_i))$, allowing distributed computation on each agent $i \in \mathcal{V}$.
	Note that we set both $\rho(\cdot)$ and $\bar{\rho}(\cdot)$ as time-varying functions for better explanability of the trigger condition \eqref{eqn_fake_trigger}, whose detailed design, i.e., explicit expression of $\rho(\cdot)$ and $\bar{\rho}(\cdot)$, for centralized and distributed scenarios are described in \cref{section_cooperative_online_learning} and \cref{section_event_triggered_learning}, respectively.
	
	\section{Cooperative Online Learning based Distributed Control with Gaussian Processes} 
	\label{section_cooperative_online_learning}
	
	\subsection{Gaussian Process Regression}
	
	A Gaussian process induces a distribution of the unknown function $f(\cdot)$ characterized by the mean function $m(\cdot): \mathbb{X} \!\rightarrow \!\mathbb{R}$ and kernel function $\kappa(\cdot, \cdot): \mathbb{X} \!\times\! \mathbb{X} \rightarrow \mathbb{R}_{0,+}$,  {i.e.}, $f(\cdot) \sim \mathcal{GP}( m(\cdot), \kappa(\cdot,\cdot) )$. Particularly, the mean function $m(\cdot)$ reflects the prior knowledge, which is set as $m(\cdot) = 0$ considering $f(\cdot)$ is fully unknown, since all known part in the dynamics \eqref{eqn_agent_dynamics} is encoded into $h(\cdot)$. And the kernel function $\kappa(\cdot,\cdot)$ indicates the covariance between two samples and is assumed to satisfy the following condition. \looseness=-1
	\begin{assumption} \label{assumption_kernel}
		The continuous function $f(\cdot)$ is sampled from a Gaussian process $\mathcal{GP}( 0, \kappa(\cdot,\cdot) )$ with stationary and Lipschitz continuous kernel function with Lipschitz constant $L_\kappa \in \mathbb{R}_+$.
		Moreover, the kernel function $\kappa(\bm{x}, \bm{x}') = \kappa( \| \bm{x} - \bm{x}' \|)$ is monotonically decreasing with respect to $\| \bm{x} - \bm{x}' \|$, and $\kappa(0) = \sigma_f^2$ with $\sigma_f \in \mathbb{R}_+$.
	\end{assumption}
	\cref{assumption_kernel} defines the prior distribution of unknown function $f(\cdot)$ by choosing suitable kernel.
	The Lipschitz continuity of $\kappa(\cdot,\cdot)$ only requires the kernel function to be continuous by considering the compact input domain $\mathbb{X}$, which holds for most kernels, such as square exponential kernel, rational quadratic kernel, and their combination\cite{williams2006gaussian}.
	Therefore, this assumption imposes no significant restrictions.
	
	Given the data set $\mathbb{D}$ with $M \in \mathbb{N}$ samples, i.e., $\mathbb{D} = \{ \bm{x}^{(\varsigma)}, y^{(\varsigma)} \}_{\varsigma = 1}^M$, satisfying \cref{assumption_dataset} under the variance of measurement noise as $\sigma_o^2$ with $\sigma_o > 0$.
	Then the posterior mean $\mu(\cdot)$ and variance $\sigma^2(\cdot)$ of the GP model are 
	\begin{align} \label{eqn_GP_prediction}
		&\mu(\bm{x}) = \bm{k}_{X}^T(\bm{x}) (\bm{K} + \sigma_o^2 \bm{I}_M)^{-1} \bm{y}, \\
		&\sigma^2(\bm{x}) = \kappa(\bm{x},\bm{x}) - \bm{k}_{X}^T(\bm{x}) (\bm{K} + \sigma_o^2 \bm{I}_M)^{-1} \bm{k}_{X}(\bm{x}), \nonumber
	\end{align}
	where $\bm{y} \!=\! [y^{(1)}, \cdots, y^{(M)}]^T$, $\bm{K} \!=\! \{ \kappa(\bm{x}^{(i)}, \bm{x}^{(j)})\}_{i,j=1,\cdots,M}$ and $\bm{k}_X(\bm{x}) \!=\! [\kappa(\bm{x}, \bm{x}^{(1)}), \!\cdots\!, \kappa(\bm{x}, \bm{x}^{(M)})]^T$.
	The posterior mean $\mu(\cdot)$ is used for the estimation of $f(\cdot)$, while the posterior variance $\sigma^2(\cdot)$ is employed to quantify the prediction error as shown in the following lemma.
	
	\begin{lemma} [$\!\!$ \cite{lederer2019uniform}] \label{lemma_GP_error_bound}
		Predict an unknown function $f(\cdot)$ sampled from $\mathcal{GP}( 0, \kappa(\cdot,\cdot) )$ satisfying \cref{assumption_kernel} using GP regression with a data set satisfying \cref{assumption_dataset}.
		Pick $\tau \in \mathbb{R}_+$ and $\delta \in (0,1) \subset \mathbb{R}$, the prediction error is upper bounded as
		\begin{align}
			| f(\bm{x}) - \mu(\bm{x}) | \leq \eta_{\delta}(\bm{x}) = \sqrt{\beta_{\delta}} \sigma(\bm{x}) + \gamma_{\delta}, ~ \forall \bm{x} \in \mathbb{X} \nonumber
		\end{align}
		with a probability of at least $1 - \delta$, where
		\begin{align}
			&\beta_{\delta} = 2 \sum\nolimits_{k=1}^{n} \log \Big( \frac{\sqrt{n}}{2 \tau} ( \bar{x}_k - \underline{x}_k ) + 1 \Big) - 2 \log \delta, \nonumber \\
			&\gamma_{\delta} = \big( \sqrt{\beta_{\delta}} L_{\sigma} + L_f + L_{\mu} \big) \tau \nonumber
		\end{align}
		with $\bar{x}_k = \max_{\bm{x} \in \mathbb{X}} x_k$ and $\underline{x}_k = \min_{\bm{x} \in \mathbb{X}} x_k$ for $x_k$ as the $k$-th dimension of $\bm{x}$, and $L_f$, $L_{\mu}$ and $L_{\sigma}$ are the Lipschitz constants for the unknown function $f(\cdot)$, the posterior mean $\mu(\cdot)$ and variance $\sigma(\cdot)$, respectively.
	\end{lemma}
	Although with conservatism, \cref{lemma_GP_error_bound} provides a calculable uniform prediction error bound on the compact domain $\mathbb{X}$.
	\ifarxiv
	\todo{
		Note that the computation of $\eta_{\delta}(\cdot)$ requires the values for Lipschitz constants $L_{\mu}$, $L_{\sigma}$ and $L_f$, where $L_{\mu}$ and $L_{\sigma}$ can be directly determined as shown in \cite{lederer2021gaussian} for kernel function $\kappa(\cdot,\cdot)$ satisfying \cref{assumption_kernel}.
		For the Lipschitz constant of unknown function $f(\cdot)$, while the existence of well-defined $L_f$ is easily proven by considering $f(\cdot)$ is continuous and its input domain $\mathbb{X}$ is compact, its exact value is usually unavailable.
		However, without the prior knowledge on $L_f$ from first principle, it can be approximated empirically or through data-driven methods \cite{lederer2019uniform}.
	}
	\fi
	Due to the computable error bound $\eta_{\delta}(\cdot)$, GP regression is widely used in safe learning-based control with guarantee \cite{umlauft2019feedback, dhiman2021control, lederer2022cooperative}.\looseness=-1
	
	\subsection{Cooperative Online Learning}
	
	In our setting, a GP model is deployed on each agent $i$ with individual data set $\mathbb{D}_i$ satisfying \cref{assumption_dataset} with measurement noise variance $\sigma_{o,i}$ and kernel function $\kappa_i(\cdot)$ under \cref{assumption_kernel} with $\kappa_i(0) = \sigma_{f,i}^2$ and $\sigma_{f,i} \in \mathbb{R}_+$.
	To obtain the prediction of $f(\cdot)$ cooperatively in MASs, each agent $i$ shares its states $\bm{x}_i$ with its bidirectional neighbors in $\bar{\mathcal{N}}_i$, such that its neighbor agents $\forall j \in \bar{\mathcal{N}}_i$ calculate the prediction $\mu_j(\bm{x}_i)$ at $\bm{x}_i$ using their own data set $\mathbb{D}_j$.
	Notably, only the predictions from agents in $\bar{\mathcal{N}}_i$ are available on agent $i$, since any agent $j$ belonging to $\mathcal{N}_i \backslash \bar{\mathcal{N}}_i$ cannot calculate the prediction $\mu_j(\bm{x}_i)$ for agent $i$.
	This is because $\bm{x}_i$ cannot send to agent $j$ due to the lack of transmission channel from $j$ to $i$, i.e., $(i,j) \notin \mathcal{E}$.
	Combining the local prediction $\mu_i(\bm{x}_i)$ and the received neighboring predictions $\{ \mu_j(\bm{x}_i) \}_{j \in \bar{\mathcal{N}}_i}$, the compensation $\hat{f}_i(\bm{x}_i)$ in \eqref{eqn_control_law} is formulated using aggregation method with a general form as
	\begin{align} \label{eqn_GP_aggregation}
		\hat{f}_i(\bm{x}_i) = \omega_{ii}(\bm{x}_i) \mu_i(\bm{x}_i) + \sum\nolimits_{j \in \bar{\mathcal{N}}_i} \omega_{ij}(\bm{x}_i) \mu_j(\bm{x}_i),
	\end{align}
	where $\omega_{ij}(\cdot): \mathbb{X} \!\to\! \mathbb{R}_{0,+}$ indicates the aggregation weight respective to $\mu_j(\boldsymbol{x}_i)$ defined for specific aggregation such as POE \cite{cao2014generalized}.
	Since online learning strategy is applied such that each agent $i$ adds newly generated data pair $\{ \bm{x}_i^{(\varsigma)}, y_i^{(\varsigma)} \}$ at $t_i^{(\varsigma)}$ into the local data set $\mathbb{D}_i$.
	The aggregated prediction error bound after the GP model update is shown as follows.
	\begin{lemma} \label{proposition_online_cooperative_prediction_error_bound}
		Let the assumptions in \cref{lemma_GP_error_bound} be satisfied for all local GP models on the agents in the MAS, and employ the aggregation method in \eqref{eqn_GP_aggregation} for agent $i \in \mathcal{V}$.
		Define $\underline{\hat{\sigma}}_i$ as the solution of the optimal problem for $i \in \mathcal{V}$ as
		\begin{align} \label{eqn_upperbound_posterior_variance_after_model_update}
			&\underline{\hat{\sigma}}_i \!=\!\! \sup_{ \bm{x}_i \in \mathbb{X} } \hat{\sigma}_i^+\!(\bm{x}) \!=\! \sup_{ \bm{x}_i \in \mathbb{X} } \!\Big(\! \omega_{ii}^+\!(\bm{x}_i) \sigma_{i}^+\!(\bm{x}_i) \!+\!\! \sum_{j \in \bar{\mathcal{N}}_i } \omega_{ij}^+\!(\bm{x}_i) \sigma_{j}(\bm{x}_i) \!\Big), \nonumber \\
			&\text{s.t.} ~ \sigma_{i}^+(\bm{x}_i) = \sigma_{o,i}  \sigma_i(\bm{x}_i) /  \sqrt{\sigma_i^2(\bm{x}_i) + \sigma_{o,i}^2},
		\end{align}
		where the aggregation weight $\omega_{ij}^+(\bm{x}_i)$ for agent $j \in \{i, \bar{\mathcal{N}}_i\}$ is evaluated at $\bm{x}_i$ after the GP model update at agent $i$.
		Moreover, let $\omega_{ij}^+(\bm{x}_i)$ maintain the property of $\sum_{j \in \{i, \bar{\mathcal{N}}_i\}} \omega_{ij}^+(\cdot) = 1$ and choose $\delta \!\in\! (0, N^{-\!1\!} )$,
		then the prediction error after GP model update by using the cooperative learning in \eqref{eqn_GP_aggregation} is bounded as \looseness=-1
		\begin{align}
			| f(\bm{x}_i) - \hat{f}_i^+(\bm{x}_i) | \le \underline{\hat{\eta}}_{\delta,i} = \sqrt{\beta_{\delta}} \underline{\hat{\sigma}}_{i} + \gamma_{\delta}, && \forall \bm{x}_i \in \mathbb{X}, \nonumber
		\end{align}
		with probability of at least $1 - N \delta$, where $\hat{f}_i^+(\cdot)$ is the aggregated prediction after local model update at agent $i$. 
	\end{lemma}
	
	\begin{IEEEproof}
		Considering each agent $i$ collects the data pairs on its own and adds them into the local data set $\mathbb{D}_i$, the posterior variance $\sigma_i^+(\bm{x})$ after GP model update agent $i$ is written as
		\begin{align} \label{eqn_updated_posterior_variance}
			(\sigma_i^+(\bm{x}))^2 =& \kappa_i(0) \nonumber \\
			&\!\!\!\!\!\!\!\!\!\!\!\!\! - \begin{bmatrix}
				\bm{k}_{X,i}(\bm{x}) \\ \kappa_i(0)
			\end{bmatrix}^{\!T\!} \!\! \begin{bmatrix}
				\bm{K}_i \!+\! \sigma^2_{o,i} \bm{I}_{M_i} & \bm{k}_{X,i}(\bm{x}) \\ \bm{k}_{X,i}^T(\bm{x}) & \kappa_i(0) \!+\! \sigma^2_{o,i}
			\end{bmatrix}^{\!-\!1\!} \!\! \begin{bmatrix}
				\bm{k}_{X,i}(\bm{x}) \\ \kappa_i(0)
			\end{bmatrix} \nonumber \\
			=& \sigma_i^2(\bm{x}) - \frac{\sigma_i^4(\bm{x})}{\sigma_i^2(\bm{x}) + \sigma_{o,i}^2} = \frac{\sigma_i^2(\bm{x}) \sigma_{o,i}^2}{\sigma_i^2(\bm{x}) + \sigma_{o,i}^2},
		\end{align}
		where $M_i$ is the number of training samples in $\mathbb{D}_i$ before adding the collected data pairs.
		The kernel vector $\bm{k}_{X,i}(\bm{x})$ and Gram matrix $\bm{K}_i$ follow the definition in \eqref{eqn_GP_prediction} using the kernel function $\kappa_i(\cdot,\cdot)$ evaluated with $\mathbb{D}_i$.
		Note that \eqref{eqn_updated_posterior_variance} is equivalent to the expression of the constraint in \eqref{eqn_upperbound_posterior_variance_after_model_update}, such that the prediction error bound at agent $i$ is written as
		\begin{align}
			\Pr \{ |\mu_i^+\!(\bm{x}) \!-\! f(\bm{x})| \!\le\! \eta_i^+\!(\bm{x}) \!=\! \sqrt{ \beta_{\delta} } \sigma_i^+\!(\bm{x}) \!+\! \gamma_{\delta}, \forall \bm{x} \!\in\! \mathbb{X} \} \!\ge\! 1 \!-\! \delta \nonumber
		\end{align}
		from \cref{lemma_GP_error_bound}, where $\mu_i^+(\cdot)$ denotes the posterior mean after model update on agent $i$.
		Recall the condition for the aggregation weights as $\sum_{j \in \{i, \bar{\mathcal{N}}_i \} } \omega_{ij}^+(\cdot) = 1$, the aggregated prediction error is bounded by
		\begin{align} \label{eqn_eta_underline}
			| f(\bm{x}_i) - \hat{f}_i^+(\bm{x}_i) | \le & \omega_{ii}^+(\bm{x}_i) | f(\bm{x}_i) - \mu_i^+(\bm{x}_i) | \\
			&+ \sum\nolimits_{j \in \bar{\mathcal{N}}_i } \omega_{ij}^+(\bm{x}_i) | f(\bm{x}_i) - \mu_j(\bm{x}_i) |. \nonumber 
		\end{align}
		Moreover, consider the prediction error bound for predictions from agents $j \in \bar{\mathcal{N}}_i$ according to \cref{lemma_GP_error_bound} as $\Pr \{ | f(\bm{x}_i) - \mu_j(\bm{x}_i) | \!\le\! \eta_j(\bm{x}_i), \forall \bm{x}_i \!\in\! \mathbb{X} \} \!\ge\! 1 \!-\! \delta$, the aggregated prediction error is bounded by
		\begin{align}
			| f(\bm{x}_i) - &\hat{f}_i^+(\bm{x}_i) |  \le \omega_{ii}^+(\bm{x}_i) \eta_i^+(\bm{x}_i) +\! \sum\nolimits_{j \in \bar{\mathcal{N}}_i } \omega_{ij}^+(\bm{x}_i) \eta_j(\bm{x}_i) \nonumber \\
			& \!=\! \sqrt{\! \beta_{\delta}} \big(\omega_{ii}^+(\bm{x}_i) \sigma_i^+\!(\bm{x}_i) \!+\! \sum\nolimits_{j \in \bar{\mathcal{N}}_i } \omega_{ij}^+(\bm{x}_i) \sigma_j(\bm{x}_i) \big) \!+\! \gamma_{\delta} \nonumber \\
			& \!=\! \sqrt{\! \beta_{\delta}} \hat{\sigma}_i^+(\bm{x}_i) + \gamma_{\delta} \le \sqrt{\! \beta_{\delta}} \underline{\hat{\sigma}}_i + \gamma_{\delta} = \underline{\hat{\eta}}_{\delta,i} \nonumber
		\end{align}
		with probability of at least $1 - N \delta \le 1 - |\bar{\mathcal{N}}_i| \delta$ by using the union bound in the first inequality \cite{lederer2021gaussian}.
	\end{IEEEproof}
	
	\ifarxiv
	\todo{
		\cref{proposition_online_cooperative_prediction_error_bound} shows the aggregated prediction error bound $\underline{\hat{\eta}}_{\delta,i}$ for each agent $i \in \mathcal{V}$ with online learning under the condition of aggregation weights as $\sum_{j \in \{i, \bar{\mathcal{N}}_i\}} \omega_{ij}(\cdot) \!=\! 1$.
		This condition is set to overcome the explosive prediction variance when leaving the training data \cite{liu2020gaussian}, and is commonly found in most well-known aggregation methods, such as mixture of experts (MOE, \cite{tresp2000mixtures}) with the form
		\begin{align} \label{eqn_GP_MOE}
			\omega_{ij}(\bm{x}_i) = \omega_{ij}, && \forall i \in \mathcal{V}, j \in \bar{\mathcal{N}}_i
		\end{align}
		and product of experts (POE, \cite{cao2014generalized}) with the form
		\begin{align} \label{eqn_GP_POE}
			\omega_{ij}(\bm{x}_i) = \frac{\omega_{ij}^* \sigma_j^{-2}(\bm{x}_i)}{\sum_{k \in \bar{\mathcal{N}}_i} \omega_{ik}^* \sigma_k^{-2}(\bm{x}_i)}, && \forall i \in \mathcal{V}, j \in \bar{\mathcal{N}}_i
		\end{align}
		with auxiliary constants $\omega_{ij}^* \in \mathbb{R}_{0,+}$.
		Note that the choice of $\omega_{ij}^*$ for POE can be arbitrary, i.e., not necessary to satisfy $\sum_{j \in \{i, \bar{\mathcal{N}}_i\}} \omega_{ij}^* \!=\! 1$, while the constants for MOE requires the condition of $\sum_{j \in \{i, \bar{\mathcal{N}}_i\}} \omega_{ij} \!=\! 1$.
		With the explicit formulations of the aggregation methods \eqref{eqn_GP_aggregation} as in \eqref{eqn_GP_MOE} and \eqref{eqn_GP_POE}, the solution $\underline{\hat{\sigma}}_i$ of \eqref{eqn_upperbound_posterior_variance_after_model_update} has the closed form calculated as follows.
		\begin{corollary} \label{corollary_underlin_hat_sigma_i}
			The solution $\underline{\hat{\sigma}}_{i}$ in \eqref{eqn_upperbound_posterior_variance_after_model_update} with aggregation methods MOE in \eqref{eqn_GP_MOE} and POE in \eqref{eqn_GP_POE} are derived as follows:
			\\ (\romannumeral1) For MOE,
			\begin{align} \label{eqn_upperbound_aggregated_posterior_variance_MOE}
				\underline{\hat{\sigma}}_{i} \!=\! \omega_{ii} (\sigma_{f,i}^{-2} \!+\! \sigma_{o,i}^{-2})^{-1 / 2} \!+\! \sum\nolimits_{j \in \bar{\mathcal{N}}_i} \omega_{ij} \sigma_{f,j};
			\end{align}
			(\romannumeral2) For POE,
			\begin{align} \label{eqn_upperbound_aggregated_posterior_variance_gPOE}
				\underline{\hat{\sigma}}_{i}^2  \!=\!  \frac{\sum_{j \in \{ i, \bar{\mathcal{N}}_i \}} \omega_{ij}^*}{\omega_{ii}^* (\sigma_{o,i}^{-2} \!+\! \sigma_{f,i}^{-2}) \!+\! \sum_{j \in \bar{\mathcal{N}}_i} \omega_{ij}^* \sigma_{f,j}^{-2}}.
			\end{align}
		\end{corollary}
		
		\begin{IEEEproof}
			(\romannumeral1) 
			Due to the constant aggregation weights $\omega_{ij}$ in \eqref{eqn_GP_MOE}, the monotonically increasing of $\hat{\sigma}_{i}$ w.r.t $\hat{\sigma}_i^+(\bm{x}_i)$ and $\sigma_{j}(\bm{x}_i)$ for $j \in \bar{\mathcal{N}}_i$ is obvious.
			Considering the definition of $\hat{\sigma}_i^+(\bm{x}_i)$ in \eqref{eqn_upperbound_posterior_variance_after_model_update}, it is derived $\hat{\sigma}_i^+(\bm{x}_i) \le (\sigma_{f,i}^{-2} + \sigma_{o,i}^{-2})^{-1/2}$, due to the monotonic increasing of $\hat{\sigma}_i^+(\bm{x}_i)$ w.r.t $\hat{\sigma}_i(\bm{x}_i)$ with $\hat{\sigma}_i(\bm{x}_i) \le \sigma_{f,i}$ for $\forall \bm{x}_i \in \mathbb{X}$ and positive $\kappa_i(\cdot,\cdot)$.
			Similarly, considering $\sigma_{j}(\bm{x}_i) \le \sigma_{f,j}$ from \eqref{eqn_GP_prediction} with positive definite kernel $\kappa_j(\cdot,\cdot)$, 
			the result in \eqref{eqn_upperbound_aggregated_posterior_variance_MOE} is straightforwardly obtained.
			\\
			(\romannumeral2)
			For POE, apply \eqref{eqn_GP_POE} into \eqref{eqn_GP_aggregation} and consider the definition of $\hat{\sigma}_i^+(\bm{x}_i)$ in \eqref{eqn_upperbound_posterior_variance_after_model_update}, the updated variance $\hat{\sigma}_i^+(\bm{x}_i)$ is written as
			\begin{align} \label{eqn_GPOE_posterior_error_variance}
				\hat{\sigma}_i^+(\bm{x}_i) = \frac{\omega_{ii}^* (\sigma_{i}^+(\bm{x}_i))^{-1} + \sum_{j \in \bar{\mathcal{N}}_i} \omega_{ij}^* \sigma_{j}^{-1}(\bm{x}_i) } {\omega_{ii}^* (\sigma_{i}^+(\bm{x}_i))^{-2} + \sum_{j \in \bar{\mathcal{N}}_i} \omega_{ij}^* \sigma_{j}^{-2}(\bm{x}_i)},
			\end{align}
			which is not monotonic w.r.t the posterior variance $\sigma_j(\cdot)$ from each agent $j \in \{ i, \bar{\mathcal{N}}_i \}$.
			Then, applying the Cauchy-Schwarz inequality on the numerator of \eqref{eqn_GPOE_posterior_error_variance}, it yields
			\begin{align}
				\Big(\!\!\!\!\!\sum_{j \in \{i, \bar{\mathcal{N}}_i\}} \!\!\!\! \omega_{ij}^* \sigma_j^{-1}(\bm{x}_i) \!\Big)^2 \!\!&\le\!\! \Big(\!\!\!\!\!\sum_{j \in \{i, \bar{\mathcal{N}}_i\}} \!\!\! \omega_{ij}^* \!\Big) \Big(\!\!\!\!\!\sum_{j \in \{i, \bar{\mathcal{N}}_i\}} \!\!\! \omega_{ij}^* \sigma_j^{-2}(\bm{x}_i) \!\Big),
			\end{align}
			such that $\hat{\sigma}_i(\bm{x}_i)$ in \eqref{eqn_GPOE_posterior_error_variance} is bounded by
			\begin{align}
				\hat{\sigma}_i(\bm{x}_i) \!\le\! \Big(\!\!\!\! \sum_{j \in \{ i, \bar{\mathcal{N}}_i \!\}} \!\!\! \omega_{ij}^* \Big)^{\!1\!/\!2\!} \!\Big(\! \omega_{ii}^* (\sigma_{j}^+(\bm{x}_i))^{\!-\!2\!} \!+\!\!\!\!\!\! \sum_{j \in \{i, \bar{\mathcal{N}}_i\}} \!\!\!\!\omega_{ij}^* \sigma_{j}^{\!-\!2\!}(\bm{x}_i) \!\Big)^{\!-\!1\!/\!2\!}, \nonumber
			\end{align}
			where the right-hand side is monotonically increasing w.r.t $\sigma_{j}(\bm{x}_i)$ for $j \in \{ i, \bar{\mathcal{N}}_i \}$.
			Therefore, the supremum of the right hand side is achieved when $\hat{\sigma}_i^+(\bm{x}) = (\sigma_{f,i}^{-2} + \sigma_{o,i}^{-2})^{-1/2}$ and $\sigma_{j}(\bm{x}) = \sigma_{f,j}$ for $j \in \bar{\mathcal{N}}_i$, leading to the result in \eqref{eqn_upperbound_aggregated_posterior_variance_gPOE}.
		\end{IEEEproof}  
		\cref{corollary_underlin_hat_sigma_i} shows exact expression of $\underline{\hat{\sigma}}_i$ with the aggregation methods using MOE \eqref{eqn_GP_MOE} and POE \eqref{eqn_GP_POE}.
		Note that the prediction error bound of online cooperative learning is non-zero due to the presence of measurement noise $\sigma_{o,i}$ and uncontrollable prediction performance from neighboring agents $j \!\in\! \bar{\mathcal{N}}_i$ resulting in conservative estimation by using $\sigma_{f,j}$.
		\looseness=-1
		
		\begin{remark} \label{remark_POE_upperbound}
			For POE, the upper bound of $\hat{\sigma}_i$ is not directly derived by using the expression in \eqref{eqn_GPOE_posterior_error_variance}, since the monotonic w.r.t each posterior variance $\sigma_i^+(\cdot)$ and $\sigma_j(\cdot), \forall j \in \bar{\mathcal{N}}_i$ cannot be ensured.
			Specifically, given a fixed $\sigma_i^+(\cdot)$ and suppose $\sigma_{f,j}(\cdot)$ are sufficiently large (compared to $\sigma_{o,i}$), then the supremum of $\hat{\sigma}_i(\cdot)$ achieves when
			\begin{align}
				&\sum{_{j \in \bar{\mathcal{N}}_i}} \omega_{ij}^* \sigma_{f,j}^{-2}(\cdot) = \sum{_{j \in \bar{\mathcal{N}}_i}} \omega_{ij}^* \sigma_{\text{aux},i}^{-2}(\cdot), \\
				&\sigma_{\text{aux},i}(\cdot) = \Big( 1 + \sqrt{1 + (\omega_{ii}^*)^{-1} \sum{_{j \in \bar{\mathcal{N}}_i}} ~ \omega_{ij}^*  } \Big) \sigma_i^+(\cdot), \nonumber
			\end{align}
			where the expression of $\sigma_{\text{aux},i}(\cdot)$ is derived by taking the derivative of $\partial \hat{\sigma}_i / \partial \sigma_{\text{aux},i}$ to be zero.
			Note that this result is not reasonable by simply considering the case with $\sigma_{f,j}(\cdot) = \sigma_{f,k}(\cdot), \forall j,k \in \bar{\mathcal{N}}_i$, since it says less accurate neighboring predictions with $\sigma_{f,j}(\cdot) > \sigma_{\text{aux},i}(\cdot)$ result in better aggregated prediction reflected by smaller $\hat{\sigma}_i(\cdot)$.
			In comparison, the derived upper bound in \eqref{eqn_upperbound_aggregated_posterior_variance_gPOE} inherits the aggregated posterior variance in \cite{cao2014generalized}, indicating the aggregation more accurate predictions improve the cooperative learning.
		\end{remark}
	}
	\fi
	
	Moreover, \cref{proposition_online_cooperative_prediction_error_bound} also allows the derivation of the overall prediction error bound using online cooperative learning, which is shown as follows.
	
	\begin{corollary} \label{corollary_overall_prediction_error_bound}
		Let all assumptions in \cref{proposition_online_cooperative_prediction_error_bound} hold, then the concatenated prediction error for all agents $i \in \mathcal{V}$ is bounded by
		\begin{align}
			\| \hat{\bm{f}}^+(\bm{x}) - \bm{f}(\bm{x}) \| \le \| \underline{\hat{\bm{\eta}}}_{\delta} \| = \| [\underline{\hat{\eta}}_{\delta,1}, \cdots, \underline{\hat{\eta}}_{\delta,N}]^T \| \nonumber
		\end{align}
		with probability of at least $1 - N^2 \delta$, where $\hat{\bm{f}}^+(\bm{x}) = [\hat{f}^+_1(\bm{x}_1), \cdots, \hat{f}^+_N(\bm{x}_N)]^T$ and $\bm{f}(\bm{x}) = [f(\bm{x}_1), \cdots, f(\bm{x}_N)]^T$.
	\end{corollary}
	\begin{IEEEproof}
		The overall prediction error is written as
		\begin{align}
			\| \hat{\bm{f}}^+(\bm{x}) \!-\! \bm{f}(\bm{x}) \|^2 \!=\! \sum_{i \in \mathcal{V}} | f(\bm{x}_i) \!-\! \hat{f}_i^+(\bm{x}_i) |^2 \le \sum_{i \in \mathcal{V}} \underline{\hat{\eta}}_{\delta,i}^2 \!=\! \| \underline{\hat{\bm{\eta}}}_{\delta} \|^2 \nonumber
		\end{align}
		with probability of at least $1 - N^2 \delta$ due to the union bound, which concludes the proof.
	\end{IEEEproof}
	\cref{corollary_overall_prediction_error_bound} shows the overall prediction error bound, which is used to determine the control performance, i.e., tracking error bound, in \cref{subsection_distributed_control}.
	\ifarxiv
	\todo{
		Moreover, to achieve the guaranteed prediction performance as shown in \cref{proposition_online_cooperative_prediction_error_bound} and \cref{corollary_overall_prediction_error_bound}, a naive methods is to add every samples into the training data set \cite{nguyen2008local, beckers2021online}.
		However, this ongoing data collection gives rise to substantial computational demands and data storage needs, inducing low data efficiency. 
		To address this limitation, a judicious data selection strategy is adopted, incorporating an event-triggered mechanism, as outlined in \cref{section_event_triggered_learning}.
	}
	\fi
	\looseness=-1
	
	\subsection{Control Performance Analysis} \label{subsection_distributed_control}
	
	In this subsection, the control performance, in particular the ultimate boundness of the formation error $\bm{\vartheta}$, is analyzed through Lyapunov theorem for the MAS with \eqref{eqn_agent_dynamics} controlled by the distributed controller \eqref{eqn_control_law} with cooperative learning in \eqref{eqn_GP_aggregation}.
	With the time-varying data sets and GP models due to the online learning, the control law \eqref{eqn_control_law} leads to a switching controller.
	This results in a closed-loop hybrid system, whose stability is analyzed through common Lyapunov theorem.
	For a high order MAS in \eqref{eqn_agent_dynamics}, the common Lyapunov candidate $V$ is usually chosen as
	\begin{align} \label{eqn_V}
		V = V_1 + V_2 && \text{with} && V_1 = \bm{r}^T \bm{P}_r \bm{r}, && V_2 = \bm{\varepsilon}^T \bm{P}_{\varepsilon} \bm{\varepsilon},
	\end{align}
	where $\bm{\varepsilon} = [\bm{\varepsilon}_1^T, \cdots, \bm{\varepsilon}_{n-1}^T]^T \in \mathbb{R}^{(n-1)N}$ is defined as the concatenated synchronization error with $\bm{\varepsilon}_k = [e_{1,k}, \cdots, e_{N,k}]^T \in \mathbb{R}^{N}, \forall k = 1, \cdots, n$.
	The positive definite matrix $\bm{P}_r$ for $\bm{r}$ is chosen as $\bm{P}$ in \cref{lemma_MAS_Lyapunov} such that $\bm{Q}_r = \bm{P}_r (\bm{\mathcal{L}} + \bm{\mathcal{B}}) + (\bm{\mathcal{L}} + \bm{\mathcal{B}})^T \bm{P}_r$.
	And the weight matrix $\bm{P}_{\varepsilon}$ for $\bm{\varepsilon}$ is calculated by $\bm{P}_{\varepsilon} = \bm{P}_{\varepsilon,s} \otimes \bm{I}_N$, where the positive definite matrix $\bm{P}_{\varepsilon,s}$ is the solution of the continuous Lyapunov equation $\bm{\Lambda}^T \bm{P}_{\varepsilon,s} + \bm{P}_{\varepsilon,s} \bm{\Lambda} = - \bm{Q}_{\varepsilon,s}$ for a given symmetric positive definite $\bm{Q}_{\varepsilon,s}$.
	Note that the existence and uniqueness of $\bm{P}_{\varepsilon,s}$ are guaranteed by considering $\bm{\Lambda}$ in \eqref{eqn_Lambda} as Hurwitz.
	Moreover, define $\bm{z} = [ \bm{r}^T , \bm{\varepsilon}^T ]^T$ and $\bm{P}_z = \mathrm{blkdiag}(\bm{P}_r, \bm{P}_{\varepsilon})$, then the Lyapunov candidate in \eqref{eqn_V} is reformulated as
	\begin{align} \label{eqn_V_in_z}
		V = \bm{z}^T \bm{P}_z \bm{z}.
	\end{align}
	Owing to the unknown system dynamics $f(\cdot)$ and considering a non-zero prediction error bound as specified in \cref{proposition_online_cooperative_prediction_error_bound}, it is intuitive that the overall tracking error denoted as $\bm{\vartheta} = \bm{x} - \bm{s} - \bm{1}_N \otimes \bm{x}_l$ cannot be nullified, instead it is upper bounded by $\bar{\vartheta}$ described in \cref{subsection_system_description}.
	This also leads to a non-zero guaranteed lower bound of $V$, confines the variable $\bm{z}$ within certain bounds considering \cref{eqn_V_in_z}.
	Specifically, the relationship between $\bm{z}$ and $\bar{\vartheta}$ is shown in the following lemma.
	
	\begin{lemma} \label{lemma_boundness_z_vartheta}
		Consider the MAS \eqref{eqn_agent_dynamics} connected through a communication network satisfying \cref{assumption_topology} and controlled by \eqref{eqn_control_law}.
		Choose the Lyapunov candidate as \eqref{eqn_V} and
		\begin{align} \label{eqn_chi}
			\chi = \| (\bm{\mathcal{L}} \!+\! \bm{\mathcal{B}})^{-1} \|\sqrt{ (1 \!+\! \|[\bm{t}, \bm{\Lambda}]\|^2) \underline{\lambda}^{-1}(\bm{P}_z) \bar{\lambda}(\bm{P}_z) } .
		\end{align}
		If there exists a positive constant $\bar{z} \in \mathbb{R}_{0,+}$, such that the negativity of $\dot{V}$ is shown for $\forall \| \bm{z} \| > \bar{z}$.
		Choose $\bar{\vartheta} = \chi \bar{z}$, then the tracking error is ultimately bounded by $\bar{\vartheta}$, i.e., $\| \bm{\vartheta} \| \le \bar{\vartheta}$.
	\end{lemma}
	\begin{IEEEproof}
		Considering the identical leader for each agent, the synchronization error in \eqref{eqn_synchronization_error} is reformulated as
		\begin{align}
			e_{i,k} = \sum\nolimits_{j \in \mathcal{N}_i} a_{ij} (\vartheta_{i,k} - \vartheta_{j,k}) + b_{ii} \vartheta_{i,k}, \nonumber
		\end{align}
		such that their concatenations $\bm{\varepsilon}_k$ for all dimensions $k \in 1, \!\cdots\! ,n-1$ are written as
		\begin{align} \label{eqn_epsilon_theta}
			\bm{\varepsilon}_k = (\bm{\mathcal{L}} + \bm{\mathcal{B}}) \tilde{\bm{\vartheta}}_k, &&
			[\bm{\varepsilon}^T, \bm{\varepsilon}_n^T]^T = (\bm{I}_n \otimes (\bm{\mathcal{L}} + \bm{\mathcal{B}})) \tilde{\bm{\vartheta}}
		\end{align}
		with $\tilde{\bm{\vartheta}} \!=\! [\tilde{\bm{\vartheta}}_1^T, \cdots, \tilde{\bm{\vartheta}}_n^T]^T$, $\tilde{\bm{\vartheta}}_k \!=\! [\vartheta_{1,k}, \cdots, \vartheta_{N,k}]^T$ and $\bm{\varepsilon}_n = [e_{1,n}, \cdots, e_{N,n}]^T$. 
		Since $\tilde{\bm{\vartheta}}$ and $\bm{\vartheta}$ are both composed of $\vartheta_{i,k}$ leading to $\| \tilde{\bm{\vartheta}} \| = \| \bm{\vartheta} \|$, the norm of $\| \bm{\vartheta} \|$ is bounded using the result in \eqref{eqn_epsilon_theta} as
		\begin{align} \label{eqn_vartheta_bound}
			\| \bm{\vartheta} \| = \| \tilde{\bm{\vartheta}} \| \le \| (\bm{\mathcal{L}} + \bm{\mathcal{B}})^{-1} \| \| [\bm{\varepsilon}_1^T, \dot{\bm{\varepsilon}}^T] \|,
		\end{align}
		considering $\| [\bm{\varepsilon}^T, \bm{\varepsilon}_n^T]^T \|  = \| [\bm{\varepsilon}_1^T, \dot{\bm{\varepsilon}}^T] \|$.
		Combining the definition of the filtered error $\bm{r}$ in \eqref{eqn_V} and synchronization error $e_{i,k}$ in \eqref{eqn_synchronization_error}, the derivative of $\bm{\varepsilon}$ denotes
		\begin{align} \label{eqn_dot_varepsilon}
			\dot{\bm{\varepsilon}} = ( \bm{\Lambda} \otimes \bm{I}_N ) \bm{\varepsilon} + ( \bm{t} \otimes \bm{I}_N ) \bm{r} = ([\bm{t}, \bm{\Lambda}] \otimes \bm{I}_N ) \bm{z}
		\end{align}
		with $\bm{t} = [0, \cdots, 0, \lambda_n^{-1}]^T \in \mathbb{R}^{n-1}$, such that the overall tracking error in \eqref{eqn_vartheta_bound} is further bounded by
		\begin{align}
			\| \bm{\vartheta} \|^2 &\le \| (\bm{\mathcal{L}} + \bm{\mathcal{B}})^{-1} \|^2 (\| \bm{\varepsilon}_1 \|^2 + \| \dot{\bm{\varepsilon}} \|^2) \nonumber \\
			&\le \| (\bm{\mathcal{L}} + \bm{\mathcal{B}})^{-1} \|^2 (1 + \| [\bm{t}, \bm{\Lambda}] \|^2 ) \| \bm{z} \|^2 \nonumber
		\end{align}
		considering $\| \bm{\varepsilon}_1 \| \le \| \bm{z} \|$, since $\bm{\varepsilon}_1$ is only one part of $\bm{z}$.
		Furthermore, considering that $V$ is in quadratic form as in \eqref{eqn_V_in_z} inducing $V \!\ge\! \underline{\lambda}(\bm{P}_z) \| \bm{z} \|^2$, the relationship between $\| \bm{\vartheta} \|$ and $V$ denotes
		\begin{align} \label{eqn_relationship_vartheta_V}
			\| \bm{\vartheta} \| &\le \| (\bm{\mathcal{L}} + \bm{\mathcal{B}})^{-1} \| \sqrt{(1 + \| [\bm{t}, \bm{\Lambda}] \|^2 ) \underline{\lambda}^{-1}(\bm{P}_z) V}.
		\end{align}
		Next, we show the upper bound of $V$. 
		Considering the case when $V \!>\! \bar{\lambda} (\bm{P}_z) \chi^{-2} \bar{\vartheta}^2$, where $\| \bm{z} \| \!>\! \chi^{-1} \bar{\vartheta} \!=\! \bar{z}$ is directly derived due to the fact that $V \!\le\! \bar{\lambda} (\bm{P}_z) \| \bm{z} \|^2$.
		Moreover, according to the setting in the lemma, $\| \bm{z} \| \!>\! \bar{z}$ leads to $\dot{V} \!<\! 0$, i.e., the decrease of $V$.
		Therefore, $V$ is ultimately bounded as $V \le \bar{\lambda} (\bm{P}_z) \chi^{-2} \bar{\vartheta}^2$.
		Apply the boundness of $V$ into \eqref{eqn_relationship_vartheta_V}, then the result in the lemma is straightforwardly derived.
	\end{IEEEproof}
	
	\cref{lemma_boundness_z_vartheta} shows the overall tracking error $\bm{\vartheta}$ is bounded, if the positive value $\bar{z}$ exists and is well defined such that $\dot{V} < 0$ when $\| \bm{z} \| > \bar{z}$.
	\ifarxiv
	\todo{
		The linear relationship between $\bar{z}$ and $\bar{\vartheta}$ denoted by $\chi$ in \eqref{eqn_chi} is related to the choice of $\lambda_k, \forall k = 1, \cdots, n$ and the communication topology $\mathcal{G}$ reflected by $\bm{\mathcal{L}}$ and $\bm{\mathcal{B}}$.
		It is obvious stronger connectivity, inducing larger singular values of the augmented graph characterized by $(\bm{\mathcal{L}} + \bm{\mathcal{B}})$, shrinks the guaranteed tracking error bound $\bar{\vartheta}$ with given $\bar{z}$, which is also revealed in \cite{lederer2022cooperative}.
		This provides a possibility to reduce the tracking error enhance through enhancing the connection among agents and the leader.
		However, increasing the connectivity cannot achieve an arbitrarily small tracking error bound, considering the minimum of $\chi$ achieved with a fully connected graph with still non-zero $\| (\bm{\mathcal{L}} + \bm{\mathcal{B}})^{-1} \|$.
		Another way is to decrease the upper bound $\bar{z}$, but at first we investigate the detailed expression of $\bar{z}$ for the controlled system \eqref{eqn_agent_dynamics} with distributed control \eqref{eqn_control_law} with cooperative online learning \eqref{eqn_GP_aggregation}.
	}
	\fi
	The expression of $\bar{z}$ and then $\bar{\vartheta}$ is derived by observing the time derivative of $V$ in \eqref{eqn_V} and using \cref{lemma_boundness_z_vartheta}, whose result is shown in the following theorem.
	
	\begin{theorem} \label{theorem_best_tracking_performance_online}
		Consider the control of MAS \eqref{eqn_agent_dynamics} with \cref{assumption_desired_trajectory} under a communicate graph satisfying \cref{assumption_topology}.
		The compensation $\hat{f}_i^+(\cdot)$ is obtained by aggregation online GP predictions satisfying \cref{assumption_dataset,assumption_dataset} and $\sum_{j \in \{i, \bar{\mathcal{N}}_i \}} \omega_{ij}(\cdot) \!=\! 1$.
		Pick $\delta \in (0,1 / N^2)$ and choose $c \in \mathbb{R}_+$ in \eqref{eqn_control_law}, such that $\bm{Q}_z \in \mathbb{R}^{nN \times nN}$ is positive definite defined as
		\begin{align} \label{eqn_Qz}
			\bm{Q}_z = \begin{bmatrix}
				c \lambda_n \bm{Q}_r - 2 \lambda_n \lambda_{n-1}^{-1} \bm{P}_r & - \bm{\Psi} \\
				- \bm{\Psi}^T & \bm{Q}_{\varepsilon}
			\end{bmatrix}
		\end{align}
		with $\bm{\lambda} =\! [\lambda_1, \cdots, \lambda_{n-1}]^T$, $\bm{\Psi} = \bm{P}_r (\bm{\lambda}^T \bm{\Lambda} \otimes \bm{I}_N ) + (\bm{t}^T \bm{P}_{\varepsilon,s}) \otimes \bm{I}_N$ and $\bm{Q}_{\varepsilon} = \bm{Q}_{\varepsilon,s} \otimes \bm{I}_N$. 
		Then, the tracking error $\| \bm{\vartheta} \|$ is bounded by $\bar{\vartheta} =  \chi \xi \| \bm{\iota} + \underline{\hat{\bm{\eta}}}_{\delta} \|$ with a probability of at least $1 - N^2 \delta$, where $\bm{\iota} \!=\! [\iota_1, \!\cdots\!, \iota_N]^T \!\in\! \mathbb{R}^N$,  $\iota_i \!=\! (1 \!-\! b_{ii}) F_l$, $\forall i \!=\! 1, \cdots, N$ with $F_l$ defined in \cref{assumption_desired_trajectory} and
		\begin{align} \label{eqn_xi}
			\xi = 2 \lambda_n \underline{\lambda}^{-1}(\bm{Q}_z) \| \bm{P}_r (\bm{\mathcal{L}} \!+\! \bm{\mathcal{B}}) \|.
		\end{align}
	\end{theorem}
	\begin{IEEEproof} 
		The tracking error bound $\bar{\vartheta}$ is derived through common Lyapunov theory, where the sign of $\dot{V} = \dot{V}_1 + \dot{V}_2$ from \eqref{eqn_V} is investigated.
		First, we show the time derivative of $V_1$ in \eqref{eqn_V}, which is written as $\dot{V}_1 = \dot{\bm{r}}^T \bm{P}_r \bm{r} + \bm{r}^T \bm{P}_r \dot{\bm{r}}$, where
		\begin{align}
			\dot{\bm{r}} = \sum\nolimits_{k=1}^{n-1} \lambda_k \dot{\bm{\varepsilon}}_k + \lambda_n \dot{\bm{\varepsilon}}_n = ( \bm{\lambda}^T \!\otimes\! \bm{I}_N ) \dot{\bm{\varepsilon}} \!+\! \lambda_n \dot{\bm{\varepsilon}}_n. \nonumber
		\end{align}
		Considering the definition of $\bm{\varepsilon}_n$ with the synchronization errors in \eqref{eqn_synchronization_error} and the system dynamics \eqref{eqn_agent_dynamics}, the derivative of $\bm{\varepsilon}_n$ is formulated as
		\begin{align} \label{eqn_dotE}
			\dot{\bm{\varepsilon}}_n  = ( \bm{\mathcal{L}} + \bm{\mathcal{B}} ) (\bm{f}(\bm{x}) + \bm{G}(\bm{x}) \bm{u} - \bm{s}_r - \bm{1}_N x_{l,r}), 
		\end{align}
		where $\bm{f}(\bm{x}) = [f(\bm{x}_1), \!\cdots,\! f(\bm{x}_N)]^T$.
		Applying the consensus law $\phi(\cdot)$ in \eqref{eqn_consensus_control_law} and compensation $\hat{\bm{f}}^+(\cdot)$ into \eqref{eqn_control_law}, the concatenated control input $\bm{u} = [u_1, \cdots, u_N]^T$ is written as
		\begin{align} \label{eqn_concatenated_control_input}
			\bm{u} = \bm{G}^{-1}(\bm{x}) ( - c \bm{r} + \bm{s}_r + \bm{\mathcal{B}} \bm{1}_N x_{l,r} - \hat{\bm{f}}^+(\bm{x}) ),
		\end{align}
		where $\bm{r} = [r_1, \cdots, r_N]^T$, $\bm{s}_r \!=\! [s_{r,1}, \!\cdots\!, s_{r,N}]^T$ and $\bm{G}(\bm{x}) \!=\! \mathrm{diag}(g(\bm{x}_1), \!\cdots\!, g(\bm{x}_N))$.
		Moreover, taking $\bm{u}$ from \eqref{eqn_concatenated_control_input} and $\dot{\bm{\varepsilon}}$ from \eqref{eqn_dot_varepsilon} into \eqref{eqn_dotE}, the derivative of $\bm{r}$ is written as
		\begin{align}
			\dot{\bm{r}} =  ( \bm{\lambda}^T \bm{t} \otimes \bm{I}_N &- c \lambda_n (\bm{\mathcal{L}} + \bm{\mathcal{B}}) ) \bm{r} \nonumber \\
			&+ ( \bm{\lambda}^T \bm{\Lambda} \otimes \bm{I}_N ) \bm{\varepsilon} + \lambda_n (\bm{\mathcal{L}} + \bm{\mathcal{B}}) \bm{\psi}^+, \nonumber
		\end{align}
		where $\bm{\psi}^+ \!=\! (\bm{\mathcal{B}} \!-\! \bm{I}_N) \bm{1}_N x_{l,r} \!+\! \bm{f}(\bm{x}) \!-\! \hat{\bm{f}}^+(\bm{x})$.
		Due to the definition of $\bm{Q}_r$ and the fact that $\bm{\lambda}^T \bm{t} = \lambda_n \lambda_{n-1}^{-1}$, the derivative of $V_1$ is written as
		\begin{align} \label{eqn_dot_V1}
			\dot{V}_1 = - \bm{r}^T ( &c \lambda_n \bm{Q}_r - 2 \lambda_n \lambda_{n-1}^{-1} \bm{P}_r ) \bm{r} \\
			&+ 2 \bm{r}^T \bm{P}_r \big( (\bm{\lambda}^T \bm{\Lambda} \otimes \bm{I}_N ) \bm{\varepsilon} + \lambda_n (\bm{\mathcal{L}} + \bm{\mathcal{B}}) \bm{\psi}^+ \big). \nonumber
		\end{align}
		Moreover, considering $\dot{\bm{\varepsilon}}$ in \eqref{eqn_dot_varepsilon} and \cref{lemma_MAS_Lyapunov}, the derivative of $V_2$ in \eqref{eqn_V} is straightforwardly derived as follows
		\begin{align} \label{eqn_dot_V2}
			\dot{V}_2 = - \bm{\varepsilon}^T \bm{Q}_{\varepsilon} \bm{\varepsilon} + 2 \bm{\varepsilon}^T ( \bm{P}_{\varepsilon,s} \bm{t} \otimes \bm{I}_N ) \bm{r}.
		\end{align}
		Combining with \eqref{eqn_dot_V1} and \eqref{eqn_dot_V2} and using the definition of $\bm{z}$, the derivative of $V$ is written as
		\begin{align} \label{eqn_dotV_trigger_prepare}
			\dot{V} = & - \bm{z}^T \bm{Q}_{z} \bm{z} + 2 \lambda_n \bm{r}^T \bm{P}_r (\bm{\mathcal{L}} \!+\! \bm{\mathcal{B}}) \bm{\psi}^+ \nonumber \\
			\le & - \underline{\lambda}(\bm{Q}_z) \| \bm{z} \|^2 + 2 \lambda_n \| \bm{z} \| \| \bm{P}_r (\bm{\mathcal{L}} \!+\! \bm{\mathcal{B}}) \| \| \bm{\psi}^+ \|\\
			\le &- \underline{\lambda}(\bm{Q}_z) \| \bm{z} \| (\| \bm{z} \| - \xi \| \bm{\psi}^+ \|), \nonumber 
		\end{align}
		where the first and second inequalities are satisfied due to the positive definite of $\bm{Q}_{z}$ from the choice of $c$ in \eqref{eqn_Qz} and the definition of $\xi$ in \eqref{eqn_xi}, respectively.
		Because of the uncertainty of unknown function $f(\cdot)$, it is not possible to compute the exact norm of $\bm{\psi}^+$.
		Instead, the upper bound of $\| \bm{\psi}^+ \|$ can be derived using both the triangular inequality and the prediction error bound described in \cref{corollary_overall_prediction_error_bound}.
		Specifically, the norm of $\bm{\psi}^+$ is bounded under \cref{assumption_desired_trajectory} and $b_{ii} \le 1, \forall i \in \mathcal{V}$ as
		\begin{align} \label{eqn_bound_Upsilon}
			\| \bm{\psi}^+ \| &\le \| (\bm{I}_N - \bm{\mathcal{B}}) \bm{1}_N | x_{l,r} | + | \bm{f}(\bm{x}) - \hat{\bm{f}}^+(\bm{x}) | \| \\
			&\le \| (\bm{I}_N - \bm{\mathcal{B}}) \bm{1}_N F_l + \hat{\bm{\eta}}_{\delta}^+(\bm{x}) \| = \| \bm{\iota} + \hat{\bm{\eta}}_{\delta}^+(\bm{x}) \| \nonumber
		\end{align}
		with a probability of at least $1 - N^2 \delta$ using the union bound, where $\hat{\bm{\eta}}_{\delta}^+(\bm{x}) = [\hat{\eta}_{\delta,1}^+(\bm{x}_1), \cdots, \hat{\eta}_{\delta,N}^+(\bm{x}_N)]^T$ with $\hat{\eta}_{\delta,i}^+(\bm{x}_i) = \sqrt{\beta_{\delta}} \sigma_i^+(\bm{x}_i) + \gamma_{\delta}$ and $\sigma_i^+(\bm{x}_i)$ defined in \cref{proposition_online_cooperative_prediction_error_bound}. 
		The vector $\bm{\iota} = (\bm{I}_N - \bm{\mathcal{B}}) \bm{1}_N F_l$ with positive entries denotes the leader misconnection term with the bounded $|x_{l,r}|$ by $F_l$ from \cref{assumption_desired_trajectory}.
		Moreover, considering the prediction error bound of the cooperative online learning in \cref{proposition_online_cooperative_prediction_error_bound}, the norm of $\bm{\psi}^+$ in \eqref{eqn_bound_Upsilon} is further bounded by $\| \bm{\psi}^+ \| \!\le\! \| \bm{\iota} \!+\! \underline{\hat{\bm{\eta}}}_{\delta} \|$, leading to the probabilistic bound as
		\begin{align} \label{eqn_dot_V3}
			\dot{V} \le - \underline{\lambda}(\bm{Q}_z) \| \bm{z} \| (\| \bm{z} \| - \xi \| \bm{\iota} + \underline{\hat{\bm{\eta}}}_{\delta} \|).
		\end{align}
		Defining $\bar{z}$ in \cref{lemma_boundness_z_vartheta} as $\bar{z} = \xi \| \bm{\iota} + \underline{\hat{\bm{\eta}}}_{\delta} \|$, the negativity of $\dot{V}$ is achieved when $\| \bm{z} \| > \bar{z}$ from \eqref{eqn_dot_V3}.
		Then, apply the result in \cref{lemma_boundness_z_vartheta}, the boundness of $\| \bm{\vartheta} \|$ in \cref{theorem_best_tracking_performance_online} is derived, inheriting the probability of at least $1 - N^2 \delta$.
	\end{IEEEproof}	
	
	\cref{theorem_best_tracking_performance_online} shows the tracking error bound $\bar{\vartheta}$ under the distributed control with cooperative online learning strategy.
	The tracking error bound $\bar{\vartheta}$ also reflected by $\bar{z}$ is related to both the prediction accuracy and connectivity with the leader reflected by $\bm{\iota} + \underline{\hat{\bm{\eta}}}_{\delta}$ with additional coefficient $\xi$.
	Despite the non-zero $\underline{\hat{\bm{\eta}}}_{\delta}$ from \cref{corollary_overall_prediction_error_bound} and non-zero $\bm{\iota}$ from leader mis-connection, the arbitrary small tracking error bound $\bar{\vartheta}$ can be achieved by choosing sufficiently large control gain $c$ such that $\underline{\lambda}(\bm{Q}_z)$ is large inducing small $\xi$ and thus small $\bar{z}$.
	\ifarxiv
	\todo{
		This conclusion is not contradictive to the previous for \cref{lemma_boundness_z_vartheta}, where $\chi$ cannot be arbitrary small.
		Moreover, considering the definition of $\xi$ in \eqref{eqn_xi}, increasing the topology connectivity for smaller $\chi$ may not efficient since $\xi$ is increasing with $\| \bm{\mathcal{L}} +\bm{\mathcal{B}} \|$, resulting in opposite effects on $\bar{\vartheta}$.
	}
	\fi
	
	However, \cref{theorem_best_tracking_performance_online} requires $\hat{\eta}_{\delta,i}(\cdot)$ to be smaller than $\underline{\hat{\eta}}_{\delta,i}$ for each agent $i \in \mathcal{V}$ without specifying the data collection strategy.
	\ifarxiv
	\todo{
		As discussed previously, collecting every new data pair is inefficient inducing large requirements on the local computation and data storage.
	}
	\fi
	In the next section, smart data selection strategies are proposed, aiming to enhance data efficiency while maintaining the desired control performance.

	\section{Event-triggered Online Learning}
	\label{section_event_triggered_learning}
	
	This section delves into the development of intelligent online data selection strategies for efficient training data storage, leveraging event-triggered mechanisms. 
	It commences with a centralized event-triggered approach in \cref{subsection_centralized_ET}, followed by the distributed event-trigger in \cref{subsection_distributed_ET}, accompanied by the theoretical performance guarantee. 
	Furthermore, in both centralized and distributed event-triggered cooperative online learning, exclusion of Zeno behavior is shown in \cref{subsection_Zeno_behavior}.
	
	Note that with event-triggered online learning mechanism, only GP models on some of agents will be updated at time $t$.
	For notational simplicity, define a time related index function $\varpi_i(\cdot): \mathbb{R}_{0,+} \to \{0,1\}$ for agent $i \in \mathcal{V}$ as
	\begin{align} \label{eqn_varpi}
		\varpi_i(t) = \begin{cases}
			1, & \text{if agent $i$ is updated at $t$} \\
			0, & \text{otherwise}
		\end{cases}.
	\end{align}
	With the index function $\varpi_i(\cdot)$, the aggregated prediction error bound for agent $i$ is defined as $\tilde{\eta}_{\delta,i}(\cdot)$with expression
	\begin{align} \label{eqn_tilde_eta}
		\tilde{\eta}_{\delta,i}(\bm{x}_i(t)) \!=\! \varpi_i(t) \hat{\eta}_{\delta,i}^+(\bm{x}_i(t)) \!+\! (1 \!-\! \varpi_i(t)) \hat{\eta}_{\delta,i}(\bm{x}_i(t))
	\end{align}
	for all $i \in \mathcal{V}$, where $\hat{\eta}_{\delta,i}(\cdot)$ and $\hat{\eta}_{\delta,i}^+(\cdot)$ are recalled as the error bound with and without online learning, respectively.
	Moreover, the concatenated prediction error bound denotes $\tilde{\bm{\eta}}_{\delta}(\bm{x}) = [\tilde{\eta}_{\delta,1}(\bm{x}_1), \cdots, \tilde{\eta}_{\delta,N}(\bm{x}_N)]^T$.
	
	\subsection{Centralized Event-triggered Online Learning}\label{subsection_centralized_ET}
	
	\ifarxiv
	\todo{
		In the centralized event-triggered scenario, the global information, i.e., the states of all agents and leader, is utilized for the evaluation of the trigger condition in \eqref{eqn_fake_trigger}. 
		This approach is typically conceptualized with the establishment of a centralized node, which collects global information and then transmits the online learning decision to each relevant agent.
		With the global information, the design of the centralized event-trigger follows the way for single agent.
		Building upon the previous research on event-triggered online learning for GP in \cite{umlauft2019feedback, jiao2022backstepping, dai2023can}, a centralized event-trigger mechanism is devised for cooperative online learning. 
		To maintain a predefined tracking error bound $\bar{\vartheta}_c \!>\! \bar{\vartheta}$ with $\bar{\vartheta}$ in \cref{theorem_best_tracking_performance_online}, the centralized trigger function and its threshold are designed as
		\looseness=-1
		\begin{align} \label{eqn_centralized_trigger}
			\rho = \| \bm{\iota} + \hat{\bm{\eta}}_{\delta}(\bm{x}) \|, && 
			\bar{\rho} = \xi^{-1} \!\max \{ \| \bm{z} \|, \chi^{-1} \bar{\vartheta}_c \}.
		\end{align}
		where the time input in \eqref{eqn_centralized_trigger} is dropped, i.e., $\rho(t) \!\to\! \rho$, for notational simplicity under the assumption of all the employed variables in the same time instances.
		Note that the condition $\bar{\vartheta}_c \!>\! \bar{\vartheta}$ indicates there exist constants $\epsilon_i \!\in\! \mathbb{R}_+$ selected for each agent $i \!\in\! \mathcal{V}$, such that $\bar{\vartheta}_c$ is reformulated as $\bar{\vartheta}_c \!=\! \xi \chi \| \bm{\iota} \!+\! \underline{\hat{\bm{\eta}}}_{\delta} + \bm{\epsilon} \|$ with $\bm{\epsilon} \!=\! [\epsilon_1, \!\cdots\!, \epsilon_N]^T$.
		The strictly positive $\epsilon_i$ excludes the Zeno behavior on agent $i$, which is shown in \cref{subsection_Zeno_behavior}.
		
		Note that the maximum operator in $\bar{\rho}(\cdot)$ divides the operation status into two cases.
		The case with $\| \bm{z} \| > \chi^{-1} \bar{\vartheta}_c$ indicating $\bar{\rho}(t) = \xi^{-1} \| \bm{z}(t) \|$ serves transient phase, since $\| \bm{\vartheta} \| > \bar{\vartheta}_c$ is also derived by using the result in \cref{lemma_boundness_z_vartheta}.
		The event-trigger condition in transient phase ensures the reduction of tracking error $\| \bm{\vartheta} \|$, when the tracking error is larger than the desired bound $\bar{\vartheta}_c$.
		In the case with $\chi^{-1} \bar{\vartheta}_c > \| \bm{z} \|$ leading to $\bar{\rho} = \chi^{-1} \bar{\vartheta}_c$, which is only related to the desired tracking error bound $\bar{\vartheta}_c$, the steady state with $\| \bm{\vartheta} \|$ upper bounded by $\bar{\vartheta}_c$ is considered.
		Then, the event-trigger ensures the tracking error will not exceed $\bar{\vartheta}_c$, making the ball set $\mathbb{B}_{\bar{\vartheta}_c} = \{ \bm{\vartheta} | \| \bm{\vartheta} \| \le \bar{\vartheta}_c \}$ be invariant.
		More detailed explanation and its effect on control performance are shown as follows.
		
		\begin{proposition} \label{proposition_centralized_ET}
			Consider the MAS \eqref{eqn_agent_dynamics} with the agents communicate through a network satisfying \cref{assumption_topology}.
			The unknown function $f(\cdot)$ is predicted by using GP regression satisfying \cref{assumption_kernel} with data sets satisfying \cref{assumption_dataset}.
			The control task is to track the reference trajectory satisfying \cref{assumption_desired_trajectory} with $\bm{x}_i(0) = \bm{s}_i(0) + \bm{x}_l(0), \forall i \in \mathcal{V}$ at initial time $t = 0$. 
			For such a task, employ the proposed control law \eqref{eqn_control_law} with \eqref{eqn_consensus_control_law} and compensation $\hat{f}_i(\cdot)$ using the aggregation strategy \eqref{eqn_GP_aggregation}.
			For better cooperative prediction performance, choose $\epsilon_i \in \mathbb{R}_+$ for each agent $i \in \mathcal{V}$ and adopt the online learning strategy with the centralized event-trigger \eqref{eqn_centralized_trigger} for $\bar{\vartheta}_c = \xi \chi \| \bm{\iota} + \underline{\hat{\bm{\eta}}}_{\delta} + \bm{\epsilon} \|$.
			Specifically, update some of local GP models by adding new data pairs into the local training data set at $t$ if $\rho(t) > \bar{\rho}(t)$, such that $\| \bm{\iota} + \tilde{\bm{\eta}}_{\delta}(\bm{x}) \| \le \| \bm{\iota} + \underline{\hat{\bm{\eta}}}_{\delta} + \bm{\epsilon} \|$.
			Pick $\delta \in (0,N^{-2})$, then the tracking error $\| \bm{\vartheta} \|$ is bounded by $\bar{\vartheta}_c$ with probability of at least $1 - N^2 \delta$.
		\end{proposition}
		
		\begin{IEEEproof}
			The proof follows the procedure for event-triggered GP update for single agent as in \cite{dai2023can}, where the sign of $\dot{V}$ for $\| \bm{z} \| > \chi^{-1} \bar{\vartheta}_c$ is investigated such that the trigger threshold in \eqref{eqn_centralized_trigger} is reformulated as $\bar{\rho}(t) = \| \bm{z}(t) \|$ at time $t$.
			Then, two cases divided by the trigger condition \eqref{eqn_centralized_trigger} is considered.
			In the case with $\rho(t) < \bar{\rho}(t)$ indicating no model update, the negativity of $\dot{V}$ is shown as
			\begin{align}
				\dot{V} < - \underline{\lambda}(\bm{Q}_z) \| \bm{z} \| ( \xi \rho - \xi \| \bm{\iota} + \hat{\bm{\eta}}_{\delta}(\bm{x}) \| ) \le 0
			\end{align}
			by considering \eqref{eqn_dotV_trigger_prepare} with \eqref{eqn_bound_Upsilon}.
			If $\rho(t) \ge \bar{\rho}(t)$, then the model update is activated, and with $\| \bm{z} \| > \chi^{-1} \bar{\vartheta}_c$ it has
			\begin{align}
				\dot{V} &< - \underline{\lambda}(\bm{Q}_z) \| \bm{z} \| ( \chi^{-1} \bar{\vartheta}_c - \xi \| \bm{\iota} + \underline{\tilde{\bm{\eta}}}_{\delta}(\bm{x}) \| ) \\
				&\le - \underline{\lambda}(\bm{Q}_z) \| \bm{z} \| \xi ( \| \bm{\iota} + \underline{\hat{\bm{\eta}}}_{\delta} + \bm{\epsilon} \| - \| \bm{\iota} + \underline{\tilde{\bm{\eta}}}_{\delta}(\bm{x}) + \bm{\epsilon} \| ) \le 0 \nonumber
			\end{align}
			where the second inequality is derived by using the definition of $\bar{\vartheta}_c$ and the guarantee of the prediction accuracy after online learning from the assumptions in \cref{proposition_centralized_ET}, i.e., $\| \bm{\iota} + \tilde{\bm{\eta}}_{\delta}(\bm{x}) \| \le \| \bm{\iota} + \underline{\hat{\bm{\eta}}}_{\delta} + \bm{\epsilon} \|$.
			Until here, the negativity of $\dot{V}$ when $\| \bm{z} \| > \chi^{-1} \bar{\vartheta}_c$ is proven, which concludes the proof by letting $\bar{z} = \chi^{-1} \bar{\vartheta}_c$ and using the result in \cref{lemma_boundness_z_vartheta}.
		\end{IEEEproof}
		
		Although the case with $\| \bm{z} \| \le \chi^{-1} \bar{\vartheta}$ is not discussed, which is equivalent to $\| \bm{\vartheta} \| \le \bar{\vartheta}$ from \cref{lemma_boundness_z_vartheta}.
		The strict negativity of $\dot{V}$ outside of $\mathbb{B}_{\bar{\vartheta}}$ ensures the compact area $\mathbb{B}_{\bar{\vartheta}}$ is an invariant set, by considering $\dot{V} \le 0$ at the margin, i.e., $\| \bm{\vartheta} \| \in \partial \mathbb{B}_{\bar{\vartheta}}$ due to the continuity of $\dot{V}$.
		Moreover, the sign of $\dot{V}$ inside of $\mathbb{B}_{\bar{\vartheta}}$ is indeterminable using the current Lyapunov theory, therefore the guaranteed control performance is characterized by the tracking error bound $\bar{\vartheta}_c$.
		
		Note that with strictly positive $\epsilon_i, \forall i \in \mathcal{V}$, the best tracking error bound in \cref{proposition_centralized_ET} is larger than the best case in \cref{theorem_best_tracking_performance_online}.
		The choice of $\epsilon_i$ effects the guaranteed minimal trigger interval on each agent $i$ for Zeno behavior avoidance, which is shown in \cref{subsection_Zeno_behavior}.
		Intuitively, smaller $\epsilon_i$ make the control performance close to the best case as in \cref{theorem_best_tracking_performance_online}, while also reduces the trigger interval inducing more computations and data storage.
		
		\begin{remark} \label{remark_centralized_update_model_selection}
			For the improvement of the overall prediction performance in \cref{proposition_centralized_ET}, it is only required $\| \bm{\iota} + \tilde{\bm{\eta}}_{\delta}(\bm{x}) + \bm{\epsilon} \| \le \| \bm{\iota} + \underline{\hat{\bm{\eta}}}_{\delta} + \bm{\epsilon} \|$ after model update but not specifies which local GP models should be updated.
			For less cost from model updates and higher data efficiency, the local GP models that need to be updated, i.e., $\varpi_i = 1$, can be selected by solving the optimization problem as
			\begin{align} \label{eqn_centralized_ET_update_agent_selection}
				\min_{\{\! \varpi_i \!\}_{i \!\in\! \mathcal{V}}} \!\sum\nolimits_{i \!\in\! \mathcal{V}} ~ c_i(\varpi_i) ~~
				\text{s.t.} ~ \| \bm{\iota} \!+\! \tilde{\bm{\eta}}_{\delta}(\bm{x}) \!+\! \bm{\epsilon} \| \!\!\le\!\! \| \bm{\iota} \!+\! \underline{\hat{\bm{\eta}}}_{\delta} \!+\! \bm{\epsilon} \|,
			\end{align}
			with $\varpi_i$ and each entry of $\tilde{\bm{\eta}}_{\delta}(\bm{x})$ defined in \eqref{eqn_varpi} and \eqref{eqn_tilde_eta} respectively, where $c_i(\varpi_i)$ represents the "cost" for model update at agent $i \in \mathcal{V}$.
			The detailed expression of $c_i(\cdot)$ is varying from different aspects, e.g., energy consumption, computation power and data storage, which is beyond the scope of this paper.
			With a sufficiently powerful centralized node, solving such binary optimization problem in \eqref{eqn_centralized_ET_update_agent_selection} in real-time is possible but still requires large computation resources.
			However, there exists a simple heuristic selection method to achieve same performance as in the constraint of \eqref{eqn_centralized_ET_update_agent_selection}, which only update the local GP model with $\hat{\eta}_{\delta,i}(\bm{x}_i) \le  \underline{\hat{\eta}}_{\delta,i}$, such that the trigger condition for each agent $i \in \mathcal{V}$ becomes
			\begin{align} \label{eqn_centralized_ET_distributed_heuristic}
				\rho_i \!\!=\! \rho \big( \hat{\eta}_{\delta,i}(\bm{x}_i) \!-\! \underline{\hat{\eta}}_{\delta,i} \big), && \bar{\rho}_i \!\!=\! \max \{ \bar{\rho}\big( \hat{\eta}_{\delta,i}(\bm{x}_i) \!-\! \underline{\hat{\eta}}_{\delta,i} \big)\!,\! 0\},
			\end{align}
			indicating agent $i$ will update its local GP model only at $t$ when both $\hat{\eta}_{\delta,i}(\bm{x}_i) > \underline{\hat{\eta}}_{\delta,i}$ and $\rho(t) > \bar{\rho}(t)$ are satisfied.
			While potentially introducing more trigger times at each agent, this heuristic selection method in \eqref{eqn_centralized_ET_distributed_heuristic} avoids solving the optimization problem \eqref{eqn_centralized_ET_update_agent_selection} and inducing faster implementation.
			Note that \eqref{eqn_centralized_ET_distributed_heuristic} cannot be distributed calculated in each agent, since the computation of $\rho$ and $\bar{\rho}$ requires the global information.
		\end{remark}
		
		While by using the centralized event-trigger in \eqref{eqn_centralized_trigger} similar control performance is reached as in \cref{theorem_best_tracking_performance_online} with potentially less model update, the evaluation of $\rho(\cdot)$ and $\bar{\rho}(\cdot)$ in \eqref{eqn_centralized_trigger} requiring all the information in the networked system is heavily rely on the perfect communication between agents and powerful centralized node.
		In practice, this structure is not robust due to accidental connection failures and not scalable to large-scale MAS systems.
		Therefore, a distributed version of \eqref{eqn_centralized_trigger} is required, which can be evaluated with only information from local agent and its neighbors.
		However, the study of centralized event-trigger is still valuable, providing us an inspiration on the design of distributed event-trigger.
		Note that the centralized trigger condition in \eqref{eqn_centralized_trigger} is equivalent to choosing \looseness=-1
		\begin{align}
			\rho^* \!=\! \| \bm{\iota} \!+\! \hat{\bm{\eta}}_{\delta}(\bm{x}) \|^2, && \bar{\rho}^* \!=\! \xi^{-2} \max \{ \| \bm{z} \|^2, \chi^{-2} \bar{\vartheta}_c^2 \} ,
		\end{align}
		where the trigger threshold is also written as
		\begin{align} \label{eqn_centralized_trigger_2}
			\bar{\rho}^* = \xi^{-2} \max \{ \| \bm{z} \|^2 - \chi^{-2} \bar{\vartheta}_c^2, 0 \} + \| \bm{\iota} + \underline{\hat{\bm{\eta}}}_{\delta} + \bm{\epsilon} \|^2.
		\end{align}
		The form in \eqref{eqn_centralized_trigger_2} is regarded as the combination of transient requirement with $\bm{z}$, desired tracking error bound $\bar{\vartheta}_c$ and guaranteed performance after model update with $\underline{\hat{\bm{\eta}}}_{\delta}$.
		Intuitively, \eqref{eqn_centralized_trigger_2} can be converted to the distributed version by decomposition of $\bm{z}$ and considering the individual performance guarantee after model update, which is detailed discussed in the next subsection.
	}
	\fi
	
	\subsection{Distributed Event-triggered Online Learning}
	\label{subsection_distributed_ET}
	
	For robustness and scalability of the information topology, the distributed event-triggered mechanism is proposed inspired from \eqref{eqn_centralized_trigger_2}.
	To achieve the tracking error bound $\bar{\vartheta}_d = \xi \chi \| \bm{\iota} + \underline{\hat{\bm{\eta}}}_{\delta} + \bm{\epsilon} \|$ with positive entries of $\bm{\epsilon}$ as $\epsilon_i, \forall i \in \mathcal{V}$, the distributed trigger condition in \eqref{eqn_fake_trigger} is written for each agent $i \in \mathcal{V}$ as
	\begin{align} \label{eqn_trigger}
		&\rho_i = \big( \hat{\eta}_{\delta,i}(\bm{x}_i) + \iota_i \big)^2 \\
		&\bar{\rho}_i = \xi^{-2} \max \big\{ \| \bm{z}_i\|^2 \!-\! \chi^{-2} \bar{\vartheta}_d^2 / N, 0 \big\} \!+\! \big( \underline{\hat{\eta}}_{\delta,i} \!+\! \iota_i \!+\! \epsilon_i \big)^2, \nonumber
	\end{align}
	where $\| \bm{z}_i \| = [r_i, e_{i,1}, \cdots, e_{i,n-1}]^T$, $\iota_i = (1 - b_{ii}) F_l$ recalled from \eqref{eqn_bound_Upsilon}.
	Similarly to \eqref{eqn_centralized_trigger}, the time input is dropped, i.e., $\rho_i(t) \to \rho_i$ and $\bar{\rho}_i(t) \to \bar{\rho}_i$, for notational simplicity.
	Note that all terms in $\rho_i$ and $\bar{\rho}_i$ in \eqref{eqn_trigger} including $\bm{z}_i$ can be calculated using only local and neighboring information, realizing distributed evaluation.
	
	\ifarxiv
	\todo{
		The trigger threshold $\bar{\rho}_i$ is related to three terms, namely the tracking error reflected by $\| \bm{z}_i \|$, desired tracking error bound $\bar{\vartheta}_d$ and upper bound of prediction error after model update represented by $(\iota_i + \underline{\hat{\eta}}_{\delta,i} + \epsilon_i)$.
		Similar as the analysis for \eqref{eqn_centralized_trigger_2}, the term $\| \bm{z}_i \|$ serves only in transition phase, i.e., $\| \bm{z}_i \| > \chi^{-1} \bar{\vartheta}_d / \sqrt{N}$, to ensure the decay of tracking error, proven in \cref{theorem_trigger}.
		Compared to \eqref{eqn_centralized_trigger_2}, the desired tracking error bound related term, i.e., $\chi^{-2} \bar{\vartheta}_d^2 / N$, is penalized by the number of agents $N$, which is due to the lack of complete information of all agents at agent $i$.
		More detailed, the division of $N$ tries to distribute the overall tracking error bound for $\| \bm{z} \|$ to the local value $\| \bm{z}_i \|$.
		The last term related to $\underline{\hat{\eta}}_{\delta,i} + \epsilon_i$ in \eqref{eqn_eta_underline} indicates the guaranteed prediction performance after GP model update on agent $i$, which serves the steady state case.
		
		\begin{remark}
			For the single agent case with $N = 1$, the distributed event-trigger \eqref{eqn_trigger} is identical to the centralized version \eqref{eqn_centralized_trigger_2} and thus \eqref{eqn_centralized_trigger}, since the agent has the access to "global" information.
			Note that the estimated local bound $\chi^{-2} \bar{\vartheta}_d^2 / N$ for $\| \bm{z}_i \|$ is not vanishing even for infinitely large-scale system with $N \to \infty$, since $\bar{\vartheta}_d$ also grows with $N$.
			Instead, $\chi^{-1} \bar{\vartheta}_d / \sqrt{N}$ represents the root mean square of the joint effects from local prediction error after model update, i.e., $\underline{\hat{\eta}}_{\delta,i} + \epsilon_i$ and from leader mis-connection reflected by $\iota_i$.
		\end{remark}
	}
	\fi
	
	Considering the maximum operator in \eqref{eqn_trigger} and comparing the values of $\rho_i$ and $\bar{\rho_i}$, the agents are divided into $4$ cases based on the criterion as \looseness=-1
	\begin{itemize}
		\item Safe set $\mathbb{S}$: Agent $i \in \mathbb{S}$ means $\| \bm{z}_i \| \leq \chi^{-1} \bar{\vartheta}_d / \sqrt{N}$;
		\item Trigger set $\mathbb{T}$: Agent $i \in \mathbb{T}$ means trigger condition with \eqref{eqn_trigger} is satisfied, i.e., $\rho_i > \bar{\rho_i}$.
	\end{itemize}
	Moreover, define $\bar{\mathbb{S}}$, $\bar{\mathbb{T}}$ as the complement sets for $\mathbb{S}$, $\mathbb{T}$ respectively, satisfying $\mathbb{S} \cup \bar{\mathbb{S}} = \mathbb{T} \cup \bar{\mathbb{T}} = \mathcal{V}$ and $\mathbb{S} \cap \bar{\mathbb{S}} = \mathbb{T} \cap \bar{\mathbb{T}} = \emptyset$.
	Note that there exists overlap between $\mathbb{S}$ and $\mathbb{T}$, such that the $4$ cases for agents are defined as $\mathbb{S} \cap \mathbb{T}$, $\bar{\mathbb{S}} \cap \mathbb{T}$, $\mathbb{S} \cap \bar{\mathbb{T}}$ and $\bar{\mathbb{S}} \cap \bar{\mathbb{T}}$ with the following properties.
	
	\begin{property} \label{property_distributed_ET}
		(\romannumeral 1) If $i \!\in\! \mathbb{T}$ or $i \!\in\! \mathbb{S} \!\cap\! \bar{\mathbb{T}}$, then $\tilde{\eta}_{\delta,i}(\bm{x}_i) \!\le\! \underline{\hat{\eta}}_{\delta,i} \!+\! \epsilon_i$;
		\\
		(\romannumeral 2) If $i \in \bar{\mathbb{S}} \cap \mathbb{T}$, then $\| \bm{z}_i \| > \chi^{-1} \bar{\vartheta}_d / \sqrt{N}$;
		\\
		(\romannumeral 3) If $i \in \bar{\mathbb{S}} \cap \bar{\mathbb{T}}$, then
		\begin{align} \label{eqn_property_no_safe_no_trigger}
			\| \bm{z}_i \|^2 \geq \chi^{-2} \bar{\vartheta}_d^2 / N + \xi^2 \big( \rho_i - (\underline{\eta}_{\delta,i} + \iota_i + \epsilon_i )^2 \big).
		\end{align}
	\end{property}
	\begin{IEEEproof}
		(\romannumeral 1) The property for agent $i \!\in\! \mathbb{T}$ is obvious, since the GP model update is activated on agent $i$ such that $\tilde{\eta}_{\delta,i}(\bm{x}_i) \!=\! \hat{\eta}_{\delta,i}^+(\bm{x}_i) \!\le\! \underline{\hat{\eta}}_{\delta,i} \!\le\! \underline{\hat{\eta}}_{\delta,i} \!+\! \epsilon_i$ by considering \cref{proposition_online_cooperative_prediction_error_bound} and positive $\epsilon_i$.
		To prove the property for agent $i \in \mathbb{S} \cap \bar{\mathbb{T}}$, the trigger condition is investigated for $i \in \bar{\mathbb{T}}$ as $\rho_i \!\le\! \bar{\rho}_i \!=\! ( \underline{\hat{\eta}}_{\delta,i} \!+\! \iota_i \!+\! \epsilon_i )^2$ due to the definition of $\mathbb{S}$, which leads to $\tilde{\eta}_{\delta,i}(\bm{x}_i) \!=\! \hat{\eta}_{\delta,i}(\bm{x}_i) \!\le\! \underline{\hat{\eta}}_{\delta,i} + \epsilon_i$ from the design of $\rho_i$ in \eqref{eqn_trigger}.
		\\
		(\romannumeral 2) This property is directly derived by the definition of $\bar{\mathbb{S}}$, which is recalled as $\| \bm{z}_i \| > \chi^{-1} \bar{\vartheta} / \sqrt{N}$.
		\\
		(\romannumeral 3) Considering the definition of $\bar{\mathbb{S}}$ and $\bar{\mathbb{T}}$, it has for agent $i \in \bar{\mathbb{S}} \cap \bar{\mathbb{T}}$ as $\rho_i \le \xi^{-2}  \| \bm{z}_i\|^2 \!-\! \chi^{-2} \bar{\vartheta}_d^2 / N \!+\! \big( \underline{\hat{\eta}}_{\delta,i} \!+\! \iota_i \!+\! \epsilon_i \big)^2$, which is equivalent to \eqref{eqn_property_no_safe_no_trigger} by putting $\| \bm{z}_i \|$ to the left-hand side.
	\end{IEEEproof}
	
	\cref{property_distributed_ET} shows relevant properties of the designed distributed event-trigger in \eqref{eqn_trigger} for stability analysis.
	Considering the formulation in \eqref{eqn_dot_V3}, where $\| \bm{z} \|$ and $\hat{\bm{\eta}}_{\delta}(\cdot)$ contribute the negative and positive part in $\dot{V}$, \cref{property_distributed_ET} ensures the positive term is sufficiently small when $i \in \mathbb{S}$ by observing \cref{property_distributed_ET} (\romannumeral 1), while the negative term is sufficiently large when $i \in \bar{\mathbb{S}}$ from \cref{property_distributed_ET} ({\romannumeral 2} \& {\romannumeral 3}).
	
	\ifarxiv
	\todo{
		\begin{remark} \label{remark_property_distributed_trigger}
			Compared to the properties for the event-trigger mechanism for single agent as in \cite{umlauft2019feedback, jiao2022backstepping, dai2023can}, which only ensures the absolute value of the negative term is larger than the positive term in $\bar{\mathbb{S}}$ to guarantee the negativity of $\dot{V}$, \cref{property_distributed_ET} includes additional properties on agents belonging to $\mathbb{S}$, i.e., \cref{property_distributed_ET} (\romannumeral 1).
			This is because, in MAS setting especially for distributed event-trigger mechanism, mere restriction on $\| \bm{z}_i \|$ for agents $i \in \bar{\mathbb{S}}$ cannot guarantee the decay of $V$ without any constraints on the behavior on agents in $\mathbb{S}$.
			This statement is intuitive by considering the positive term related to $\hat{\eta}_{\delta,i}(\cdot), \forall i \in \mathbb{S}$ is sufficiently large without constraints as in \cref{property_distributed_ET} (\romannumeral 1) such that any negative term with $\bm{x}_j \in \mathbb{X}$, $\forall i \in \mathcal{V}$ cannot compensate.
			Then, the negativity of $\dot{V}$ cannot be derived resulting in no guarantee on control performance.
		\end{remark}
	}
	\fi
	
	Before the rigorous proof of the control performance with distributed event-triggered online cooperative learning with \eqref{eqn_trigger}, an auxiliary lemma to \cref{lemma_boundness_z_vartheta} for distribution of the error $\| \bm{z} \|$ to each single agent is given as follows.
	
	\begin{lemma} \label{lemma_zi}
		For a MAS with $N$ agents, if $\| \bm{z} \| > \bar{z}$, then at least one agent $i \in \mathcal{V}$ satisfying $\| \bm{z}_i \| > \bar{z} / \sqrt{N}$.
	\end{lemma}
	\begin{IEEEproof}
		See appendix.
	\end{IEEEproof}
	
	\cref{lemma_zi} distributes the condition on $\| \bm{z} \|$ to each agent on $\| \bm{z}_i \|$, which serves as an extension to \cref{lemma_boundness_z_vartheta} such that the boundness of $\| \bm{\vartheta} \|$ is related to individual error $\| \bm{z}_i \|$.
	The complete proof for the tracking error bound is shown in the following theorem. 
	
	\begin{theorem} \label{theorem_trigger}
		Consider the MAS \eqref{eqn_agent_dynamics}, where agents communicate through a network satisfying \cref{assumption_topology}.
		The unknown function $f(\cdot)$ is predicted by using GP regression satisfying \cref{assumption_dataset} and \ref{assumption_kernel} and the aggregation method \eqref{eqn_GP_aggregation}.
		The control task is to track the reference trajectory satisfying \cref{assumption_desired_trajectory} and $\bm{x}_{i}(0) = \bm{s}_i(0) + \bm{x}_{l}(0)$, $\forall i \in \mathcal{V}$ at initial time $t = 0$. 
		For such a task, employ the proposed control law \eqref{eqn_control_law} with \eqref{eqn_consensus_control_law} and the distributed event-trigger mechanism for online learning with \eqref{eqn_trigger} and $\epsilon_i \in \mathbb{R}_+, \forall i \in \mathcal{V}$.
		Pick $\delta \in (0,N^{-2})$, and then the tracking error $\| \bm{\vartheta} \|$ is bounded by $\bar{\vartheta}_d = \xi \chi \| \bm{\iota} + \underline{\hat{\bm{\eta}}}_{\delta} + \bm{\epsilon}\|$ with probability of at least $1 - N^2 \delta$.
	\end{theorem}
	
	\begin{IEEEproof}
		Given a desired tracking error bound $\bar{\vartheta}$ and using the inverse result in \cref{lemma_boundness_z_vartheta}, it requires to prove $\dot{V} < 0$ if $\| \bm{z} \| > \chi^{-1} \bar{\vartheta}$.
		Dividing the agents according to the maximum operator and trigger condition into $\mathbb{S} \cap \mathbb{T}$, $\bar{\mathbb{S}} \cap \mathbb{T}$, $\mathbb{S} \cap \bar{\mathbb{T}}$ and $\bar{\mathbb{S}} \cap \bar{\mathbb{T}}$ with properties shown in \cref{property_distributed_ET}, the concatenated effects of the prediction error and leader misconnection, i.e., $\bm{\iota} + \tilde{\bm{\eta}}_{\delta}(\bm{x}) + \bm{\epsilon}$, is upper bounded by
		\begin{align}
			\| \bm{\iota} \!+ &\tilde{\bm{\eta}}_{\delta}(\bm{x}) \!+\! \bm{\epsilon} \|^2 \!\!=\!\!\! \sum_{i \in \mathbb{T} }  ( \iota_i \!+\! \hat{\eta}_{\delta,i}^+(\bm{x}_i) \!+\! \epsilon_i )^2 \!\!+\!\! \sum_{i \in \mathbb{S} \cap \bar{\mathbb{T}} } \!\! \rho_i  \!+\!\! \sum_{i \in \bar{\mathbb{S}} \cap \bar{\mathbb{T}} } \!\! \rho_i  \nonumber \\
			&\le \sum_{i \in \mathbb{T} } (\iota_i + \underline{\hat{\eta}}_{\delta,i} \!+\! \epsilon_i )^2 \!+\! \!\! \sum_{i \in \mathbb{S} \cap \bar{\mathbb{T}} } \!\! (\iota_i \!+\! \underline{\hat{\eta}}_{\delta,i} \!+\! \epsilon_i )^2  \!+\! \!\! \sum_{i \in \bar{\mathbb{S}} \cap \bar{\mathbb{T}} } \rho_i \nonumber \\
			&= \sum_{i \in \mathcal{V} } (\iota_i + \underline{\hat{\eta}}_{\delta,i} \!+\! \epsilon_i )^2 \!+\! \sum_{i \in \bar{\mathbb{S}} \cap \bar{\mathbb{T}} } (\rho_i \!-\! (\iota_i \!+\! \underline{\hat{\eta}}_{\delta,i} \!+\! \epsilon_i)^2). \nonumber
		\end{align}
		Moreover, considering $\sum_{i \in \mathcal{V} } (\iota_i + \tilde{\underline{\eta}}_{\delta,i} \!+\! \epsilon_i )^2 = \| \bm{\iota} \!+\! \underline{\hat{\bm{\eta}}}_{\delta} + \bm{\epsilon} \|^2 = (\xi \chi)^{-2} \bar{\vartheta}_d^2 / N$ from the condition of $\bar{\vartheta}_d$ in \cref{theorem_trigger}, the upper bound for $\| \bm{\iota} + \tilde{\bm{\eta}}_{\delta}(\bm{x}) + \bm{\epsilon} \|^2$ is further written as
		\begin{align} \label{eqn_bound_iota_eta_epsilon}
			\| \bm{\iota} + \hat{\bm{\eta}}_{\delta}^+(\bm{x}) + \bm{\epsilon} \|^2 <& (\xi \chi)^{-2} \bar{\vartheta}_d^2 / N \\
			&+ \sum\nolimits_{i \in \bar{\mathbb{S}} \cap \bar{\mathbb{T}} } (\rho_i - (\iota_i + \tilde{\underline{\eta}}_{\delta,i} + \epsilon_i)^2). \nonumber
		\end{align}
		Furthermore, considering \cref{property_distributed_ET} the concatenated error $\bm{z}$ is lower bounded by
		\begin{align} \label{eqn_bound_z}
			\| \bm{z} \|^2 \!&= \sum_{i \in \mathcal{V}} \! \| \bm{z}_i \|^2 \geq \!\!\! \sum_{i \in \bar{\mathbb{S}} \cap \mathbb{T} } \!\! \| \bm{z}_i \|^2 \!+\! \!\! \sum_{i \in \bar{\mathbb{S}} \cap \bar{\mathbb{T}} } \!\! \| \bm{z}_i \|^2 \\
			& > \!\!\! \sum_{i \in \bar{\mathbb{S}} \cap \mathbb{T} } \!\! \chi^{\!-\!2\!} \bar{\vartheta}_d^2 \!/\! N \!+ \!\!\! \sum_{i \in \bar{\mathbb{S}} \cap \bar{\mathbb{T}} } \!\! (\chi^{\!-\!2\!} \bar{\vartheta}_d^2 \!/\! N \!+\! \xi^2 ( \rho_i \!-\! (\underline{\eta}_{\delta,i} \!\!+\! \iota_i \!+\! \epsilon_i )^2 )) \nonumber \\
			&= \chi^{-2} \bar{\vartheta}_d^2 | \bar{\mathbb{S}} | / N \!+\! \!\! \sum\nolimits_{i \in \bar{\mathbb{S}} \cap \bar{\mathbb{T}} } \xi^2 ( \rho_i \!-\! (\underline{\eta}_{\delta,i} \!+\! \iota_i \!+\! \epsilon_i )^2 ). \nonumber
		\end{align}
		Apply \eqref{eqn_bound_iota_eta_epsilon} and \eqref{eqn_bound_z} into \eqref{eqn_dot_V3}, then $\dot{V}$ is upper bounded by
		\begin{align} \label{eqn_dot_V_distributed_ET}
			\dot{V} &\le - \frac{\underline{\lambda}(\bm{Q}_z) \| \bm{z} \|}{\| \bm{z} \| + \xi \| \bm{\iota} + \underline{\hat{\bm{\eta}}}_{\delta} \|} \left(\| \bm{z} \|^2 - \xi^2 \| \bm{\iota} + \underline{\hat{\bm{\eta}}}_{\delta} \|^2 \right) \\
			&< - \frac{\underline{\lambda}(\bm{Q}_z) \| \bm{z} \| \bar{\vartheta}^2}{N \chi^2 (\| \bm{z} \| + \xi \| \bm{\iota} + \underline{\hat{\bm{\eta}}}_{\delta} \|)} \left(| \bar{\mathbb{S}} | - 1 \right). \nonumber
		\end{align}
		Next, we recall the case with $\| \bm{z} \| > \chi^{-1} \bar{\vartheta}_d$ indicating at least one agent satisfying $\| \bm{z}_i \| > \chi^{-1} \bar{\vartheta}_d / \sqrt{N}$ from \cref{lemma_zi},  i.e., $| \bar{\mathbb{S}} | \geq 1$.
		Then, the negativity of $\dot{V}$ in \eqref{eqn_dot_V_distributed_ET} is directly obtained for $\| \bm{z} \| > \chi^{-1} \bar{\vartheta}_d$.
		Let $\bar{z} = \chi^{-1} \bar{\vartheta}_d$ and apply the result in \cref{lemma_boundness_z_vartheta}, the boundness of $\| \bm{\vartheta} \|$ by $\bar{\vartheta}_d$ is derived.
	\end{IEEEproof}
	
	\cref{theorem_trigger} shows the bounded tracking error $\| \bm{\vartheta} \|$ with the proposed distributed event-trigger \eqref{eqn_trigger}.
	\ifarxiv
	\todo{
		Indeed, \cref{property_distributed_ET} plays an important role in stability analysis as discussed in \cref{remark_property_distributed_trigger}, ensuring the negative term related to $\| \bm{z}_i \|$ in \eqref{eqn_dot_V3} is sufficiently large outside of $\mathbb{B}_{\bar{\vartheta}_d} = \{ \bm{\vartheta} | \| \bm{\vartheta} \| \le \bar{\vartheta}_c \}$ for the decay of Lyapunov function $V$.
	}
	\fi
	The stability analysis in \cref{theorem_trigger} provides a guideline for devising the distributed event-trigger, which means any design satisfying \cref{property_distributed_ET} guarantee the tracking error bound with $\bar{\vartheta}_d$.
	
	\ifarxiv
	\todo{
		\begin{remark}
			Recall that \cref{proposition_centralized_ET} and \cref{theorem_trigger} show the bounded tracking error $\bm{\vartheta}$ for the entire MAS.
			Due to the coupled relationship in the synchronization error $e_{i,k}$, the tracking error $\bm{\vartheta}_i = \bm{x}_i - \bm{s}_i - \bm{x}_l$ cannot separately bounded for each agent $i$, however by considering $\| \bm{\vartheta}_i \| \le \| \bm{\vartheta} \|$ it is directly derived $\| \bm{\vartheta}_i \| \le \bar{\vartheta}_c$ and $\| \bm{\vartheta}_i \| \le \bar{\vartheta}_d$ for centralized and distributed event-triggers respectively.
		\end{remark}
		
		Note that with same choice of $\epsilon_i, \forall i \in \mathcal{V}$ in centralized and distributed event-trigger, i.e., \eqref{eqn_centralized_trigger} and \eqref{eqn_trigger}, leads to identical guaranteed tracking error bound $\bar{\vartheta}_c = \bar{\vartheta}_d = \bar{\vartheta}_e$, where $\bar{\vartheta}_e$ is defined for notational simplicity in the subsequent analysis in \cref{subsection_Zeno_behavior}.
		It is intuitive that choosing smaller $\epsilon_i$, in particular $\epsilon_i \to 0, \forall i \in \mathcal{V}$, induces closer control performance as $\bar{\vartheta}$ in \cref{theorem_best_tracking_performance_online} but also potential Zeno behavior, since in the worst case the prediction error bound after local model update equals $\underline{\hat{\eta}}_{\delta,i}$.
		For Zeno behavior avoidance, strict positivity of $\epsilon_i$ is necessary, and its relationship to the minimal trigger interval is shown in the next subsection. 
	}
	\fi
	
	\subsection{Zeno Behavior for Event-triggered Cooperative Learning}
	\label{subsection_Zeno_behavior}
	
	The Zeno behavior is an essential problem for event-triggered strategies \cite{fan2015self}, meaning infinite triggers in finite time.
	In this subsection, we discuss the Zeno behavior on each agent $i$ using the proposed centralized and distributed event-trigger, i.e., \eqref{eqn_centralized_trigger} and \eqref{eqn_trigger} respectively.
	Note that the exclusion of the Zeno behavior on each agent $i \in \mathcal{V}$ prohibits the jump of $\bm{x}_i$, which means it	requires limited changing rate of the agent states, i.e., $\dot{\bm{x}}_i$, under the proposed controller \eqref{eqn_control_law}.
	The upper bound of $\dot{\bm{x}}_i$ for $\forall i \in \mathcal{V}$ is shown in the following lemma.
	
	\begin{lemma} \label{lemma_Fi}
		Let all the assumptions in \cref{proposition_online_cooperative_prediction_error_bound} with centralized event-trigger mechanism with \eqref{eqn_centralized_trigger} and any update model selection methods, or \cref{theorem_trigger} with distributed event-trigger strategy in \eqref{eqn_trigger} hold.
		Choose $\epsilon_i \in \mathbb{R}_+$ for each agent $i \in \mathcal{V}$, such that $\bar{\vartheta}_e = \xi \chi \| \bm{\iota} + \underline{\hat{\bm{\eta}}}_{\delta} + \bm{\epsilon} \|$.
		Pick $\delta \in (0, N^{-2})$, then the state changing rate $\dot{\bm{x}}_i(t)$ for each agent $i$ is probabilistic bounded, i.e., $\Pr \{ | \dot{\bm{x}}_i(t) | \le F_i, \forall t \in \mathbb{R}_{0,+} \} \le 1 - N^2 \delta$ for $F_i \in \mathbb{R}_+$ defined as
		\begin{align} \label{eqn_Fi}
			F_i \!=\! F_{r,i} \!+\! \big(1 \!+\! \sqrt{2} c (l_{ii} \!+\! b_{ii}) \lambda^* \big) \bar{\vartheta}_e \!+\! (1 \!-\! b_{ii}) F_l \!+\!  \bar{\hat{\eta}}_{\delta, i},
		\end{align}
		where $\lambda^* = \max_{k = 1, \cdots, n} \lambda_k$ and $\bar{\hat{\eta}}_{\delta, i}$ is obtained by solving
		\begin{align} \label{eqn_max_aggregated_prediction_bound}
			\bar{\hat{\eta}}_{\delta,i} \!=\! \sup_{\bm{x}_i \in \mathbb{X}} \Big( \omega_{ii}(\bm{x}_i) \sigma_i(\bm{x}_i) \!+\! \sum\nolimits_{j \in \bar{\mathcal{N}}_i} \omega_{ij}(\bm{x}_i) \sigma_j(\bm{x}_i) \Big).
		\end{align}
		The real positive constants $F_{r,i}$ and $F_l$ are the upper bounds for $\| \dot{\bm{x}}_{l} + \dot{\bm{s}}_{i} \|$ and $| x_{l,r} |$ respectively as in \cref{assumption_desired_trajectory}.
	\end{lemma}
	
	\begin{IEEEproof}
		The bound of $\| \dot{\bm{x}}_i \|$ is derived by finding the supremum of the tracking error $\| \dot{\bm{\vartheta}}_i \|$ for all $i \in \mathcal{V}$.
		Apply the control law \eqref{eqn_control_law} with \eqref{eqn_consensus_control_law} and \eqref{eqn_GP_aggregation}, then the error dynamics w.r.t $\bm{\vartheta}_i = [\vartheta_{i,1}, \cdots, \vartheta_{i,n}]^T$ with $\vartheta_{i,k} = x_{i,k} - s_{i,k} - x_{l,k}$ for the controlled system yields
		\begin{align}
			&\dot{\vartheta}_{i,k} = \vartheta_{i,k + 1}, ~ \forall k = 1, \cdots, n-1, \nonumber \\
			&\dot{\vartheta}_{i,n} = - c r_i - (1 - b_{ii}) x_{l,r} + f(\bm{x}_i) - \tilde{f}_i(\bm{x}_i), \nonumber
		\end{align}
		where $\tilde{f}_i(\cdot) = \varpi_i \hat{f}_i^+(\cdot) + (1 - \varpi_i) \hat{f}_i(\cdot)$ represents the applied aggregated prediction under event-trigger mechanism using the notation $\varpi_i$ in \cref{section_event_triggered_learning}.
		Then, $\| \dot{\bm{\vartheta}}_i \| = [\dot{\vartheta}_{i,1}, \cdots, \dot{\vartheta}_{i,n}]^T$ is bounded using the triangular inequality by
		\begin{align} \label{eqn_dot_vartheta_i_1}
			\| \dot{\bm{\vartheta}}_i \| \!\!\le& \|\! [\dot{\vartheta}_{i,1}, \!\cdots\!, \dot{\vartheta}_{i,n-1}]^{\!T\!} \| + |\dot{\vartheta}_{i,n}| \\
			\le& \|\! [\vartheta_{i,2}, \!\cdots\!, \vartheta_{i,n}]^{\!T\!} \| \!\!+\! c |r_i| \!\!+\!\! (1 \!-\! b_{ii}) |x_{l,r}| \!\!+\!\! |f(\bm{x}_i) \!\!-\!\! \tilde{f}_i(\bm{x}_i)|. \nonumber
		\end{align}
		Consider \cref{assumption_desired_trajectory} and recall $\tilde{\eta}_{\delta,i}(\cdot)$ in \cref{section_event_triggered_learning} as the probabilistic prediction error bound of $|f(\cdot) - \tilde{f}_i(\cdot)|$ with probability of at least $1 - N \delta$ inherited from \cref{proposition_online_cooperative_prediction_error_bound}, \eqref{eqn_dot_vartheta_i_1} is reformulated by using $\| [\vartheta_{i,2}, \!\cdots\!, \vartheta_{i,n}]^{\!T\!} \| \le \| \bm{\vartheta}_i \| \le \| \bm{\vartheta} \|$ as
		\begin{align} \label{eqn_dot_vartheta_i_2}
			\| \dot{\bm{\vartheta}}_i \| \le& \| \bm{\vartheta} \| + c |r_i| + (1 - b_{ii}) F_l + \tilde{\eta}_{\delta, i}(\bm{x}_i),
		\end{align}
		which holds with probability of at least $1 - N \delta$.
		Due to the definition of the filtered error $r_i$ in \eqref{eqn_consensus_control_law}, the norm of $r_i$ is bounded by using Cauchy-Schwarz inequality and considering non-negative $\lambda_k$ as $|r_i|^2 \le \sum_{k=1}^n \lambda_k^2 | e_{i,k} |^2$, where the synchronization errors $| e_{i,k} |$ from \eqref{eqn_synchronization_error} is bounded by
		\begin{align} \label{eqn_eik_bound}
			| e_{i,k} |^2  =& \left| \big(b_{ii} + \sum\nolimits_{j \in \mathcal{N}_i} a_{ij} \big) \vartheta_{i,k} + \sum\nolimits_{j \in \mathcal{N}_i} a_{ij} \vartheta_{j,k} \right|^2 \nonumber \\
			\le& (b_{ii} + l_{ii} )^2 \vartheta_{i,k}^2 + \sum\nolimits_{j \in \mathcal{N}_i} a_{ij}^2 \vartheta_{j,k}^2 \\
			&+ 2 (b_{ii} + l_{ii} )  \left(\sum\nolimits_{j \in \mathcal{N}_i} a_{ij} | \vartheta_{i,k} | | \vartheta_{j,k} | \right),  \nonumber
		\end{align}
		where $l_{ii}$ is recalled as the $i$-th entry of the diagonal of Laplacian matrix $\bm{\mathcal{L}}$ such that $l_{ii} \ge a_{ij}, \forall j \in \mathcal{V}$ due to the non-negative of $a_{ij}$.
		Moreover, considering $b_{ii} \ge 0$, it has $0 \le a_{ij} \le b_{ii} + l_{ii}, \forall j \in \mathcal{V}$.
		Using the Young's inequality on $| \vartheta_{i,k} | | \vartheta_{j,k} |$, \eqref{eqn_eik_bound} is further bounded by
		\begin{align} \label{eqn_eik_bound2}
			| e_{i,k} |^2  \le& (b_{ii} + l_{ii} )^2 \sum\nolimits_{j \in \{ i, \mathcal{N}_i \}} \vartheta_{j,k}^2 \nonumber \\
			&+ (b_{ii} + l_{ii} ) \left(\sum\nolimits_{j \in \mathcal{N}_i} a_{ij} (\vartheta_{i,k}^2 + \vartheta_{j,k}^2 ) \right) \\
			\le& 2 (b_{ii} + l_{ii} )^2 \sum\nolimits_{j \in \mathcal{V}} ~ \vartheta_{j,k}^2 , \nonumber
		\end{align}
		considering $\{ i, \mathcal{N}_i \} \subseteq \mathcal{V}$ and $a_{ij} = 0, \forall j \notin \mathcal{N}_i$.
		Apply the boundness of $| e_{i,k} |$ in \eqref{eqn_eik_bound2}, the upper bound of the filtered error $r_i$ is written as
		\begin{align}
			|r_i|^2 &\le 2 ( l_{ii} + b_{ii} )^2 \sum\nolimits_{k=1}^n \lambda_k^2 \sum\nolimits_{j \in \mathcal{V}} \vartheta_{j,k}^2 \nonumber \\
			& \le 2 ( l_{ii} + b_{ii} )^2 (\lambda^*)^2 \sum\nolimits_{k=1}^n \sum\nolimits_{j \in \mathcal{V}} \vartheta_{j,k}^2. \nonumber
		\end{align}
		Note that $\| \bm{\vartheta} \|^2 = \sum_{k=1}^n \sum_{j = 1}^N |\vartheta_{j,k}|^2$, then it is derived that $|r_i| \le \sqrt{2} ( l_{ii} + b_{ii} ) \lambda^* \| \bm{\vartheta} \|$.
		Next, we investigate the upper bound of $\tilde{\eta}_{\delta, i}(\cdot)$ in \eqref{eqn_dot_vartheta_i_2}.
		Since the optimization problem \eqref{eqn_max_aggregated_prediction_bound} for agent $i$ covers the case in \eqref{eqn_upperbound_posterior_variance_after_model_update} by removing the constraint on $\sigma_i(\cdot)$, it has $\bar{\hat{\eta}}_{\delta,i} \ge \underline{\hat{\eta}}_{\delta,i}$ such that
		\begin{align} \label{eqn_upperbound_tilde_eta_i}
			\tilde{\eta}_{\delta,i}(\bm{x}_i) \le \varpi \underline{\hat{\eta}}_{\delta,i} + (1 - \varpi) \bar{\hat{\eta}}_{\delta,i} \le \bar{\hat{\eta}}_{\delta,i}, ~ \forall \bm{x}_i \in \mathbb{X}.
		\end{align}
		Apply the upper bound for $|r_i|$ and \eqref{eqn_upperbound_tilde_eta_i} into \eqref{eqn_dot_vartheta_i_2}, then $\| \dot{\bm{\vartheta}}_i \|$ is upper bounded by
		\begin{align}
			\| \dot{\bm{\vartheta}}_i \| \!\le\! \big(1 \!+\! \sqrt{2} c ( l_{ii} \!+\! b_{ii} ) \lambda^* \big) \| \bm{\vartheta} \| \!+\! (1 \!-\! b_{ii}) F_l \!+\! \bar{\hat{\eta}}_{\delta, i}. \nonumber
		\end{align}
		Moreover, considering $\| \dot{\bm{x}}_i \| = \| \dot{\bm{x}}_l + \dot{\bm{s}}_i + \dot{\bm{\vartheta}}_i \| \le \| \dot{\bm{x}}_l + \dot{\bm{s}}_i \| + \| \dot{\bm{\vartheta}}_i \|$ and the boundness of the reference in \cref{assumption_desired_trajectory} as well as the tracking error bound $\bar{\vartheta}_e$ from \cref{proposition_centralized_ET} or \cref{theorem_trigger}, the upper bound $F_i$ in \eqref{eqn_Fi} of state changing rate for agent $i$, i.e., $\| \dot{\bm{x}}_i \|$, is derived.
		Furthermore, the probability as at least $1 - N^2 \delta$ is inherited by considering the usage of all the local predictions in $\bar{\vartheta}_e$.
	\end{IEEEproof}
	
	\ifarxiv
	\todo{
		\cref{lemma_Fi} shows the state changing rate is bounded by a non-zero constant $F_i$ for each agent $i$.
		Note that the value of $F_i$ relies on connectivity of agent $i$ reflected by $l_{ii}$ and $b_{ii}$.
		In particular, strong connectivity with higher $l_{ii}$ and $b_{ii}$ induces larger changing rate $F_i$, which is intuitive since it allows agent $i$ react faster to the consensus error by using more neighboring information and tracking error\cite{lewis2013cooperative}, respectively.
		Moreover, $F_i$ also depends on the smoothness of the reference with $F_{r,i}$, which is straightforward since the state changes rapidly for fast changing reference with bounded tracking error by $\bar{\vartheta}$.
		The relationship between local prediction accuracy $\bar{\hat{\eta}}_{\delta, i}$ and $F_i$ is also shown in \cref{lemma_Fi}, indicating smaller prediction error reduce the conservatism for the approximation of $F_i$.
	}
	\fi
	
	Besides the bounded state changing rate, the exclusion of the Zeno behavior also requires the continuity of the prediction performance reflected by its error bound $\hat{\eta}_{\delta,i}(\cdot)$ w.r.t. $\bm{x}_i$, which means the derivative of $\hat{\eta}_{\delta,i}(\bm{x}_i)$ , i.e., $\nabla \hat{\eta}_{\delta,i}(\bm{x}_i) = \mathrm{d} \hat{\eta}_{\delta,i}(\bm{x}_i) / \mathrm{d} \bm{x}_i$, is bounded.
	However, considering the event-triggered online learning, $\nabla \hat{\eta}_{\delta,i}(\bm{x}_i(t_i^{(\varsigma)}))$ is not continuous at $t_i^{(\varsigma)}$ due to the change of the GP model with updated data set.
	Therefore, we intend to bound $\nabla \hat{\eta}_{\delta,i}(\bm{x}_i(t))$ on agent $i$ in the time interval without model update locally or on the neighbors, i.e., $t_i \in [\underline{t}_i, \bar{t}_i)$, where $\bar{t}$ and $\underline{t}$ are defined as
	\begin{align} \label{eqn_t_lower_and_upper_bound_for_no_update}
		&\underline{t}_i = \max_{j \in \{ i, \bar{\mathcal{N}}_i \}, \varsigma \in \mathbb{N}_+} \left\{t_j^{(\varsigma)} \in \mathbb{R}_{0,+} : t_j^{(\varsigma)} \le t_i \right\}, \\
		&\bar{t}_i = \min_{j \in \{ i, \bar{\mathcal{N}}_i \}, \varsigma \in \mathbb{N}_+} \left\{t_j^{(\varsigma)} \in \mathbb{R}_{0,+} : t_j^{(\varsigma)} > t_i \right\}.
	\end{align}
	Note that considering the definition of $\hat{\eta}_{\delta,i}(\cdot)$ as $\hat{\eta}_{\delta,i}(\cdot) = \sqrt{\beta_{\delta}} \hat{\sigma}_{i}(\cdot) + \gamma_{\delta}$, the boundness of $\nabla \hat{\eta}_{\delta,i}(\bm{x}_i)$ only requires the bounded $\nabla \hat{\sigma}_{i}(\bm{x}_i)$.
	The existence of a well-defined upper bound for $\| \nabla \hat{\sigma}_{i}(\bm{x}_i) \|$ is easy to see by assuming the aggregation weight $\omega_{ij}(\bm{x}_i), \forall j \in \{ i, \bar{\mathcal{N}}_i \}$ is continuous.
	Then, $\omega_{ij}(\bm{x}_i)$ is Lipschitz continuous within the compact domain $\mathbb{X}$ with the Lipschitz constant defined as $L_{\omega,ij} \in \mathbb{R}_{0,+}$ for $j \in \{ i, \bar{\mathcal{N}}_i \}$ and $i \in \mathcal{V}$.
	Moreover, the individual posterior variance $\sigma_i(\cdot)$ is also Lipschitz continuous with Lipschitz constant $L_{\sigma,i}$ as shown in \cref{lemma_GP_error_bound} under \cref{assumption_kernel}, whose detailed expression in \cite{lederer2019uniform, lederer2021gaussian}.
	With the Lipschitz aggregation weights and individual posterior variances and the choice of $\omega_{ij} \in \mathbb{R}_{0,+}$ such that $\sum_{j \in \{ i, \bar{\mathcal{N}}_i \}} \omega_{ij}(\cdot) = 1$, the upper bound of $\| \nabla \hat{\sigma}_{i}(\bm{x}_i) \|$ is obtained for $\forall i \in \mathcal{V}$ by considering
	\begin{align}
		\| \nabla \hat{\sigma}_{i\!}(\bm{x}_i\!) \| \!\le& \!\!\!\!\sum_{j \in \{\!i,\bar{\mathcal{N}}_i\!\}} \!\!\!\!\! \left(\| \nabla \omega_{ij}(\bm{x}_i) \| \sigma_j(\bm{x}_i) \!+\! \omega_{ij}(\bm{x}_i) \| \nabla \sigma_j(\bm{x}_i) \|  \right) \nonumber \\
		\le& \sum\nolimits_{j \in \{i,\bar{\mathcal{N}}_i\}} \left(L_{\omega, ij} \sigma_j(\bm{x}_i) \!+\! \omega_{ij}(\bm{x}_i) L_{\sigma,j} \right) \nonumber \\
		\le& \sum\nolimits_{j \in \{i,\bar{\mathcal{N}}_i\}} \left(L_{\omega, ij} \sigma_{f,j} \!+\! L_{\sigma,j} \right)  \nonumber
	\end{align}
	for all $\bm{x}_i \!\in\! \mathbb{X}$, where the third inequality is derived by considering $\sigma_j(\cdot) \!\le\! \sigma_{f,j}$ from \eqref{eqn_GP_prediction} and $\omega_{ij}(\cdot) \!\le\! 1$.
	For notational simplicity, define the upper bound for $\| \nabla \hat{\sigma}_{i}(\bm{x}_i) \|$, i.e., Lipschitz constant for $\hat{\sigma}_{i}(\bm{x}_i)$, as $\hat{L}_{\sigma,i} \!=\! \sum_{j \in \{i,\bar{\mathcal{N}}_i\}} (L_{\omega, ij} \sigma_{f,j} \!+\! L_{\sigma,j} )$.
	The boundness in whole time domain $\mathbb{R}_{0,+}$ is directly derived by considering $\| \nabla \hat{\sigma}_{i}(\cdot) \| \!\le\! \hat{L}_{\sigma,i}$ in $[\underline{t}_i, \bar{t}_i)$ with \eqref{eqn_t_lower_and_upper_bound_for_no_update} for $\forall t_i \!\in\! \mathbb{R}_{0,+}$.
	\ifarxiv
	\todo{
		Specifically for MOE \eqref{eqn_GP_MOE} and POE \eqref{eqn_GP_POE}, the upper bound $\hat{L}_{\sigma,i}$ is tighter due to the given aggregation structure, which is shown in the following corollary.
		
		\begin{corollary} \label{corollary_derivative_variance}
			The Lipschitz constants for $\hat{\sigma}_i(\cdot)$ for MOE \eqref{eqn_GP_MOE} and POE \eqref{eqn_GP_POE} are derived as follows:
			\begin{flalign} \label{eqn_MOE_L_sigma}
				\text{(\romannumeral 1) For MOE, }~\hat{L}_{\sigma,i} = \omega_{ii} L_{\sigma,i} + \sum{_{j \in \bar{\mathcal{N}}_i}} \omega_{ij} L_{\sigma,j}; &&
			\end{flalign}
			(\romannumeral 2) For POE with $\hat{\sigma}_i^2(\cdot) \!\!=\!\!  (\sum_{j \!\in\! \{\! i, \bar{\mathcal{N}}_i \!\}} \omega_{ij}^*)\!(\sum_{j \!\in\! \{\! i, \bar{\mathcal{N}}_i \!\}} \omega_{ij}^* \sigma_{j}^{\!-\!2\!}(\cdot))^{\!-\!1\!}$,
			\begin{align} \label{eqn_POE_L_sigma}
				\hat{L}_{\sigma,i\!} \!=\! \big(\sum{_{j \in \{ i, \bar{\mathcal{N}}_i \}}} \omega_{ij}^* \big)^{\!\frac{1}{2}} \! \sum{_{j \in \{ i, \bar{\mathcal{N}}_i \}}} \big(\omega_{ij}^*\big)^{-\!\frac{1}{2}} L_{\sigma,j}.
			\end{align}
		\end{corollary}
		
		\begin{IEEEproof}
			(\romannumeral 1)
			For MOE with aggregation weights recalled as $\omega_{ij}(\bm{x}) \!=\! \omega_{ij} \!\in\! \mathbb{R}_{0,+}$, the derivative $\nabla \hat{\sigma}_i(\bm{x}_i)$ is bounded by using triangular inequality as
			\begin{align} \label{eqn_derivative_variance_MOE}
				\| \nabla \hat{\sigma}_i(\bm{x}_i) \| \!\le\! \!\!\!\! \sum_{j \in \{ i, \bar{\mathcal{N}}_i \}} \!\!\!\! \omega_{ij} \| \nabla \sigma_j(\bm{x}_i) \| \!\le\! \!\!\!\! \sum_{j \in \{ i, \bar{\mathcal{N}}_i \}} \!\!\!\! \omega_{ij} L_{\sigma,j},
			\end{align}
			for any $\bm{x}_i \in \mathbb{X}$ and $i \in \mathcal{V}$, which is identical as \eqref{eqn_MOE_L_sigma}.
			\\
			(\romannumeral 2)
			For POE with the expression of $\hat{\sigma}_i(\cdot)$ in the setting, define an auxiliary function as $\check{\sigma}_i(\bm{x}_i) \!=\! \hat{\sigma}_i^{-2}(\bm{x}_i) (\sum_{j \in \{ i, \bar{\mathcal{N}}_i \}} \omega_{ij}^*) $.
			Using the chain rule, the derivative of $\hat{\sigma}_i(\cdot)$ is written as $\nabla \hat{\sigma}_i(\bm{x}_i) = \left(\partial \hat{\sigma}_i(\bm{x}_i) / \partial \check{\sigma}_i(\bm{x}_i) \right) \nabla \check{\sigma}_i(\bm{x}_i)$, where
			\begin{align}
				&\frac{\partial \hat{\sigma}_i(\bm{x}_i)}{\partial \check{\sigma}_i(\bm{x}_i)} \!=\! -\frac{1}{2} \sqrt{\frac{\sum_{j \in \{ i, \bar{\mathcal{N}}_i \}} \omega_{ij}^*}{\check{\sigma}_i^3(\bm{x}_i)}} \!=\! -\frac{1}{2} \frac{\hat{\sigma}_i^{3}(\bm{x}_i)}{\sum_{j \in \{ i, \bar{\mathcal{N}}_i \}} \omega_{ij}^*}, \\
				&\nabla \check{\sigma}_i(\bm{x}_i) = -2 \sum_{j \in \{ i, \bar{\mathcal{N}}_i \}} \omega_{ij}^* \sigma_j^{-3}(\bm{x}_i) \frac{\mathrm{d} \sigma_i(\bm{x}_i)}{\mathrm{d} \bm{x}_i}.
			\end{align}
			Then, the norm of $\nabla \hat{\sigma}_i(\bm{x}_i)$ is upper bounded as
			\begin{align}
				\| \nabla \hat{\sigma}_i(\bm{x}_i) \| &\!=\! \Big\| \frac{\hat{\sigma}_i^{3}(\bm{x}_i)}{\sum_{j \in \{ i, \bar{\mathcal{N}}_i \}} \omega_{ij}^*} \sum_{j \in \{ i, \bar{\mathcal{N}}_i \}} \omega_{ij}^* \sigma_j^{-3}(\bm{x}_i) \frac{\mathrm{d} \sigma_i(\bm{x}_i)}{\mathrm{d} \bm{x}_i} \Big\| \nonumber \\
				&\!\le\! \sum_{j \in \{ i, \bar{\mathcal{N}}_i \}} \frac{\omega_{ij}^*}{\sum_{k \in \{ i, \bar{\mathcal{N}}_i \}} \omega_{ik}^*} \frac{\hat{\sigma}_i^{3}(\bm{x}_i)}{\sigma_j^{3}(\bm{x}_i)} \Big\|\! \frac{\mathrm{d} \sigma_i(\bm{x}_i)}{\mathrm{d} \bm{x}_i} \!\Big\| \\
				&\!=\! \Big(\sum_{j \in \{ i, \bar{\mathcal{N}}_i \}} \omega_{ij}^* \Big)^{\frac{1}{2}} \!\!\! \sum_{j \in \{ i, \bar{\mathcal{N}}_i \}} \omega_{ij}^*  \frac{\check{\sigma}_i^{-\frac{3}{2}}(\bm{x}_i)}{\sigma_j^{3}(\bm{x}_i)} \Big\|\! \frac{\mathrm{d} \sigma_i(\bm{x}_i)}{\mathrm{d} \bm{x}_i} \!\Big\|, \nonumber
			\end{align}
			where the second equality is obtained by considering the definition of $\check{\sigma}_i(\cdot)$.
			Moreover, $\check{\sigma}_i(\cdot)$ is also written as
			\begin{align}
				\check{\sigma}_i(\cdot) = \sum{_{j \in \{ i, \bar{\mathcal{N}}_i \}}} \omega_{ij}^* \sigma_{j}^{-2}(\cdot) \ge \omega_{ij}^* \sigma_{j}^{-2}(\cdot)
			\end{align}
			for all $j \in \{ i, \bar{\mathcal{N}}_i \}$ due to the non-negative aggregation weights $\omega_{ij}^*$ and posterior variances $\sigma_{j}(\cdot)$.
			This also indicates $\check{\sigma}_i^{-3/2}(\bm{x}_i) \le (\omega_{ij}^*)^{-3/2} \sigma_{j}^{3}(\bm{x}_i)$, such that
			\begin{align}
				\| \nabla \hat{\sigma}_i(\bm{x}_i) \| &\le \Big(\sum_{j \in \{ i, \bar{\mathcal{N}}_i \}} \omega_{ij}^* \Big)^{\frac{1}{2}} \!\!\! \sum_{j \in \{ i, \bar{\mathcal{N}}_i \}} \Big(\omega_{ij}^*\Big)^{-\frac{1}{2}} L_{\sigma,j}, 
			\end{align}
			using the Lipschitz constants $L_{\sigma,j}$ for $\sigma_{j}(\cdot)$, $\forall j \in \{ i, \bar{\mathcal{N}}_i \}$.
		\end{IEEEproof}
		
		\cref{corollary_derivative_variance} shows the existence of the Lipschitz constants $\hat{L}_{\sigma,i}$ for the derivative of $\hat{\sigma}_i(\cdot)$ by using the common aggregation strategies, in particular for MOE and POE.
		The expression of Lipschitz constant $L_{\sigma,i}$ for the posterior variance of the $i$-th local GP model depends on the choice of the kernel function $\kappa_i(\cdot,\cdot)$, and specifically for the widely applied square exponential (SE) kernel the calculation of $L_{\sigma,i}$ the detailed expression of $L_{\sigma,i}$ refer to \cite{lederer2021gaussian}.
		
		\begin{remark}
			With similar reason as in \cref{remark_POE_upperbound}, the expression in \eqref{eqn_GPOE_posterior_error_variance} is not chosen to evaluate $\hat{\sigma}_i(\cdot)$ due to its abnormal behavior discussed in \cref{remark_POE_upperbound} and infinite derivative when $\sigma_{j}(\cdot) \to 0$ for any $j \in \{ i, \bar{\mathcal{N}}_i \}$.
			Instead, the same formulation as \eqref{eqn_upperbound_aggregated_posterior_variance_gPOE} is used in \cref{corollary_derivative_variance} to inherit the property of aggregated posterior variance in \cite{cao2014generalized}.
		\end{remark}
	}
	\fi
	
	With the bounded state changing rate as in \cref{lemma_Fi} and bounded change of prediction performance analyzed above, the exclusion of the Zeno behavior for the both proposed centralized and distributed event-trigger mechanism shown in \eqref{eqn_centralized_trigger} and \eqref{eqn_trigger} is proven as follows.
	
	\begin{proposition} \label{lemma_distributed_ET_Zeno}
		Let all assumptions in \cref{lemma_Fi} hold and use the event-triggered online cooperative learning with either centralized mechanism in \eqref{eqn_centralized_trigger} with heuristic update model selection in \eqref{eqn_centralized_ET_distributed_heuristic} or distributed mechanism in \eqref{eqn_trigger}, in which $\epsilon_i \in \mathbb{R}_+$ is chosen for each agent $i \in \mathcal{V}$.
		Pick $\delta \in (0, N^{-2})$, then the inter-event time $\Delta_i^{(\varsigma)} = t_i^{(\varsigma + 1)} - t_i^{(\varsigma)}, \forall \varsigma \in \mathbb{N}$ for each agent $i \in \mathcal{V}$ is lower bounded by $\underline{\Delta}_i \!\in\! \mathbb{R}_+$ as
		\begin{align} \label{eqn_underline_Delta_i}
			\underline{\Delta}_i = \big(\sqrt{\beta_{\delta}} F_i \hat{L}_{\sigma,i} \big)^{-1} \epsilon_i
		\end{align}
		with probability of at least $1 \!-\! N^2 \delta$.
	\end{proposition}
	
	\begin{IEEEproof}
		From the design of trigger condition in \eqref{eqn_centralized_trigger_2} and \eqref{eqn_trigger} with the heuristic model selection strategy in \eqref{eqn_centralized_ET_distributed_heuristic}, it is obvious the model update will be activated on agent $i$ at $t_i^{(\varsigma + 1)}$ if $\rho_i(t_i^{(\varsigma + 1)}) > \bar{\rho}_i(t_i^{(\varsigma + 1)})$ indicating $\hat{\eta}_{\delta,i}(\bm{x}_i(t_i^{(\varsigma + 1)})) > \underline{\hat{\eta}}_{\delta,i} + \epsilon_i$.
		Moreover, considering the model update on the agent $i$ occurs at $t_i^{(\varsigma)}$ such that the aggregated prediction error bound after model update is upper bounded as $\hat{\eta}_{\delta,i}^+(\bm{x}_i(t_i^{(\varsigma)})) \le \underline{\hat{\eta}}_{\delta,i}$ from \cref{proposition_online_cooperative_prediction_error_bound}.
		Note that no new data pairs are added into the data set $\mathbb{D}_i$ in $(t_i^{(\varsigma)}, t_i^{(\varsigma + 1)})$, such that the prediction model on agent $i$ maintains unchanged.
		With the same prediction model, the difference between $\hat{\eta}_{\delta,i}^+(\bm{x}_i(t_i^{(\varsigma)}))$ and $\hat{\eta}_{\delta,i}(\bm{x}_i(t_i^{(\varsigma + 1)}))$ is lower bounded by $\epsilon_i$, which is also written as
		\begin{align} \label{eqn_Delta_i_1}
			\epsilon_i &< \hat{\eta}_{\delta,i}(\bm{x}_i(t_i^{(\varsigma + 1)})) - \hat{\eta}_{\delta,i}^+(\bm{x}_i(t_i^{(\varsigma)})) \\
			&\leq \int\nolimits_{t_i^{(\varsigma)}}^{t_i^{(\varsigma + 1)}}  \!\!\! | \dot{\hat{\eta}}_i(\bm{x}_i(\tau)) | d \tau = \sqrt{\beta_{\delta}} \int\nolimits_{t_i^{(\varsigma)}}^{t_i^{(\varsigma + 1)}}  \!\!\! | \dot{\hat{\sigma}}_i(\bm{x}_i(\tau)) | d \tau. \nonumber
		\end{align}
		Using the chain rule, the derivative of $\hat{\sigma}_i(\bm{x}_i)$ is bounded by
		\begin{align}
			| \dot{\hat{\sigma}}_i(&\bm{x}_i) | \!=\! \Big| \dot{\bm{x}}_i^T \frac{\mathrm{d} \hat{\sigma}_i(\bm{x}_i)}{\mathrm{d} \bm{x}_i} \Big| \!\le\! \| \dot{\bm{x}}_i \| \Big\| \frac{\mathrm{d} \hat{\sigma}_i(\bm{x}_i)}{\mathrm{d} \bm{x}_i} \Big\| \!\le\! F_i \hat{L}_{\sigma,i} \nonumber
		\end{align}
		with probability of at least $1 - N^2 \delta$, due the result for $F_i$ and $\hat{L}_{\sigma,i}$ from \cref{lemma_Fi}.
		Then, \eqref{eqn_Delta_i_1} is reformulated by
		\begin{align}
			\epsilon_i < \sqrt{\beta_{\delta}} F_i \hat{L}_{\sigma,i} \Delta_i^{(\varsigma)}, \nonumber
		\end{align}
		indicating the boundness as in \eqref{eqn_underline_Delta_i} holds for $\forall \varsigma \in \mathbb{N}$.
	\end{IEEEproof}
	
	\cref{lemma_distributed_ET_Zeno} shows the minimal trigger interval for each agent is non-zero by using both centralized and distributed event-triggered mechanism for cooperative online learning, indicating the exclusion of the Zeno behavior.
	Note that although the proof for centralized event-trigger only works for heuristic update model selection strategy in \eqref{eqn_centralized_ET_distributed_heuristic}, the Zeno behavior exclusion for other selection methods is easily derived by using the similar methods in \cite{umlauft2019feedback,jiao2022backstepping}, which considers the minimal trigger interval for the whole MAS.
	Moreover, \cref{lemma_distributed_ET_Zeno} directly shows, only choosing strict positive $\epsilon_i$ leads to guaranteed non-zero minimal trigger interval $\underline{\Delta}_i$, resulting in Zeno behavior exclusion.
	Combining with the tracking error bound in \cref{proposition_centralized_ET} and \cref{theorem_trigger}, there exists a trade-off between the control performance and update frequency, i.e., larger $\epsilon_i$ induces worse tracking performance but allowing low model update rate.
	
	\ifarxiv
	\todo{
		\begin{remark}
			While the trigger condition for each agent varies, the overall tracking error is bounded by $\bar{\vartheta}_e$ for any choice of $\epsilon_1, \cdots, \epsilon_N$ such that $\bar{\vartheta}_e = \xi \chi \| \bm{\iota} + \underline{\hat{\bm{\eta}}}_{\delta} + \bm{\epsilon} \|$.
			This free choice of $\epsilon_i$ provides a possibility to design the performance-guaranteed frequency allocation method.
			Intuitively, choosing large $\epsilon_i$ for the strongly connected agents can reduce their trigger frequency, while by simultaneously decreasing $\epsilon_i$ for weakly connected agents a desired $\bar{\vartheta}_e$ can still be achieved.
		\end{remark}
	}
	\fi
	
	\section{Numerical Simulations} \label{section_simulation}
	
	\subsection{Toy Example}
	
	\subsubsection{Simulation Setting}
	
	In this section\footnote{The code is available at \url{https://drive.google.com/drive/folders/1cOdToV_h__VWfHNKmi-ib-VioJSpKCGq?usp=sharing}.}, we consider a multi-agent system including $N \!=\! 4$ agents.
	Each agent $i \in \mathcal{V} = \{1,2,3,4\}$ follows the dynamics in \eqref{eqn_agent_dynamics} with $n \!=\! 2$, $h(\bm{x}_i) \!=\! 1$, $g(\bm{x}_i) \!=\! 1$ and the unknown function \looseness=-1
	\begin{align}
		f(\bm{x}_i) = 5 \sin (10 x_{i,1}) + 0.5 / (1 + \exp (x_{i,2} / 10)) + 10, \nonumber
	\end{align}
	\ifarxiv
	\todo{
	where the input domain is set as $ \mathbb{X} = [-1.5,1.5] \times [-0.6,0.6]$ as shown in \cref{figure_UnknownFunction}.
	The agents are connected with a directed communication network defined by the edge set as $\mathcal{E} = \{ (2,3), (3,1), (3,2),$ $(3,4), (4,1), (4,2), (4,3) \}$ and $a_{ij} = 1, \forall (j,i) \in \mathcal{E}$.
	Moreover, only agent $1$ and $3$ have the access to the leader, i.e., $b_{ii} = 1$ for $i = 1, 3$ and $b_{ii} = 0$ for $i = 2,4$.
	The topology is visualized as in \cref{figure_topology}.
	}
	\fi
	\ifarxiv
	\todo{
		The prediction of $f(\cdot)$ is obtained by \eqref{eqn_GP_aggregation} using POE \cite{liu2020gaussian}.
	}
	\fi
	Moreover, consider $\underline{\hat{\sigma}}_{i}^2  =  (|\bar{\mathcal{N}}_i| + 1) ((\sigma_{o,i}^{-2} + \sigma_{f,i}^{-2}) + \sum_{j \in \bar{\mathcal{N}}_i} \sigma_{f,j}^{-2})^{-1}$ as the solution of \eqref{eqn_upperbound_posterior_variance_after_model_update}, such that it is easy to see $\hat{L}_{\sigma,i} = (|\bar{\mathcal{N}}_i| + 1 )^{\frac{1}{2}} \sum_{j \in \{ i, \bar{\mathcal{N}}_i \}} L_{\sigma,j}$.
	The kernel functions $\kappa(\cdot,\cdot)$ in all local Gaussian processes are identical, which is in square exponential form as
	\begin{align}
		\kappa_i(\bm{x}_i, \bm{x}'_i) = \sigma_{f,i}^2 \exp (- 0.5 l_i^{-2} \| \boldsymbol{x}_i - \bm{x}'_i\|^2), ~\forall i \!\in\! \mathcal{V} \nonumber
	\end{align}
	with $\sigma_{f,i} = 1, l_i = 0.1$.
	The data pairs for the prediction satisfy \cref{assumption_dataset} with the measurements $y_i$ perturbed by Gaussian noise with $\sigma_{o,i} = 0.01$, $\forall i \in \mathcal{V}$.
	The initial data sets vary depending on the applied learning methods, and are discussed later.
	The grid factor and probability for uniform error bound in \cref{lemma_GP_error_bound} are set as $\tau = 10^{-6}$ and $\delta = 0.001$, respectively.
	The control objective is to track the leader trajectory with
	\begin{align} \label{eqn_experiment_xl}
		x_{l,1} = \sin (2 t / 5), &&
		x_{l,2} = \dot{x}_{l,1}, &&
		x_{l,r} = \dot{x}_{l,2},
	\end{align}
	and maintain relative states for each agent $i$ as
	\begin{align} \label{eqn_experiment_si}
		s_{i,1} \!=\! 0.01 \sin (6 t + 2 \pi i / N), &&
		s_{i,2} \!=\! \dot{s}_{i,1}, &&
		s_{i,r} \!=\! \dot{s}_{i,2},
	\end{align}
	such that the individual reference trajectories are shown in \cref{figure_dataset_reference}.
	The control law \eqref{eqn_control_law} is designed for each agent $i$, where the control gains in \eqref{eqn_consensus_control_law} are set as $c \!=\! 20$ and $\lambda_1 \!=\! \lambda_2 \!=\! 1$.
	Moreover, select $\bm{Q}_{\varepsilon} \!=\! 1$ and then the matrix $\bm{Q}_z$ in \eqref{eqn_Qz} is ensured to be positive definite.
	The simulation time is set to $40$.
	
	\ifarxiv
	\begin{figure}[t]
		\centering  
		\begin{subfigure}{0.4\columnwidth}
			\centering  
			\begin{tikzpicture}[scale=0.5,every node/.append style={transform shape}]
				\tikzstyle{agent_collectable} = [very thick, circle, minimum width = 0.5cm, minimum height=1.2cm,text centered, draw = black ,fill= white ]
				\tikzstyle{agent} = [very thick, circle, minimum width = 0.5cm, minimum height=1.2cm,text centered, draw = black ,fill= black!20 ]
				\tikzstyle{agent_empty} = [very thick, circle, minimum width = 0.5cm, minimum height=1.2cm,text centered, draw = white ,fill= white, fill opacity=0 ]
				\tikzstyle{arrow1} = [very thick, <->,>=stealth]
				\tikzstyle{arrow} = [very thick, ->,>=stealth]
				
				\node (leader) [agent_collectable, 	xshift=2cm, yshift=5cm] {};
				\node (agent1) [agent_collectable, 	xshift=0cm, yshift=3cm] {\large 1};
				\node (agent2) [agent_collectable, 	xshift=0cm, yshift=0cm] {\large 2};
				\node (agent3) [agent_collectable, 	xshift=4cm, yshift=3cm] {\large 3};
				\node (agent4) [agent_collectable, 	xshift=4cm, yshift=0cm] {\large 4};
				\node at (2,6) {\Large Leader};
				
				\draw [arrow]    (leader) to  (agent1);
				\draw [arrow]    (leader) to  (agent3);
				\draw [arrow]    (agent3) to  (agent1);
				\draw [arrow]    (agent4) to  (agent2);
				\draw [arrow1]   (agent1) to  (agent4);
				\draw [arrow1]   (agent2) to  (agent3);
				\draw [arrow1]   (agent3) to  (agent4);
				
			\end{tikzpicture}
			\caption{}
			\label{figure_topology}
		\end{subfigure}
		\begin{subfigure}{0.55\columnwidth}
			\def\file{fig/UnknownFunction.txt}
			\begin{tikzpicture}
				\begin{axis}[xlabel={$x_1$},ylabel={$x_2$},zlabel={$f(\bm{x})$},
					view={-37.5}{80},
					xmin=-1.5, ymin = -0.6, xmax = 1.5,ymax=0.6, zmin=1, zmax=11,legend columns=1,
					width=5.5cm,height=5cm,legend style={at={(1.05,0.5)},anchor=west},
					clip mode=individual]
					\addplot3[surf,	opacity=0.6, domain=5:10,domain y=5:10,mesh/cols=10, colormap/jet, line width=0.01pt]    table[x = X1_f , y  = X2_f, z = Y_f ]{\file};
				\end{axis}
			\end{tikzpicture}
			\caption{}
			\label{figure_UnknownFunction}
		\end{subfigure}
		\vspace{-0.3cm}
		\caption{
			\todo{
			(a) Communication topology among agents and leader; 
			(b) The manifold of the unknown function $f(\cdot)$ evaluated on the compact domain $\mathbb{X}$. 
			}
		}
		\vspace{-0.4cm}
	\end{figure}
	\fi
	
	\begin{figure}[t] 
		\centering
		\def\file{fig/reference.txt}
		\begin{tikzpicture}
			\begin{axis}[xlabel={$x_1$},ylabel={$x_2$},
				xmin=-1.5, ymin = -0.6, xmax = 1.5,ymax=0.6,legend columns=1,
				width=7cm,height=4.2cm,legend style={at={(1.05,0.5)},anchor=west},
				clip mode=individual]
				\addplot[only marks, mark=o, black!40]    table[x = dataset_x1 , y  = dataset_x2 ]{\file};
				
				\addplot[paried_red!80, thick]    table[x = s_1_1_set_tikz , y  = s_1_2_set_tikz ]{\file};
				\addplot[paried_blue!80, thick]    table[x = s_2_1_set_tikz , y  = s_2_2_set_tikz ]{\file};
				\addplot[paried_green!80, thick,]    table[x = s_3_1_set_tikz , y  = s_3_2_set_tikz ]{\file};
				\addplot[paried_orange!80, thick]    table[x = s_4_1_set_tikz , y  = s_4_2_set_tikz ]{\file};
				
				\addplot[black, thick, dashed]    table[x = x_l_1_set_tikz , y  = x_l_2_set_tikz ]{\file};
				\legend{Data set, $\bm{x}_l \!+\! \bm{s}_1$, $\bm{x}_l \!+\! \bm{s}_2$, $\bm{x}_l \!+\! \bm{s}_3$, $\bm{x}_l \!+\! \bm{s}_4$, $\bm{x}_l$}
			\end{axis}
		\end{tikzpicture}
		\vspace{-0.4cm}
		\caption{
			Uniformly distributed initial data set for offline learning for instance at agent $1$ and the references, including the trajectory for leader $\bm{x}_l$ and $\bm{x}_l + \bm{s}_i$ for each agent $i \in \mathcal{V}$.
		}
		\vspace{-0.3cm}
		\label{figure_dataset_reference}
	\end{figure}
	
	To demonstrate the effectiveness of the proposed event-triggered online learning method, we compare the tracking error $\| \bm{\vartheta}(t) \|$ by using the following learning strategies:
	\begin{enumerate}
		\item $\!$Offline cooperative learning \cite{yang2021distributed}: 
		Each agent $i \in \mathcal{V}$ provides the prediction $\sigma_i(\cdot)$ by using the initial offline data set, which contains $200$ random samples uniformly distributed in $\mathbb{X}$ as shown in \cref{figure_dataset_reference};
		\item $\!$Centralized event-trigger (CET): The trigger in \eqref{eqn_centralized_trigger} with heuristic update model selection in \eqref{eqn_centralized_ET_distributed_heuristic}. 
		\item $\!$Distributed event-trigger (DET): The trigger in \eqref{eqn_trigger}. 
		\item $\!$Time-trigger (TT) \cite{beckers2021online}: Update the local GPs on each agent every $0.015$ time interval.
		\item $\!$Exact model: Let $\hat{f}_i(\cdot) = f(\cdot)$ in \eqref{eqn_control_law} for $\forall i \in \mathcal{V}$.
	\end{enumerate}
	Note that the trigger interval for time-triggered learning, i.e., $0.015$, is set such that it is similar to the minimal trigger interval by using centralized and distributed event-triggered learning as shown in \cref{figure_trigger}.
	Moreover, for both centralized and distributed event-triggered online learning, the initial data sets for each agent are set as empty, and $\epsilon_i$ in \eqref{eqn_eta_underline} are chosen as $\epsilon_i \!=\! \sigma_{o,i} - (\sigma_{f,i}^{-2} + \sigma_{o,i}^{-2})^{-1/2}$, $\forall i \!\in\! \mathcal{V}$.
	The effectiveness of the above learning methods is shown in the following subsections.
	
	\subsubsection{Performance with Event-triggered Online Learning} \label{subsection_experiment_method}
	
	First, we observe the effectiveness of different cooperative learning strategies for MAS with same initial condition, i.e., $\bm{x}_i(0) = \bm{x}_l(0) + \bm{s}_i(0), \forall i \in \mathcal{V}$.
	\ifarxiv
	\todo{
		The states $\bm{x}_i$ and tracking error $\| \bm{\vartheta}_i(t) \|$ for each agent $i \in \mathcal{V}$ are shown in \cref{figure_Agent_StateAndError}, and the overall tracking error $\| \bm{\vartheta}(t) \|$ w.r.t time is shown in \cref{figure_error}.
	}
	\fi
	\ifarxiv
	\todo{
		While with offline learning the tracking error is much larger than with other methods especially in some period, e.g., $t \in (23, 28)$, due to the lack of sufficient data in the domain around references, online learning performs better with lower $\| \bm{\vartheta}(t) \|$.
		Among the online learning methods, time-triggered mechanism behaviors best with tracking error curve almost the same as with exact model, where the tracking error does not tend to $0$ due to the non-fully connection between the agents and leader inducing non-zeros $\bm{\iota}$ in \cref{theorem_best_tracking_performance_online}.
		Moreover, both event-triggered learning strategies perform merely slightly worse than the time-triggered mechanism after $t = 20$, due to the underestimation of the achievable prediction performance as in \cref{proposition_online_cooperative_prediction_error_bound} as well as the positive $\epsilon_i$ inducing larger guaranteed tracking error bound.
		However, the performance between centralized and distributed event-triggered mechanisms is similar, indicating almost no performance loss from the distributed computation.
	}
	\fi
	
	\ifarxiv
	\begin{figure}[t]
		\centering
		\begin{tikzpicture}
			\def\fileCET{fig/AgentStateError_centralized.txt}
			\def\fileDET{fig/AgentStateError_distributed.txt}
			\def\fileTT{fig/AgentStateError_continuous0015.txt}
			\def\fileOFF{fig/AgentStateError_NoTrigger.txt}
			\def\fileEXA{fig/AgentStateError_exact.txt}
			\def\fileREF{fig/AgentStateError_Reference.txt}
			
			\begin{axis}[xlabel={},ylabel={Agent $1$},
				xmin=1, ymin = -1.1, xmax = 39,ymax=2.6,legend columns=2,
				width=0.26\textwidth,height=4cm,legend style={at={(0.02,0.8)},anchor=west},
				xticklabels={,,}]
				\addplot[pink!80, very thick, dashed]      table[x = t_set_norminal , y  = Agent1_r1 ]{\fileREF};
				\addplot[paried_red!80, semithick, dashed]      table[x = t_set_norminal , y  = Agent1_x1 ]{\fileCET};
				\addplot[paried_blue!80, semithick, dashed]     table[x = t_set_norminal , y  = Agent1_x1 ]{\fileDET};
				\addplot[paried_orange!80, semithick, dashed]    table[x = t_set_norminal , y  = Agent1_x1 ]{\fileTT};
				\addplot[paried_green!80, semithick, dashed]    table[x = t_set_norminal , y  = Agent1_x1 ]{\fileOFF};
				\addplot[black!50, semithick, dashed]    table[x = t_set_norminal , y  = Agent1_x1 ]{\fileEXA};
				
				\addplot[pink!80, very thick, ]      table[x = t_set_norminal , y  = Agent1_r2 ]{\fileREF};
				\addplot[paried_red!80, semithick, ]      table[x = t_set_norminal , y  = Agent1_x2 ]{\fileCET};
				\addplot[paried_blue!80, semithick, ]     table[x = t_set_norminal , y  = Agent1_x2 ]{\fileDET};
				\addplot[paried_orange!80, semithick, ]    table[x = t_set_norminal , y  = Agent1_x2 ]{\fileTT};
				\addplot[paried_green!80, semithick, ]    table[x = t_set_norminal , y  = Agent1_x2 ]{\fileOFF};
				\addplot[black!50, semithick, ]    table[x = t_set_norminal , y  = Agent1_x2 ]{\fileEXA};
				
				\legend{$\!x_{l,\!1\!}\!+\!\! s_{1,\!1\!}$, , , , , $\!x_{1,\!1\!}\!\!$, $\!x_{l,\!2\!}\!+\!\! s_{1,\!2\!}$, , , , , $\!x_{1,\!2\!}\!\!$}
			\end{axis}
			\begin{semilogyaxis}[xlabel={},ylabel={},
				xmin=1, ymin = 0.6e-3, xmax = 39,ymax=2e-1,legend columns=5,
				width=0.26\textwidth,height=4cm,legend style={at={(-1.6,1.12)},anchor=west},,
				xticklabels={,,},
				xshift=4cm]
				\addplot[paried_red!80, thick, ]      table[x = t_set_norminal , y  = Agent1_e ]{\fileCET};
				\addplot[paried_blue!80, thick, ]     table[x = t_set_norminal , y  = Agent1_e ]{\fileDET};
				\addplot[paried_orange!80, thick, ]    table[x = t_set_norminal , y  = Agent1_e ]{\fileTT};
				\addplot[paried_green!80, thick, ]    table[x = t_set_norminal , y  = Agent1_e ]{\fileOFF};
				\addplot[black!50, thick, ]    table[x = t_set_norminal , y  = Agent1_e ]{\fileEXA};
				\legend{CET, DET, TT, Offline learning, Exact model}
			\end{semilogyaxis}
			
			\begin{axis}[xlabel={},ylabel={Agent $2$},
				xmin=1, ymin = -1.1, xmax = 39,ymax=2.6,legend columns=2,
				width=0.26\textwidth,height=4cm,legend style={at={(0.02,0.8)},anchor=west},
				xticklabels={,,},
				yshift=-2.5cm]
				\addplot[pink!80, very thick, dashed]      table[x = t_set_norminal , y  = Agent2_r1 ]{\fileREF};
				\addplot[paried_red!80, semithick, dashed]      table[x = t_set_norminal , y  = Agent2_x1 ]{\fileCET};
				\addplot[paried_blue!80, semithick, dashed]     table[x = t_set_norminal , y  = Agent2_x1 ]{\fileDET};
				\addplot[paried_orange!80, semithick, dashed]    table[x = t_set_norminal , y  = Agent2_x1 ]{\fileTT};
				\addplot[paried_green!80, semithick, dashed]    table[x = t_set_norminal , y  = Agent2_x1 ]{\fileOFF};
				\addplot[black!50, semithick, dashed]    table[x = t_set_norminal , y  = Agent2_x1 ]{\fileEXA};
				
				\addplot[pink!80, very thick, ]      table[x = t_set_norminal , y  = Agent2_r2 ]{\fileREF};
				\addplot[paried_red!80, semithick, ]      table[x = t_set_norminal , y  = Agent2_x2 ]{\fileCET};
				\addplot[paried_blue!80, semithick, ]     table[x = t_set_norminal , y  = Agent2_x2 ]{\fileDET};
				\addplot[paried_orange!80, semithick, ]    table[x = t_set_norminal , y  = Agent2_x2 ]{\fileTT};
				\addplot[paried_green!80, semithick, ]    table[x = t_set_norminal , y  = Agent2_x2 ]{\fileOFF};
				\addplot[black!50, semithick, ]    table[x = t_set_norminal , y  = Agent2_x2 ]{\fileEXA};
				
				\legend{$\!x_{l,\!1\!}\!+\!\! s_{2,\!1\!}$, , , , , $\!x_{2,\!1\!}\!\!$, $\!x_{l,\!2\!}\!+\!\! s_{2,\!2\!}$, , , , , $\!x_{2,\!2\!}\!\!$}
			\end{axis}
			\begin{semilogyaxis}[xlabel={},ylabel={},
				xmin=1, ymin = 0.6e-3, xmax = 39,ymax=2e-1,legend columns=5,
				width=0.26\textwidth,height=4cm,legend style={at={(0,1.15)},anchor=west},,
				xticklabels={,,},
				yshift=-2.5cm,xshift=4cm]
				\addplot[paried_red!80, thick, ]      table[x = t_set_norminal , y  = Agent2_e ]{\fileCET};
				\addplot[paried_blue!80, thick, ]     table[x = t_set_norminal , y  = Agent2_e ]{\fileDET};
				\addplot[paried_orange!80, thick, ]    table[x = t_set_norminal , y  = Agent2_e ]{\fileTT};
				\addplot[paried_green!80, thick, ]    table[x = t_set_norminal , y  = Agent2_e ]{\fileOFF};
				\addplot[black!50, thick, ]    table[x = t_set_norminal , y  = Agent2_e ]{\fileEXA};
			\end{semilogyaxis}
			
			\begin{axis}[xlabel={},ylabel={Agent $3$},
				xmin=1, ymin = -1.1, xmax = 39,ymax=2.6,legend columns=2,
				width=0.26\textwidth,height=4cm,legend style={at={(0.02,0.8)},anchor=west},
				xticklabels={,,},
				yshift=-5.0cm]
				\addplot[pink!80, very thick, dashed]      table[x = t_set_norminal , y  = Agent3_r1 ]{\fileREF};
				\addplot[paried_red!80, semithick, dashed]      table[x = t_set_norminal , y  = Agent3_x1 ]{\fileCET};
				\addplot[paried_blue!80, semithick, dashed]     table[x = t_set_norminal , y  = Agent3_x1 ]{\fileDET};
				\addplot[paried_orange!80, semithick, dashed]    table[x = t_set_norminal , y  = Agent3_x1 ]{\fileTT};
				\addplot[paried_green!80, semithick, dashed]    table[x = t_set_norminal , y  = Agent3_x1 ]{\fileOFF};
				\addplot[black!50, semithick, dashed]    table[x = t_set_norminal , y  = Agent3_x1 ]{\fileEXA};
				
				\addplot[pink!80, very thick, ]      table[x = t_set_norminal , y  = Agent3_r2 ]{\fileREF};
				\addplot[paried_red!80, semithick, ]      table[x = t_set_norminal , y  = Agent3_x2 ]{\fileCET};
				\addplot[paried_blue!80, semithick, ]     table[x = t_set_norminal , y  = Agent3_x2 ]{\fileDET};
				\addplot[paried_orange!80, semithick, ]    table[x = t_set_norminal , y  = Agent3_x2 ]{\fileTT};
				\addplot[paried_green!80, semithick, ]    table[x = t_set_norminal , y  = Agent3_x2 ]{\fileOFF};
				\addplot[black!50, semithick, ]    table[x = t_set_norminal , y  = Agent3_x2 ]{\fileEXA};
				
				\legend{$\!x_{l,\!1\!}\!+\!\! s_{3,\!1\!}$, , , , , $\!x_{3,\!1\!}\!\!$, $\!x_{l,\!2\!}\!+\!\! s_{3,\!2\!}$, , , , , $\!x_{3,\!2\!}\!\!$}
			\end{axis}
			\begin{semilogyaxis}[xlabel={},ylabel={},
				xmin=1, ymin = 0.6e-3, xmax = 39,ymax=2e-1,legend columns=5,
				width=0.26\textwidth,height=4cm,legend style={at={(0,1.15)},anchor=west},,
				xticklabels={,,},
				yshift=-5.0cm,xshift=4cm]
				\addplot[paried_red!80, thick, ]      table[x = t_set_norminal , y  = Agent3_e ]{\fileCET};
				\addplot[paried_blue!80, thick, ]     table[x = t_set_norminal , y  = Agent3_e ]{\fileDET};
				\addplot[paried_orange!80, thick, ]    table[x = t_set_norminal , y  = Agent3_e ]{\fileTT};
				\addplot[paried_green!80, thick, ]    table[x = t_set_norminal , y  = Agent3_e ]{\fileOFF};
				\addplot[black!50, thick, ]    table[x = t_set_norminal , y  = Agent3_e ]{\fileEXA};
			\end{semilogyaxis}
			
			\begin{axis}[xlabel={{\color{white} 1111111} $t$},ylabel={Agent $4$},
				xmin=1, ymin = -1.1, xmax = 39,ymax=2.6,legend columns=2,
				width=0.26\textwidth,height=4cm,legend style={at={(0.02,0.8)},anchor=west},
				xlabel style={text width=2.5cm},
				yshift=-7.5cm]
				\addplot[pink!80, very thick, dashed]      table[x = t_set_norminal , y  = Agent4_r1 ]{\fileREF};
				\addplot[paried_red!80, semithick, dashed]      table[x = t_set_norminal , y  = Agent4_x1 ]{\fileCET};
				\addplot[paried_blue!80, semithick, dashed]     table[x = t_set_norminal , y  = Agent4_x1 ]{\fileDET};
				\addplot[paried_orange!80, semithick, dashed]    table[x = t_set_norminal , y  = Agent4_x1 ]{\fileTT};
				\addplot[paried_green!80, semithick, dashed]    table[x = t_set_norminal , y  = Agent4_x1 ]{\fileOFF};
				\addplot[black!50, semithick, dashed]    table[x = t_set_norminal , y  = Agent4_x1 ]{\fileEXA};
				
				\addplot[pink!80, very thick, ]      table[x = t_set_norminal , y  = Agent4_r2 ]{\fileREF};
				\addplot[paried_red!80, semithick, ]      table[x = t_set_norminal , y  = Agent4_x2 ]{\fileCET};
				\addplot[paried_blue!80, semithick, ]     table[x = t_set_norminal , y  = Agent4_x2 ]{\fileDET};
				\addplot[paried_orange!80, semithick, ]    table[x = t_set_norminal , y  = Agent4_x2 ]{\fileTT};
				\addplot[paried_green!80, semithick, ]    table[x = t_set_norminal , y  = Agent4_x2 ]{\fileOFF};
				\addplot[black!50, semithick, ]    table[x = t_set_norminal , y  = Agent4_x2 ]{\fileEXA};
				
				\legend{$\!x_{l,\!1\!}\!+\!\! s_{4,\!1\!}$, , , , , $\!x_{4,\!1\!}\!\!$, $\!x_{l,\!2\!}\!+\!\! s_{4,\!2\!}$, , , , , $\!x_{4,\!2\!}\!\!$}
			\end{axis}
			\begin{semilogyaxis}[xlabel={{\color{white} 1111111} $t$},ylabel={},
				xmin=1, ymin = 0.6e-3, xmax = 39,ymax=2e-1,legend columns=5,
				width=0.26\textwidth,height=4cm,legend style={at={(0,1.15)},anchor=west},
				xlabel style={text width=2.5cm},
				yshift=-7.5cm,xshift=4cm]
				\addplot[paried_red!80, thick, ]      table[x = t_set_norminal , y  = Agent4_e ]{\fileCET};
				\addplot[paried_blue!80, thick, ]     table[x = t_set_norminal , y  = Agent4_e ]{\fileDET};
				\addplot[paried_orange!80, thick, ]    table[x = t_set_norminal , y  = Agent4_e ]{\fileTT};
				\addplot[paried_green!80, thick, ]    table[x = t_set_norminal , y  = Agent4_e ]{\fileOFF};
				\addplot[black!50, thick, ]    table[x = t_set_norminal , y  = Agent4_e ]{\fileEXA};
			\end{semilogyaxis}
		\end{tikzpicture}
		\vspace{-0.4cm}
		\caption{
			\todo{
				System states and tracking error of each agent over time.
			}
		}
		\vspace{-0.3cm}
		\label{figure_Agent_StateAndError}
	\end{figure}
	\fi
	
	\begin{figure}[t] 
		\centering
		\def\file{fig/Error.txt}
		\begin{tikzpicture}
			\begin{semilogyaxis}[xlabel={$t$},ylabel={$\| \bm{\vartheta}(t) \|$},
				xmin=0, ymin =0.003, xmax = 39.9,ymax=3, legend columns=3,
				width=0.48\textwidth,height=4.5cm,legend pos= north east,
				clip mode=individual]
				
				\addplot[paried_red!80, thick]    table[x = t_set_tikz , y  = norm_vartheta_set_tikz_2 ]{\file};
				\addplot[paried_blue!80, thick]    table[x = t_set_tikz , y  = norm_vartheta_set_tikz_3 ]{\file};
				\addplot[paried_orange!80, thick]    table[x = t_set_tikz , y  = norm_vartheta_set_tikz_4 ]{\file};
				\addplot[paried_green!80, thick]    table[x = t_set_tikz , y  = norm_vartheta_set_tikz_1 ]{\file};
				\addplot[black!50, very thick, dashed]    table[x = t_set_tikz , y  = norm_vartheta_set_tikz_5 ]{\file};
				\legend{CET, DET, TT, Offline learning, Exact model}
			\end{semilogyaxis}
		\end{tikzpicture}
		\vspace{-0.4cm}
		\caption{
			Overall tracking error with respect to time.
		}
		\vspace{-0.3cm}
		\label{figure_error}
	\end{figure}
	
	Besides the tracking error, the number of triggers reflect the data efficiency, where we only consider the triggers for event-triggered online learning mechanisms shown in \cref{figure_trigger}.
	It is obvious that most trigger occurs before $t = 18$ due to the periodic references as in \eqref{eqn_experiment_xl} and \eqref{eqn_experiment_si}, indicating the collected data set around $t = 18$ is sufficient for the guaranteed tracking performance in \cref{proposition_online_cooperative_prediction_error_bound}.
	Only few triggers happens after $t = 20$ for distributed event-trigger due to its slight conservatism, which results in near $20 \%$ more triggers compared to the centralized trigger.
	
	\begin{figure}[t] 
		\centering
		\def\file{fig/trigger.txt}
		\begin{tikzpicture}
			\begin{axis}[xlabel={},ylabel={CET},
				xmin=0, ymin = 0.5, xmax = 39.9,ymax=4.5,legend columns=1,
				width=0.48\textwidth,height=3cm,legend pos= north east,
				ytick={1,2,3,4},
				yticklabels={{Agent 1}, {Agent 2}, {Agent 3}, {Agent 4}},
				xticklabels={,,,,}]
				\addplot[only marks, mark=x, paried_red!80]    table[x = time_centralized_1 , y  = trigger_centralized_1 ]{\file};
				\addplot[only marks, mark=x, paried_red!80]    table[x = time_centralized_2 , y  = trigger_centralized_2 ]{\file};
				\addplot[only marks, mark=x, paried_red!80]    table[x = time_centralized_3 , y  = trigger_centralized_3 ]{\file};
				\addplot[only marks, mark=x, paried_red!80]    table[x = time_centralized_4 , y  = trigger_centralized_4 ]{\file};
				\addplot[only marks, mark=x, blue]    coordinates{(-1,-1)};
				
			\end{axis}
			\begin{axis}[xlabel={$t$},ylabel={DET},
				xmin=0, ymin = 0.5, xmax = 39.9,ymax=4.5,legend columns=1,
				width=0.48\textwidth,height=3cm,legend pos= south east,
				ytick={1,2,3,4},
				yticklabels={{Agent 1}, {Agent 2}, {Agent 3}, {Agent 4}},
				yshift=-1.5cm]
				\addplot[only marks, mark=x, red]    coordinates{(-1,-1)};
				\addplot[only marks, mark=x, paried_blue!80]    table[x = time_distributed_1 , y  = trigger_distributed_1 ]{\file};
				\addplot[only marks, mark=x, paried_blue!80]    table[x = time_distributed_2 , y  = trigger_distributed_2 ]{\file};
				\addplot[only marks, mark=x, paried_blue!80]    table[x = time_distributed_3 , y  = trigger_distributed_3 ]{\file};
				\addplot[only marks, mark=x, paried_blue!80]    table[x = time_distributed_4 , y  = trigger_distributed_4 ]{\file};
				
			\end{axis}
		\end{tikzpicture}
		\vspace{-0.4cm}
		\caption{
			Trigger times and instances for centralized and distributed event-trigger mechanisms for each agent.
			Specifically, for centralized version $131$, $130$, $122$ and $120$ samples are collected for agent $1$ to $4$.
			The minimal trigger interval for each agent denotes $0.036$, $0.027$, $0.020$ and $0.020$.
			With distributed event-trigger, each agent collects $154$, $150$, $132$ and $131$ data pairs under minimal trigger interval $0.025$, $0.024$, $0.016$ and $0.021$.
		}
		\vspace{-0.3cm}
		\label{figure_trigger}
	\end{figure}
	
	\subsubsection{Monte Carlo Test}
	
	To test the generalization of the proposed event-triggered cooperative online learning methods, a Mont-Carlo test is employed, in which each algorithm is repeated for $100$ times by using different initial states, i.e., uniformly distributed $\bm{x}_i(0) \in \mathbb{X}$, uniformly distributed initial data set for offline learning and normally distributed random measurement noise $w^{\varsigma}_i$ for each agent $i \in \mathcal{V}$.
	
	\ifarxiv
	\todo{
		The control performance for each learning algorithm is reflected by the ultimate tracking error bound as shown in \cref{figure_AgentError_MonteCarlo}, \cref{figure_ultimate_tracking_error} and \cref{figure_error_time_MonteCarlo}.
	}
	\fi
	In practice, we consider the maximal tracking error in $t \!\in\! [20,40]$, in order to neglect larger $\| \bm{\vartheta} \|$ from poor initial states $\bm{x}(0)$ and focus on the steady states.
	\ifarxiv
		Similarly as the result in \cref{figure_error}, the time-triggered online learning achieves the best control performance in statistic, close to the performance with exact model.
		Offline learning performs worst with larger median error, larger variance and more outliers compared to the online learning methods, which is a result from the lack of adaption for the data set.
		Both event-triggered learning strategies perform slightly worse than time-triggered learning, because the tracking error bound $\bar{\vartheta}_d$ used in distributed event-trigger design in \eqref{eqn_trigger} is determined only by one-point GP accuracy as in \cref{proposition_online_cooperative_prediction_error_bound}, underestimating the performance of GP update.
		This means, more samples are collected online with time-triggered learning by considering the time-trigger interval as the minimal trigger interval in the event-triggered cases, resulting in a smaller prediction performance as expected in \cref{lemma_GP_error_bound} and therefore a smaller tracking error.
		Both event-triggered online learning methods perform much better than offline learning, reflected by smaller tacking error with acceptable variance.
		Only few outliers for event-triggered mechanism exists due to the fact that the ultimate boundness in \cref{proposition_centralized_ET} or \cref{theorem_trigger} is probabilistic.
		It is also seen from \cref{figure_ultimate_tracking_error}, that the proposed distributed event-trigger performs slightly better than the centralized method.
		This phenomenon is intuitive since more conservatism is included when considering the property of distributed computation.
		This conservatism underestimates the performance of the entire multi-agent system and leads to more triggers on the individual agent, and then results in better control performance with the same reason for time-triggered case.
		Although conservatism benefits the tracking performance, it brings side effects such as the requirement of larger local data storage and high energy consumption for more frequent model updates.
		Therefore, we also compare the trigger times to show the conservatism degree and the resulted data efficiency.
		\looseness=-1
	\fi
	
	\ifarxiv
	\begin{figure}[t]
		\centering
		\begin{subfigure}{1\textwidth}
			\begin{tikzpicture}
				\def\fileCET{fig/AgentError_MonteCarlo_centralized.txt}
				\def\fileDET{fig/AgentError_MonteCarlo_distributed.txt}
				\def\fileTT{fig/AgentError_MonteCarlo_continuous0015.txt}
				\def\fileOFF{fig/AgentError_MonteCarlo_NoTrigger.txt}
				\def\fileEXA{fig/AgentError_MonteCarlo_exact.txt}
				
				\begin{semilogyaxis}[xlabel={},ylabel={$\| \bm{\vartheta}_1(t) \|$},
					xmin=1, ymin = 2e-3, xmax = 39,ymax=0.3,legend columns=4,
					width=0.48\textwidth,height=4cm,legend style={at={(-0.1,1.3)},anchor=west},
					xticklabels={,,}]
					\addplot[paried_red!80, thick]    table[x = t_set_nominal , y  = Mean_Agent1 ]{\fileCET};
					\addplot+[name path=max_1,black,no markers, draw=none] table[x = t_set_nominal , y  = Max_Agent1 ]{\fileCET};
					\addplot+[name path=min_1,black,no markers, draw=none] table[x = t_set_nominal , y  = Min_Agent1 ]{\fileCET};
					\addplot[paried_red!80, fill opacity=0.2] fill between[of=max_1 and min_1];
					
					\addplot[paried_blue!80, thick]    table[x = t_set_nominal , y  = Mean_Agent1 ]{\fileDET};
					\addplot+[name path=max_2,black,no markers, draw=none] table[x = t_set_nominal , y  = Max_Agent1 ]{\fileDET};
					\addplot+[name path=min_2,black,no markers, draw=none] table[x = t_set_nominal , y  = Min_Agent1 ]{\fileDET};
					\addplot[paried_blue!80, fill opacity=0.2] fill between[of=max_2 and min_2];
					
					\addplot[paried_orange!80, thick]    table[x = t_set_nominal , y  = Mean_Agent1 ]{\fileTT};
					\addplot+[name path=max_3,black,no markers, draw=none] table[x = t_set_nominal , y  = Max_Agent1 ]{\fileTT};
					\addplot+[name path=min_3,black,no markers, draw=none] table[x = t_set_nominal , y  = Min_Agent1 ]{\fileTT};
					\addplot[paried_orange!80, fill opacity=0.2] fill between[of=max_3 and min_3];
					
					\addplot[paried_green!80, thick]    table[x = t_set_nominal , y  = Mean_Agent1 ]{\fileOFF};
					\addplot+[name path=max_4,black,no markers, draw=none] table[x = t_set_nominal , y  = Max_Agent1 ]{\fileOFF};
					\addplot+[name path=min_4,black,no markers, draw=none] table[x = t_set_nominal , y  = Min_Agent1 ]{\fileOFF};
					\addplot[paried_green!80, fill opacity=0.2] fill between[of=max_4 and min_4];
					
					\addplot[black!50, very thick, dashed]    table[x = t_set_nominal , y  = Mean_Agent1 ]{\fileEXA};
					\addplot+[name path=max_5,black,no markers, draw=none] table[x = t_set_nominal , y  = Max_Agent1 ]{\fileEXA};
					\addplot+[name path=min_5,black,no markers, draw=none] table[x = t_set_nominal , y  = Min_Agent1 ]{\fileEXA};
					\addplot[black!50, fill opacity=0.2] fill between[of=max_5 and min_5];
					
					\legend{
						CET,,,Variance,
						DET,,,Variance,
						TT,,,Variance,
						Offline learning,,,Variance,
						Exact model,,,Variance
					};
				\end{semilogyaxis}
			\end{tikzpicture}
		\end{subfigure}
		\vspace{-0.1cm}
		\begin{subfigure}{1\textwidth}
			\begin{tikzpicture}
				\def\fileCET{fig/AgentError_MonteCarlo_centralized.txt}
				\def\fileDET{fig/AgentError_MonteCarlo_distributed.txt}
				\def\fileTT{fig/AgentError_MonteCarlo_continuous0015.txt}
				\def\fileOFF{fig/AgentError_MonteCarlo_NoTrigger.txt}
				\def\fileEXA{fig/AgentError_MonteCarlo_exact.txt}
				
				\begin{semilogyaxis}[xlabel={},ylabel={$\| \bm{\vartheta}_2(t) \|$},
					xmin=1, ymin = 2e-3, xmax = 39,ymax=0.3,legend columns=4,
					width=0.48\textwidth,height=4cm,legend style={at={(0.02,0.8)},anchor=west},
					xticklabels={,,},
					]
					\addplot[paried_red!80, thick]    table[x = t_set_nominal , y  = Mean_Agent2 ]{\fileCET};
					\addplot+[name path=max_1,black,no markers, draw=none] table[x = t_set_nominal , y  = Max_Agent2 ]{\fileCET};
					\addplot+[name path=min_1,black,no markers, draw=none] table[x = t_set_nominal , y  = Min_Agent2 ]{\fileCET};
					\addplot[paried_red!80, fill opacity=0.2] fill between[of=max_1 and min_1];
					
					\addplot[paried_blue!80, thick]    table[x = t_set_nominal , y  = Mean_Agent2 ]{\fileDET};
					\addplot+[name path=max_2,black,no markers, draw=none] table[x = t_set_nominal , y  = Max_Agent2 ]{\fileDET};
					\addplot+[name path=min_2,black,no markers, draw=none] table[x = t_set_nominal , y  = Min_Agent2 ]{\fileDET};
					\addplot[paried_blue!80, fill opacity=0.2] fill between[of=max_2 and min_2];
					
					\addplot[paried_orange!80, thick]    table[x = t_set_nominal , y  = Mean_Agent2 ]{\fileTT};
					\addplot+[name path=max_3,black,no markers, draw=none] table[x = t_set_nominal , y  = Max_Agent2 ]{\fileTT};
					\addplot+[name path=min_3,black,no markers, draw=none] table[x = t_set_nominal , y  = Min_Agent2 ]{\fileTT};
					\addplot[paried_orange!80, fill opacity=0.2] fill between[of=max_3 and min_3];
					
					\addplot[paried_green!80, thick]    table[x = t_set_nominal , y  = Mean_Agent2 ]{\fileOFF};
					\addplot+[name path=max_4,black,no markers, draw=none] table[x = t_set_nominal , y  = Max_Agent2 ]{\fileOFF};
					\addplot+[name path=min_4,black,no markers, draw=none] table[x = t_set_nominal , y  = Min_Agent2 ]{\fileOFF};
					\addplot[paried_green!80, fill opacity=0.2] fill between[of=max_4 and min_4];
					
					\addplot[black!50, very thick, dashed]    table[x = t_set_nominal , y  = Mean_Agent2 ]{\fileEXA};
					\addplot+[name path=max_5,black,no markers, draw=none] table[x = t_set_nominal , y  = Max_Agent2 ]{\fileEXA};
					\addplot+[name path=min_5,black,no markers, draw=none] table[x = t_set_nominal , y  = Min_Agent2 ]{\fileEXA};
					\addplot[black!50, fill opacity=0.2] fill between[of=max_5 and min_5];
				\end{semilogyaxis}
			\end{tikzpicture}
		\end{subfigure}
		\vspace{-0.1cm}
		\begin{subfigure}{1\textwidth}
			\begin{tikzpicture}
				\def\fileCET{fig/AgentError_MonteCarlo_centralized.txt}
				\def\fileDET{fig/AgentError_MonteCarlo_distributed.txt}
				\def\fileTT{fig/AgentError_MonteCarlo_continuous0015.txt}
				\def\fileOFF{fig/AgentError_MonteCarlo_NoTrigger.txt}
				\def\fileEXA{fig/AgentError_MonteCarlo_exact.txt}
				
				\begin{semilogyaxis}[xlabel={},ylabel={$\| \bm{\vartheta}_3(t) \|$},
					xmin=1, ymin = 2e-3, xmax = 39,ymax=0.3,legend columns=4,
					width=0.48\textwidth,height=4cm,legend style={at={(0.02,0.8)},anchor=west},
					xticklabels={,,},
					]
					\addplot[paried_red!80, thick]    table[x = t_set_nominal , y  = Mean_Agent3 ]{\fileCET};
					\addplot+[name path=max_1,black,no markers, draw=none] table[x = t_set_nominal , y  = Max_Agent3 ]{\fileCET};
					\addplot+[name path=min_1,black,no markers, draw=none] table[x = t_set_nominal , y  = Min_Agent3 ]{\fileCET};
					\addplot[paried_red!80, fill opacity=0.2] fill between[of=max_1 and min_1];
					
					\addplot[paried_blue!80, thick]    table[x = t_set_nominal , y  = Mean_Agent3 ]{\fileDET};
					\addplot+[name path=max_2,black,no markers, draw=none] table[x = t_set_nominal , y  = Max_Agent3 ]{\fileDET};
					\addplot+[name path=min_2,black,no markers, draw=none] table[x = t_set_nominal , y  = Min_Agent3 ]{\fileDET};
					\addplot[paried_blue!80, fill opacity=0.2] fill between[of=max_2 and min_2];
					
					\addplot[paried_orange!80, thick]    table[x = t_set_nominal , y  = Mean_Agent3 ]{\fileTT};
					\addplot+[name path=max_3,black,no markers, draw=none] table[x = t_set_nominal , y  = Max_Agent3 ]{\fileTT};
					\addplot+[name path=min_3,black,no markers, draw=none] table[x = t_set_nominal , y  = Min_Agent3 ]{\fileTT};
					\addplot[paried_orange!80, fill opacity=0.2] fill between[of=max_3 and min_3];
					
					\addplot[paried_green!80, thick]    table[x = t_set_nominal , y  = Mean_Agent3 ]{\fileOFF};
					\addplot+[name path=max_4,black,no markers, draw=none] table[x = t_set_nominal , y  = Max_Agent3 ]{\fileOFF};
					\addplot+[name path=min_4,black,no markers, draw=none] table[x = t_set_nominal , y  = Min_Agent3 ]{\fileOFF};
					\addplot[paried_green!80, fill opacity=0.2] fill between[of=max_4 and min_4];
					
					\addplot[black!50, very thick, dashed]    table[x = t_set_nominal , y  = Mean_Agent3 ]{\fileEXA};
					\addplot+[name path=max_5,black,no markers, draw=none] table[x = t_set_nominal , y  = Max_Agent3 ]{\fileEXA};
					\addplot+[name path=min_5,black,no markers, draw=none] table[x = t_set_nominal , y  = Min_Agent3 ]{\fileEXA};
					\addplot[black!50, fill opacity=0.2] fill between[of=max_5 and min_5];
				\end{semilogyaxis}
			\end{tikzpicture}
		\end{subfigure}
		\vspace{-0.1cm}
		\begin{subfigure}{1\textwidth}
			\begin{tikzpicture}
				\def\fileCET{fig/AgentError_MonteCarlo_centralized.txt}
				\def\fileDET{fig/AgentError_MonteCarlo_distributed.txt}
				\def\fileTT{fig/AgentError_MonteCarlo_continuous0015.txt}
				\def\fileOFF{fig/AgentError_MonteCarlo_NoTrigger.txt}
				\def\fileEXA{fig/AgentError_MonteCarlo_exact.txt}
				
				\begin{semilogyaxis}[xlabel={{\color{white} 1111111}$t$},ylabel={$\| \bm{\vartheta}_4(t) \|$},
					xmin=1, ymin = 2e-3, xmax = 39,ymax=0.3,legend columns=4,
					width=0.48\textwidth,height=4cm,legend style={at={(0.02,0.8)},anchor=west},
					xlabel style={text width=2.5cm},
					]
					\addplot[paried_red!80, thick]    table[x = t_set_nominal , y  = Mean_Agent4 ]{\fileCET};
					\addplot+[name path=max_1,black,no markers, draw=none] table[x = t_set_nominal , y  = Max_Agent4 ]{\fileCET};
					\addplot+[name path=min_1,black,no markers, draw=none] table[x = t_set_nominal , y  = Min_Agent4 ]{\fileCET};
					\addplot[paried_red!80, fill opacity=0.2] fill between[of=max_1 and min_1];
					
					\addplot[paried_blue!80, thick]    table[x = t_set_nominal , y  = Mean_Agent4 ]{\fileDET};
					\addplot+[name path=max_2,black,no markers, draw=none] table[x = t_set_nominal , y  = Max_Agent4 ]{\fileDET};
					\addplot+[name path=min_2,black,no markers, draw=none] table[x = t_set_nominal , y  = Min_Agent4 ]{\fileDET};
					\addplot[paried_blue!80, fill opacity=0.2] fill between[of=max_2 and min_2];
					
					\addplot[paried_orange!80, thick]    table[x = t_set_nominal , y  = Mean_Agent4 ]{\fileTT};
					\addplot+[name path=max_3,black,no markers, draw=none] table[x = t_set_nominal , y  = Max_Agent4 ]{\fileTT};
					\addplot+[name path=min_3,black,no markers, draw=none] table[x = t_set_nominal , y  = Min_Agent4 ]{\fileTT};
					\addplot[paried_orange!80, fill opacity=0.2] fill between[of=max_3 and min_3];
					
					\addplot[paried_green!80, thick]    table[x = t_set_nominal , y  = Mean_Agent4 ]{\fileOFF};
					\addplot+[name path=max_4,black,no markers, draw=none] table[x = t_set_nominal , y  = Max_Agent4 ]{\fileOFF};
					\addplot+[name path=min_4,black,no markers, draw=none] table[x = t_set_nominal , y  = Min_Agent4 ]{\fileOFF};
					\addplot[paried_green!80, fill opacity=0.2] fill between[of=max_4 and min_4];
					
					\addplot[black!50, very thick, dashed]    table[x = t_set_nominal , y  = Mean_Agent4 ]{\fileEXA};
					\addplot+[name path=max_5,black,no markers, draw=none] table[x = t_set_nominal , y  = Max_Agent4 ]{\fileEXA};
					\addplot+[name path=min_5,black,no markers, draw=none] table[x = t_set_nominal , y  = Min_Agent4 ]{\fileEXA};
					\addplot[black!50, fill opacity=0.2] fill between[of=max_5 and min_5];
				\end{semilogyaxis}
			\end{tikzpicture}
		\end{subfigure}
		\vspace{-0.4cm}
		\caption{
			\todo{
				Tracking error and its variance of each agent over time from $100$ times Monte Carlo tests.
			}
		}
		\vspace{-0.3cm}
		\label{figure_AgentError_MonteCarlo}
	\end{figure}
	\fi
	
	\pgfplotsset{
		boxplot/box extend=0.48,
		boxplot/every whisker/.style={thick,blue},
		boxplot/every box/.style={thick,blue},
		boxplot/every median/.style={red}
	}
	\begin{figure}[t] 
		\centering
		\begin{tikzpicture}
			\def\file{fig/Method_max_e.txt}
			\begin{semilogyaxis}[
				boxplot/draw direction = y,
				ylabel style={align=center},ylabel={$\max_{t \!\in\! [20,\! 40]} \| \bm{\vartheta}(t) \|$},
				ymin=0.028, ymax=2.5,legend columns=2,
				xmin=0.65,xmax=5.35,
				width=0.48\textwidth,height=5cm,
				xtick={1,2,3,4,5},xticklabel style={align=center}, 
				xticklabels = {{Offline\\learning},{Centralized\\trigger},{Distributed\\trigger},{Time\\trigger}, {Exact\\model}}]
				\addplot+[boxplot, mark=triangle] table[y=NoTrigger_max_e] {\file};
				\addplot+[boxplot, mark=triangle] table[y=centralized_max_e] {\file};
				\addplot+[boxplot, mark=triangle] table[y=distributed_max_e] {\file};
				\addplot+[boxplot, mark=triangle] table[y=continuous0015_max_e ] {\file};
				\addplot+[boxplot, mark=triangle] table[y=exact_max_e] {\file};
			\end{semilogyaxis}
			
			\def\file{fig/Method_max_e.txt}
			\begin{axis}[
				boxplot/draw direction = y,
				ylabel style={align=center},ylabel={},
				ymin=0.035, ymax=0.075,legend columns=2,
				xmin=0.6,xmax=3.4,
				width=7cm,height=3cm,
				xtick={1,2,3,4},xticklabel style={align=center}, 
				xticklabels = {{Centralized\\trigger},{Distributed\\trigger},{Time\\trigger}, {Exact\\model}},
				xshift=1.5cm,yshift=1.7cm,
				axis background/.style={fill=white}]
				\addplot+[boxplot, mark=triangle] table[y=centralized_max_e] {\file};
				\addplot+[boxplot, mark=triangle] table[y=distributed_max_e] {\file};
				\addplot+[boxplot, mark=triangle] table[y=continuous0015_max_e] {\file};
			\end{axis}
		\end{tikzpicture}
		\vspace{-0.3cm}
		\caption{
			Maximal tracking error in the steady state.
		}
		\vspace{-0.3cm}
		\label{figure_ultimate_tracking_error}
	\end{figure}
	
	\ifarxiv
	\begin{figure}[t] 
		\centering
		\def\file{fig/Method_e_over_t.txt}
		\begin{tikzpicture}
			\begin{semilogyaxis}[xlabel={$t$},ylabel={$\| \bm{\vartheta}(t) \|$},
				xmin=0, ymin =0.003, xmax = 39.9,ymax=200, legend columns=4,
				width=0.48\textwidth,height=5cm,legend style={at={(-0.1,1.2)},anchor=west},
				clip mode=individual]
				\addplot[paried_red!80, thick]    table[x = t_set_nominal , y  = centralized_mean_e_t ]{\file};
				\addplot+[name path=max_1,black,no markers, draw=none] table[x = t_set_nominal , y  = centralized_max_e_t ]{\file};
				\addplot+[name path=min_1,black,no markers, draw=none] table[x = t_set_nominal , y  = centralized_min_e_t ]{\file};
				\addplot[paried_red!80, fill opacity=0.2] fill between[of=max_1 and min_1];
				
				\addplot[paried_blue!80, thick]    table[x = t_set_nominal , y  = distributed_mean_e_t ]{\file};
				\addplot+[name path=max_2,black,no markers, draw=none] table[x = t_set_nominal , y  = distributed_max_e_t ]{\file};
				\addplot+[name path=min_2,black,no markers, draw=none] table[x = t_set_nominal , y  = distributed_min_e_t ]{\file};
				\addplot[paried_blue!80, fill opacity=0.2] fill between[of=max_2 and min_2];
				
				\addplot[paried_orange!80, thick]    table[x = t_set_nominal , y  = continuous0015_mean_e_t ]{\file};
				\addplot+[name path=max_3,black,no markers, draw=none] table[x = t_set_nominal , y  = continuous0015_max_e_t ]{\file};
				\addplot+[name path=min_3,black,no markers, draw=none] table[x = t_set_nominal , y  = continuous0015_min_e_t ]{\file};
				\addplot[paried_orange!80, fill opacity=0.2] fill between[of=max_3 and min_3];
				
				\addplot[paried_green!80, thick]    table[x = t_set_nominal , y  = NoTrigger_mean_e_t ]{\file};
				\addplot+[name path=max_4,black,no markers, draw=none] table[x = t_set_nominal , y  = NoTrigger_max_e_t ]{\file};
				\addplot+[name path=min_4,black,no markers, draw=none] table[x = t_set_nominal , y  = NoTrigger_min_e_t ]{\file};
				\addplot[paried_green!80, fill opacity=0.2] fill between[of=max_4 and min_4];
				
				\addplot[black!50, very thick, dashed]    table[x = t_set_nominal , y  = exact_mean_e_t ]{\file};
				\addplot+[name path=max_5,black,no markers, draw=none] table[x = t_set_nominal , y  = exact_max_e_t ]{\file};
				\addplot+[name path=min_5,black,no markers, draw=none] table[x = t_set_nominal , y  = exact_min_e_t ]{\file};
				\addplot[black!50, fill opacity=0.2] fill between[of=max_5 and min_5];
				\legend{
					CET,,,Variance,
					DET,,,Variance,
					TT,,,Variance,
					Offline learning,,,Variance,
					Exact model,,,Variance
				}
			\end{semilogyaxis}
			
			\begin{axis}[
				ylabel style={align=center},ylabel={},
				ymin=0.005, ymax=0.05,legend columns=2,
				xmin=31.1,xmax=37.9,
				width=7.3cm,height=3cm,
				xshift=1.25cm,yshift=1.65cm,
				axis background/.style={fill=white}]
				\addplot[paried_red!80, thick]    table[x = t_set_nominal , y  = centralized_mean_e_t ]{\file};
				\addplot+[name path=max_1,black,no markers, draw=none] table[x = t_set_nominal , y  = centralized_max_e_t ]{\file};
				\addplot+[name path=min_1,black,no markers, draw=none] table[x = t_set_nominal , y  = centralized_min_e_t ]{\file};
				\addplot[paried_red!80, fill opacity=0.2] fill between[of=max_1 and min_1];
				
				\addplot[paried_blue!80, thick]    table[x = t_set_nominal , y  = distributed_mean_e_t ]{\file};
				\addplot+[name path=max_2,black,no markers, draw=none] table[x = t_set_nominal , y  = distributed_max_e_t ]{\file};
				\addplot+[name path=min_2,black,no markers, draw=none] table[x = t_set_nominal , y  = distributed_min_e_t ]{\file};
				\addplot[paried_blue!80, fill opacity=0.2] fill between[of=max_2 and min_2];
				
				\addplot[paried_orange!80, thick]    table[x = t_set_nominal , y  = continuous0015_mean_e_t ]{\file};
				\addplot+[name path=max_3,black,no markers, draw=none] table[x = t_set_nominal , y  = continuous0015_max_e_t ]{\file};
				\addplot+[name path=min_3,black,no markers, draw=none] table[x = t_set_nominal , y  = continuous0015_min_e_t ]{\file};
				\addplot[paried_orange!80, fill opacity=0.2] fill between[of=max_3 and min_3];
				
				\addplot[paried_green!80, thick]    table[x = t_set_nominal , y  = NoTrigger_mean_e_t ]{\file};
				\addplot+[name path=max_4,black,no markers, draw=none] table[x = t_set_nominal , y  = NoTrigger_max_e_t ]{\file};
				\addplot+[name path=min_4,black,no markers, draw=none] table[x = t_set_nominal , y  = NoTrigger_min_e_t ]{\file};
				\addplot[paried_green!80, fill opacity=0.2] fill between[of=max_4 and min_4];
				
				\addplot[black!50, very thick, dashed]    table[x = t_set_nominal , y  = exact_mean_e_t ]{\file};
				\addplot+[name path=max_5,black,no markers, draw=none] table[x = t_set_nominal , y  = exact_max_e_t ]{\file};
				\addplot+[name path=min_5,black,no markers, draw=none] table[x = t_set_nominal , y  = exact_min_e_t ]{\file};
				\addplot[black!50, fill opacity=0.2] fill between[of=max_5 and min_5];
			\end{axis}
		\end{tikzpicture}
		\vspace{-0.4cm}
		\caption{
			\todo{
				Mean and variance of overall tracking error with respect to time from $100$ times Monte Carlo tests.
			}
		}
		\vspace{-0.3cm}
		\label{figure_error_time_MonteCarlo}
	\end{figure}
	\fi
	
	The number of triggers reflecting the data efficiency is depicted in \cref{figure_Method_TriggerTimes} with the maximal number of training samples in each data set.
	It is obvious that both event-triggered online learning mechanisms results in the data set with slightly less than $200$ samples, which is close to the designed initial data set for offline learning.
	Combined with the comparison of control performance in \cref{figure_ultimate_tracking_error}, where event-triggered online cooperative learning mechanism achieves smaller ultimate tracking error bound, the improved data efficiency is shown by choosing more informative data related to the control task using event-triggered online learning.
	\ifarxiv
	\todo{
		Moreover, in comparison to centralized methods, the distributed approach allows agents to collect only a slightly larger amount of data without the need for an extra computational center.
		Combining with the observation in \cref{figure_trigger}, limited data is eventually added into individual data set for each agent, while time-triggered method intends to collect data without stop resulting in an infinitely growing data set linear with operating time and high demand on the computational resources.
	}
	\fi
	
	\begin{figure}[t] 
		\centering
		\begin{tikzpicture}
			\def\file{fig/Method_M.txt}
			\begin{axis}[ylabel style={align=center},ylabel={Maximum Size \\ of the Data Set},
				ymin=0, ymax=350,legend columns=2,
				xmin=0.5,xmax=4.5,
				width=0.48\textwidth,height=4cm,legend pos= north east,
				ylabel shift = -0.1cm,  xshift=0cm,
				ybar,
				xtick={1,2,3,4},xticklabel style={align=center}, 
				xticklabels = {{Agent 1},{Agent 2},{Agent 3},{Agent 4}},
				bar width=0.3cm,ybar=0.0cm,
				legend image code/.code={
					\draw [#1] (0cm,-0.1cm) rectangle (0.4cm,0.1cm); },]
				\addplot[fill=pariedLight_red!50]    table[x = AgentNrSet , y  = centralized_M_mean ]{\file};
				\addplot[fill=pariedLight_blue!50]    table[x = AgentNrSet , y  = distributed_M_mean ]{\file};
				\addplot[fill=pariedLight_green!50]    table[x = AgentNrSet , y  = NoTrigger_M_mean ]{\file};
				\addplot[fill=pariedLight_orange!50]    table[x = AgentNrSet , y  = continuous0015_T_max ]{\file};
				\addplot [paried_red, only marks, mark=.] plot [error bars/.cd, y dir=both, y explicit relative]
				table [x = AgentNrSet , y  = centralized_M_mean, y error plus=centralized_M_var, y error minus=centralized_M_var] {\file};
				\addplot [paried_blue, only marks, mark=.] plot [error bars/.cd, y dir=both, y explicit relative]
				table [x = AgentNrSet , y  = distributed_M_mean, y error plus=distributed_M_var, y error minus=distributed_M_var] {\file};
				\addplot [black, only marks, mark=.] plot [error bars/.cd, y dir=both, y explicit relative]
				table [x = AgentNrSet , y  = NoTrigger_M_mean, y error plus=NoTrigger_M_var, y error minus=NoTrigger_M_var] {\file};
				\addplot [black, only marks, mark=.] plot [error bars/.cd, y dir=both, y explicit relative]
				table [x = AgentNrSet , y  = continuous0015_T_mean, y error plus=continuous0015_T_var, y error minus=continuous0015_T_var] {\file};
				\legend{{Centralized trigger},{Distributed trigger},{Offline learning},{Time trigger}}
			\end{axis}
		\end{tikzpicture}
		\vspace{-0.3cm}
		\caption{
			Maximal size of data set, which for offline learning is $200$ from the size of initial data set and for time-triggered learning is $4000$ due to the fixed trigger interval $0.01$.
			With centralized event-triggered mechanism, the maximal size of the data set for each agent $1$ to $4$ denotes $157 \pm 13$, $152 \pm 13$, $145 \pm 14$ and $135 \pm 12$, respectively.
			The maximal numbers of training samples collected through distributed even-triggered mechanism are $177 \pm 16$, $175 \pm 15$, $158 \pm 17$ and $153 \pm 15$ for agent $1$ to $4$, respectively.
		}
		\vspace{-0.4cm}
		\label{figure_Method_TriggerTimes}
	\end{figure}
	
	\subsection{Manipulators with 2 Degrees of Freedoms}
	
	In this subsection, a multi-agent system consisting of $N = 6$ manipulators with 2 degrees of freedoms (DoFs) is considered, where the communication topology is described as $\mathcal{G} = \{ \mathcal{V}, \mathcal{E} \}$ with $\mathcal{V} = \{ 1, \cdots, 6\}$ and $\mathcal{E} \!=\! \{ (1,2),\! (1,6),\! (2,3),\! (3,2),\! (4,3),\! (4,5),\! (5,6),\! (6,5) \}$.
	Each manipulator follows a second order dynamics in \eqref{eqn_agent_dynamics} with $\bm{g}(\bm{x}_i) \!\!=\!\! \bm{M}^{-\!1\!}(\bm{x}_{i,1}) \bm{T}(\bm{x}_{i,\!1})$, $\bm{h}(\bm{x}_i) \!\!=\!\! \bm{M}^{-\!1\!}(\bm{x}_{i,\!1}) \bm{f}_h(\bm{x}_i)$ and $n \!=\! 2$, $p \!=\! q \!=\! 2$, where $\bm{x}_{i,1}, \bm{x}_{i,2}$ represent the absolute position and translational velocity of end effector for each manipulator $i$ in inertial coordinate with the compact domain $\mathbb{X} \!=\! [-1.5, 1.5]^4$.
	The mass matrix $\bm{M}(\bm{x}_{i,1})$, Jacobian matrix $\bm{T}(\bm{x}_{i,1})$ and generalized force $\bm{f}_h(\bm{x}_i)$ are known and calculated in appendix with corresponding manipulators' properties.
	The unknown function $\bm{f}(\cdot)$ represents the friction effect directly affected on the end effector and is written as
	\begin{align}
		\bm{f}(\bm{x}_i) = \begin{bmatrix}
			5 \sin(x_{i,1,1}) + 3 \cos(x_{i,1,1}) + x_{i,2,1}^2 + 6 \\
			3 \cos(x_{i,1,2}) + 5 \cos(x_{i,1,2}) + x_{i,2,2}^2 + 10
		\end{bmatrix}, \nonumber
	\end{align}
	where $x_{i,j,k}$ denotes the $k$-th entry of $\bm{x}_{i,j}$ for $\forall k \!=\! 1, \!\cdots\!, p$.
	Moreover, only agents $1$ and $4$ have access to the leader, indicating $b_{11} \!=\! b_{44} \!=\! 1$ and $b_{ii} \!=\! 0$ for $\forall i \!\in\! \mathcal{V} \backslash \{1, 4\}$.
	The prediction of $\bm{f}(\cdot)$ on each agent is obtained using Gaussian process regression with squared exponential kernel as $\kappa_i(\bm{x}_i, \bm{x}_i') = \exp( 2 \| \bm{x}_i - \bm{x}_i' \|^2 )$ for $\forall i \in \mathcal{V}$ and POE as aggregation strategy.
	The training data pairs satisfy \cref{assumption_dataset} with the standard deviation of the measurement noise as $\sigma_{o,i} = 0.01$ for $\forall i \in \mathcal{V}$.
	The grid factor and probability for calculating the uniform prediction error bound in \cref{lemma_GP_error_bound} are set as $\tau = 10^{-8}$ and $\delta = 0.1$, respectively.
	The control objective is designed such that the end effector of each manipulator tracks a leader trajectory defined as
	\begin{align}
		\bm{x}_{l,1}(t) \!\!=\!\! [\sin(0.5 t),\! \cos(0.5 t)]^T\! ,\! \bm{x}_{l,2}(t) \!\!=\!\! \dot{\bm{x}}_{l,1}(t),\! \bm{x}_{l,r}(t) \!\!=\!\! \dot{\bm{x}}_{l,2}(t) \nonumber
	\end{align}
	with relative states for each manipulator $i$ as
	\begin{align}
		&\bm{s}_{i,1}(t) = 0.2 [\cos(1.5 t + i \pi / 3), \sin(1.5 t + i \pi / 3)]^T, \nonumber \\ 
		&\bm{s}_{i,2}(t) = \dot{\bm{s}}_{i,1}(t), \qquad \bm{s}_{i,r}(t) = \dot{\bm{s}}_{i,2}(t). \nonumber
	\end{align}
	The control parameters in \eqref{eqn_control_law} are set as $c = 10$, $\bm{\lambda} = [1, 1]^T$.
	Moreover, choose $\bm{Q}_{\varepsilon} = \bm{I}_2$ inducing positive definite $\bm{Q}_z$ in \eqref{eqn_Qz}.
	The simulation time is set from $0$ to $40$.
	
	To demonstrate the effectiveness of the proposed event-triggered online learning strategies, the tracking error $\| \bm{\vartheta}(t) \|$ and the number of trigger events are compared among the learning strategies in \cref{subsection_simulation_toy_example_setting}.
	Note that the number of initial data set for offline learning increases to $350$ due to the expansion of the system dimension $p$.
	The comparison results are shown in the following subsections. 
	
	\subsubsection{Performance with Event-triggered Online Learning}
	
	To show the effectiveness of different cooperative learning strategies, the tracking errors from the same initial states, i.e., $\bm{x}_1(0) \!=\! [0.8147,\! 0.9058,\! 0.1270,\! 0.9134]^T$, $\bm{x}_2(0) \!=\! [0.6324,\! 0.0975,\! 0.2785,\! 0.5469]^T$, $\bm{x}_3(0) \!=\! [0.9575,\! 0.9649,\! $ $0.1576,\! 0.9706]^T$, $\bm{x}_4(0) \!=\! [0.9572,\! 0.4854,\! 0.8003,\! 0.1419]^T$, $\bm{x}_5(0) \!=\! [0.4218,\! 0.9157,\! 0.7922,\! 0.9595]^T$, $\bm{x}_6(0) \!=\! [0.6557,\!$ $ 0.0357,\! 0.8491,\! 0.9340]^T$, are compared.
	The tracking error $\| \bm{\vartheta}_i(t) \|$ for each agent $i \in \mathcal{V}$ are shown in \cref{figure_Manipulator_Agent_Error}.
	It is obvious to see online learning performs better with lower $\| \bm{\vartheta}_i(t) \|$ for $\forall i \in \mathcal{V}$, while with offline learning the tracking error is much larger than with other methods especially due to the lack of sufficient data in the domain around references.
	Note that the tracking error does not tend to $0$ due to the non-fully connection between the agents and leader inducing non-zeros $\bm{\iota}$ in \cref{theorem_best_tracking_performance_online}.
	However, the tracking performance of CET, DET and TT is similar, showing the efficiency of the event-triggered strategies.
	\looseness=-1
	
	\begin{figure}[t]
		\centering
		\begin{tikzpicture}
			\def\fileCET{fig/Manipulator_AgentError_centralized.txt}
			\def\fileDET{fig/Manipulator_AgentError_distributed.txt}
			\def\fileTT{fig/Manipulator_AgentError_time.txt}
			\def\fileOFF{fig/Manipulator_AgentError_offline.txt}
			\def\fileEXA{fig/Manipulator_AgentError_exact.txt}
			
			\begin{semilogyaxis}[xlabel={},ylabel={$\| \bm{\vartheta}_1(t) \|$},
				xmin=1, ymin = 1.2e-3, xmax = 39,ymax=0.8e0,legend columns=5,
				width=0.27\textwidth,height=2.8cm,legend style={at={(-1.6,1.12)},anchor=west},
				xticklabels={,,}, ylabel shift=-0.2cm,]
				\addplot[paried_red!80, thick, ]      table[x = t_set , y  = Agent1_e ]{\fileCET};
				\addplot[paried_blue!80, thick, ]     table[x = t_set , y  = Agent1_e ]{\fileDET};
				\addplot[paried_orange!80, thick, ]   table[x = t_set , y  = Agent1_e ]{\fileTT};
				\addplot[paried_green!80, thick, ]    table[x = t_set , y  = Agent1_e ]{\fileOFF};
				\addplot[black!50, thick, ]           table[x = t_set , y  = Agent1_e ]{\fileEXA};
			\end{semilogyaxis}
			\begin{semilogyaxis}[xlabel={},ylabel={$\| \bm{\vartheta}_2(t) \|$},
				xmin=1, ymin = 1.2e-3, xmax = 39,ymax=0.8e0,legend columns=5,
				width=0.27\textwidth,height=2.8cm,legend style={at={(-0.98,1.25)},anchor=west},,
				xticklabels={,,}, ylabel shift=-0.2cm, yticklabels={,,}, ylabel shift=-0.2cm,
				xshift=4cm]
				\addplot[paried_red!80, thick, ]      table[x = t_set , y  = Agent2_e ]{\fileCET};
				\addplot[paried_blue!80, thick, ]     table[x = t_set , y  = Agent2_e ]{\fileDET};
				\addplot[paried_orange!80, thick, ]   table[x = t_set , y  = Agent2_e ]{\fileTT};
				\addplot[paried_green!80, thick, ]    table[x = t_set , y  = Agent2_e ]{\fileOFF};
				\addplot[black!50, thick, ]           table[x = t_set , y  = Agent2_e ]{\fileEXA};
				\legend{CET, DET, TT, Offline, Exact}
			\end{semilogyaxis}
			
			\begin{semilogyaxis}[xlabel={},ylabel={$\| \bm{\vartheta}_3(t) \|$},
				xmin=1, ymin = 1.2e-3, xmax = 39,ymax=0.8e0,legend columns=5,
				width=0.27\textwidth,height=2.8cm,legend style={at={(-1.6,1.12)},anchor=west},,
				xticklabels={,,}, ylabel shift=-0.2cm,
				yshift=-1.3cm]
				\addplot[paried_red!80, thick, ]      table[x = t_set , y  = Agent3_e ]{\fileCET};
				\addplot[paried_blue!80, thick, ]     table[x = t_set , y  = Agent3_e ]{\fileDET};
				\addplot[paried_orange!80, thick, ]   table[x = t_set , y  = Agent3_e ]{\fileTT};
				\addplot[paried_green!80, thick, ]    table[x = t_set , y  = Agent3_e ]{\fileOFF};
				\addplot[black!50, thick, ]           table[x = t_set , y  = Agent3_e ]{\fileEXA};
			\end{semilogyaxis}
			\begin{semilogyaxis}[xlabel={},ylabel={$\| \bm{\vartheta}_4(t) \|$},
				xmin=1, ymin = 1.2e-3, xmax = 39,ymax=0.8e0,legend columns=5,
				width=0.27\textwidth,height=2.8cm,legend style={at={(0,1.15)},anchor=west},,
				xticklabels={,,}, ylabel shift=-0.2cm, yticklabels={,,}, ylabel shift=-0.2cm,
				yshift=-1.3cm,xshift=4cm]
				\addplot[paried_red!80, thick, ]      table[x = t_set , y  = Agent4_e ]{\fileCET};
				\addplot[paried_blue!80, thick, ]     table[x = t_set , y  = Agent4_e ]{\fileDET};
				\addplot[paried_orange!80, thick, ]   table[x = t_set , y  = Agent4_e ]{\fileTT};
				\addplot[paried_green!80, thick, ]    table[x = t_set , y  = Agent4_e ]{\fileOFF};
				\addplot[black!50, thick, ]           table[x = t_set , y  = Agent4_e ]{\fileEXA};
			\end{semilogyaxis}
			
			\begin{semilogyaxis}[xlabel={{\color{white} 1111111} $t$},ylabel={$\| \bm{\vartheta}_5(t) \|$},
				xmin=1, ymin = 1.2e-3, xmax = 39,ymax=0.8e0,legend columns=5,
				width=0.27\textwidth,height=2.8cm,legend style={at={(0,1.15)},anchor=west},
				xlabel style={text width=2.5cm}, ylabel shift=-0.2cm,
				yshift=-2.6cm]
				\addplot[paried_red!80, thick, ]      table[x = t_set , y  = Agent5_e ]{\fileCET};
				\addplot[paried_blue!80, thick, ]     table[x = t_set , y  = Agent5_e ]{\fileDET};
				\addplot[paried_orange!80, thick, ]   table[x = t_set , y  = Agent5_e ]{\fileTT};
				\addplot[paried_green!80, thick, ]    table[x = t_set , y  = Agent5_e ]{\fileOFF};
				\addplot[black!50, thick, ]           table[x = t_set , y  = Agent5_e ]{\fileEXA};
			\end{semilogyaxis}
			\begin{semilogyaxis}[xlabel={{\color{white} 1111111} $t$},ylabel={$\| \bm{\vartheta}_6(t) \|$},
				xmin=1, ymin = 1.2e-3, xmax = 39,ymax=0.8e0,legend columns=5,
				width=0.27\textwidth,height=2.8cm,legend style={at={(0,1.15)},anchor=west},
				xlabel style={text width=2.5cm}, ylabel shift=-0.2cm,
				yticklabels={,,}, ylabel shift=-0.2cm,
				yshift=-2.6cm,xshift=4cm]
				\addplot[paried_red!80, thick, ]      table[x = t_set , y  = Agent6_e ]{\fileCET};
				\addplot[paried_blue!80, thick, ]     table[x = t_set , y  = Agent6_e ]{\fileDET};
				\addplot[paried_orange!80, thick, ]   table[x = t_set , y  = Agent6_e ]{\fileTT};
				\addplot[paried_green!80, thick, ]    table[x = t_set , y  = Agent6_e ]{\fileOFF};
				\addplot[black!50, thick, ]           table[x = t_set , y  = Agent6_e ]{\fileEXA};
			\end{semilogyaxis}
		\end{tikzpicture}
		\vspace{-0.3cm}
		\caption{
			Tracking error of each agent over time.
		}
		\vspace{-0.3cm}
		\label{figure_Manipulator_Agent_Error}
	\end{figure}
	
	The time instances for each trigger in centralized and distributed event-triggered online learning are shown in \cref{figure_Manipulator_trigger} for each agent.
	It is seen that total number of data pairs in each agent is smaller than $350$, which is the size of initial data set for offline learning.
	Moreover, no trigger occurs after $t = 25$, indicating sufficiently accurate GP prediction is achieved for the entire MAS and desired tracking performance.
	
	\begin{figure}[t] 
		\centering
		\def\file{fig/Manipulator_trigger.txt}
		\begin{tikzpicture}
			\begin{axis}[xlabel={},ylabel={CET},
				xmin=0.1, ymin = 0.5, xmax = 39.9,ymax=6.5,legend columns=1,
				width=0.48\textwidth,height=3.5cm,legend pos= north east,
				ytick={1,2,3,4,5,6},
				yticklabels={{Agent 1}, {Agent 2}, {Agent 3}, {Agent 4}, {Agent 5}, {Agent 6}},
				xticklabels={,,,,}]
				\addplot[only marks, mark=x, paried_red!80]    table[x = time_centralized_1 , y  = trigger_centralized_1 ]{\file};
				\addplot[only marks, mark=x, paried_red!80]    table[x = time_centralized_2 , y  = trigger_centralized_2 ]{\file};
				\addplot[only marks, mark=x, paried_red!80]    table[x = time_centralized_3 , y  = trigger_centralized_3 ]{\file};
				\addplot[only marks, mark=x, paried_red!80]    table[x = time_centralized_4 , y  = trigger_centralized_4 ]{\file};
				\addplot[only marks, mark=x, paried_red!80]    table[x = time_centralized_4 , y  = trigger_centralized_5 ]{\file};
				\addplot[only marks, mark=x, paried_red!80]    table[x = time_centralized_4 , y  = trigger_centralized_6 ]{\file};
				\addplot[only marks, mark=x, blue]    coordinates{(-1,-1)};
				
			\end{axis}
			\begin{axis}[xlabel={$t$},ylabel={DET},
				xmin=0.1, ymin = 0.5, xmax = 39.9,ymax=6.5,legend columns=1,
				width=0.48\textwidth,height=3.5cm,legend pos= south east,
				ytick={1,2,3,4,5,6},
				yticklabels={{Agent 1}, {Agent 2}, {Agent 3}, {Agent 4}, {Agent 5}, {Agent 6}},
				yshift=-2cm]
				\addplot[only marks, mark=x, red]    coordinates{(-1,-1)};
				\addplot[only marks, mark=x, paried_blue!80]    table[x = time_distributed_1 , y  = trigger_distributed_1 ]{\file};
				\addplot[only marks, mark=x, paried_blue!80]    table[x = time_distributed_2 , y  = trigger_distributed_2 ]{\file};
				\addplot[only marks, mark=x, paried_blue!80]    table[x = time_distributed_3 , y  = trigger_distributed_3 ]{\file};
				\addplot[only marks, mark=x, paried_blue!80]    table[x = time_distributed_4 , y  = trigger_distributed_4 ]{\file};
				\addplot[only marks, mark=x, paried_blue!80]    table[x = time_distributed_3 , y  = trigger_distributed_5 ]{\file};
				\addplot[only marks, mark=x, paried_blue!80]    table[x = time_distributed_4 , y  = trigger_distributed_6 ]{\file};
			\end{axis}
		\end{tikzpicture}
		\vspace{-0.3cm}
		\caption{
			Trigger instances for CET and DET for each agent.
			Specifically, for CET $250$, $238$, $273$, $263$, $277$ and $260$ samples are collected for agent $1$ to $6$.
			With DET, each agent collects $274$, $258$, $303$, $288$, $303$ and $286$ data pairs.
		}
		\vspace{-0.3cm}
		\label{figure_Manipulator_trigger}
	\end{figure}
	
	\subsubsection{Monte Carlo Test}
	
	The statistical property of the derived theorems are demonstrated through Monte Carlo Test, where the simulations are repeated for $100$ times with random initial states $\bm{x}(0)$ sampled from uniform distribution in $[0,1]^{nNp}$, normally distributed measurement noise and random initial data set with $\bm{x}_i^{(\iota)} \in \mathbb{X}$, $\forall i \in \mathcal{V}$, $\forall \iota = 1, \cdots, 350$ for offline learning.
	The control performance for each learning algorithm is reflected by the ultimate tracking error as shown in \cref{figure_Manipulator_Monte_Carlo_error}.
	Specifically, the overall tracking error $\| \bm{\vartheta}(t) \|$ over time $t$ is shown in \cref{figure_Manipulator_error_time_MonteCarlo}.
	Moreover, \cref{figure_Manipulator_ultimate_tracking_error} indicates the ultimate tracking error by considering the maximal $\| \bm{\vartheta}(t) \|$ after $t = 20$, such that the initial behavior of the systems is excluded.
	
	\pgfplotsset{
		boxplot/box extend=0.48,
		boxplot/every whisker/.style={thick,blue},
		boxplot/every box/.style={thick,blue},
		boxplot/every median/.style={red}
	}
	\begin{figure}[t] 
		\centering
		\begin{subfigure}{\columnwidth}
			\def\file{fig/Manipulator_Method_e_over_t.txt}
			\begin{tikzpicture}
				\begin{semilogyaxis}[xlabel={$t$},ylabel={$\| \bm{\vartheta}(t) \|$},
					xmin=0.1, ymin =0.02, xmax = 39.9,ymax=39.9, legend columns=4,
					width=\textwidth,height=4cm,legend pos=north east, 
					clip mode=individual]
					\addplot[paried_red!80, thick]    table[x = t_set , y  = centralized_mean_e_t ]{\file};
					\addplot+[name path=max_1,black,no markers, draw=none] table[x = t_set , y  = centralized_max_e_t ]{\file};
					\addplot+[name path=min_1,black,no markers, draw=none] table[x = t_set , y  = centralized_min_e_t ]{\file};
					\addplot[paried_red!80, fill opacity=0.2] fill between[of=max_1 and min_1];
					
					\addplot[paried_blue!80, thick]    table[x = t_set , y  = distributed_mean_e_t ]{\file};
					\addplot+[name path=max_2,black,no markers, draw=none] table[x = t_set , y  = distributed_max_e_t ]{\file};
					\addplot+[name path=min_2,black,no markers, draw=none] table[x = t_set , y  = distributed_min_e_t ]{\file};
					\addplot[paried_blue!80, fill opacity=0.2] fill between[of=max_2 and min_2];
					
					\addplot[paried_orange!80, thick]    table[x = t_set , y  = time_mean_e_t ]{\file};
					\addplot+[name path=max_3,black,no markers, draw=none] table[x = t_set , y  = time_max_e_t ]{\file};
					\addplot+[name path=min_3,black,no markers, draw=none] table[x = t_set , y  = time_min_e_t ]{\file};
					\addplot[paried_orange!80, fill opacity=0.2] fill between[of=max_3 and min_3];
					
					\addplot[paried_green!80, thick]    table[x = t_set , y  = offline_mean_e_t ]{\file};
					\addplot+[name path=max_4,black,no markers, draw=none] table[x = t_set , y  = offline_max_e_t ]{\file};
					\addplot+[name path=min_4,black,no markers, draw=none] table[x = t_set , y  = offline_min_e_t ]{\file};
					\addplot[paried_green!80, fill opacity=0.2] fill between[of=max_4 and min_4];
					
					\addplot[black!50, very thick, dashed]    table[x = t_set , y  = exact_mean_e_t ]{\file};
					\addplot+[name path=max_5,black,no markers, draw=none] table[x = t_set , y  = exact_max_e_t ]{\file};
					\addplot+[name path=min_5,black,no markers, draw=none] table[x = t_set , y  = exact_min_e_t ]{\file};
					\addplot[black!50, fill opacity=0.2] fill between[of=max_5 and min_5];
					\legend{
						CET,,,Variance,
						DET,,,Variance,
						TT,,,Variance,
						Offline,,,Variance,
						Exact,,,Variance
					}
				\end{semilogyaxis}
			\end{tikzpicture}
			\vspace{-0.2cm}
			\caption{
				Mean and variance of overall tracking error with respect to time from $100$ times Monte Carlo tests.
			}
			\label{figure_Manipulator_error_time_MonteCarlo}
		\end{subfigure}
		\begin{subfigure}{\columnwidth}
			\begin{tikzpicture}
				\def\file{fig/Manipulator_Method_max_e.txt}
				\begin{semilogyaxis}[
					boxplot/draw direction = y,
					ylabel style={align=center},ylabel={$\max\limits_{t \!\in\! [20,\! 40]} \| \bm{\vartheta}(t) \|$},
					ymin=0.058, ymax=1.8,legend columns=2,
					xmin=0.65,xmax=5.35,
					width=\textwidth,height=4cm,
					xtick={1,2,3,4,5},xticklabel style={align=center}, 
					xticklabels = {Offline,CET,DET,TT, Exact}]
					\addplot+[boxplot, mark=triangle] table[y=offline_max_e] {\file};
					\addplot+[boxplot, mark=triangle] table[y=centralized_max_e] {\file};
					\addplot+[boxplot, mark=triangle] table[y=distributed_max_e] {\file};
					\addplot+[boxplot, mark=triangle] table[y=time_max_e ] {\file};
					\addplot+[boxplot, mark=triangle] table[y=exact_max_e] {\file};
				\end{semilogyaxis}
				
				\def\file{fig/Manipulator_Method_max_e.txt}
				\begin{axis}[
					boxplot/draw direction = y,
					ylabel style={align=center},ylabel={},
					ymin=0.07, ymax=0.119,legend columns=2,
					xmin=0.6,xmax=3.4,
					width=6.3cm,height=2.7cm,
					xtick={1,2,3,4},xticklabel style={align=center}, 
					xticklabels = {CET,DET,TT, Exact},
					xshift=2.2cm,yshift=1.2cm,
					axis background/.style={fill=white}]
					\addplot+[boxplot, mark=triangle] table[y=centralized_max_e] {\file};
					\addplot+[boxplot, mark=triangle] table[y=distributed_max_e] {\file};
					\addplot+[boxplot, mark=triangle] table[y=time_max_e] {\file};
				\end{axis}
			\end{tikzpicture}
			\vspace{-0.2cm}
			\caption{
				Maximal tracking error in the steady state.
			}
			\label{figure_Manipulator_ultimate_tracking_error}
		\end{subfigure}
		\vspace{-0.3cm}
		\caption{
			Tracking error from different learning strategy.
		}
		\vspace{-0.4cm}
		\label{figure_Manipulator_Monte_Carlo_error}
	\end{figure}
	
	The online learning efficiency is reflected by the maximal size of eventual training data set on each agent shown in \cref{figure_Manipulator_Method_TriggerTimes}.
	It is easy to see that the maximal number of data pairs in the training data set for each agent with both event-triggered online learning mechanisms results is less than $350$ samples, which is the designed size of initial data set for offline learning.
	Moreover, in comparison to centralized methods, the distributed approach allows agents to collect only a slightly larger amount of data without the need for an extra computational center.
	Combining with the observation in \cref{figure_Manipulator_trigger}, limited data is eventually added into individual data set for each agent with event-triggered strategies.
	In comparison, time-triggered method intends to collect data without termination, which results in an infinitely growing data set linear with operating time and high demand on the computational resources.

	\begin{figure}[t] 
		\centering
		\begin{tikzpicture}
			\def\file{fig/Manipulator_Method_M.txt}
			\begin{axis}[ylabel style={align=center},ylabel={Maximum Size \\ of the Data Set},
				ymin=1, ymax=720,legend columns=2,
				xmin=0.5,xmax=6.5,
				width=0.48\textwidth,height=3.5cm,legend pos= north east,
				ylabel shift = -0.1cm,  xshift=0cm,
				ybar,
				xtick={1,2,3,4,5,6},xticklabel style={align=center}, 
				xticklabels = {{Agent 1},{Agent 2},{Agent 3},{Agent 4},{Agent 5},{Agent 6}},
				bar width=0.25cm,ybar=0.0cm,
				legend image code/.code={
					\draw [#1] (0cm,-0.1cm) rectangle (0.4cm,0.1cm); },]
				\addplot[fill=pariedLight_red!50]    table[x = AgentNrSet , y  = centralized_M_mean ]{\file};
				\addplot[fill=pariedLight_blue!50]    table[x = AgentNrSet , y  = distributed_M_mean ]{\file};
				\addplot[fill=pariedLight_green!50]    table[x = AgentNrSet , y  = offline_M_mean ]{\file};
				\addplot[fill=pariedLight_orange!50]    table[x = AgentNrSet , y  = time_M_mean ]{\file};
				\addplot [paried_red, only marks, mark=.] plot [error bars/.cd, y dir=both, y explicit relative]
				table [x = AgentNrSet , y  = centralized_M_mean, y error plus=centralized_M_var, y error minus=centralized_M_var] {\file};
				\addplot [paried_blue, only marks, mark=.] plot [error bars/.cd, y dir=both, y explicit relative]
				table [x = AgentNrSet , y  = distributed_M_mean, y error plus=distributed_M_var, y error minus=distributed_M_var] {\file};
				\addplot [black, only marks, mark=.] plot [error bars/.cd, y dir=both, y explicit relative]
				table [x = AgentNrSet , y  = offline_M_mean, y error plus=offline_M_var, y error minus=offline_M_var] {\file};
				\addplot [black, only marks, mark=.] plot [error bars/.cd, y dir=both, y explicit relative]
				table [x = AgentNrSet , y  = time_M_mean, y error plus=time_M_var, y error minus=time_M_var] {\file};
				\legend{{Centralized trigger},{Distributed trigger},{Offline learning},{Time trigger}}
			\end{axis}
		\end{tikzpicture}
		\vspace{-0.3cm}
		\caption{
			Maximal size of data set, which for offline learning is $350$ from the size of initial data set and for TT is $2667$ due to the fixed trigger interval $0.015$.
			With CET, the maximal size of the data set for each agent $1$ to $6$ denotes $246 \!\pm\! 5$, $251 \!\pm\! 9$, $265 \!\pm\! 12$, $268 \!\pm\! 10$, $271 \!\pm\! 8$ and $266 \!\pm\! 6$, respectively.
			The maximal numbers of training samples collected through DET are $268 \!\pm\! 7$, $277 \!\pm\! 10$, $290 \!\pm\! 13$, $294 \!\pm\! 12$, $298 \!\pm\! 8$ and $295 \!\pm\! 7$ for agent $1$ to $6$, respectively.
		}
		\vspace{-0.4cm}
		\label{figure_Manipulator_Method_TriggerTimes}
	\end{figure}
	
	\ifarxiv
	Until here, all effects from the proposed distributed event-triggered online learning are observed through the actual tracking performance and trigger times, which demonstrate the effectiveness of the proposed method.
	\fi
	
	\section{Conclusions} \label{section_conclusion}
	
	In this paper, we consider a distributed control framework for leader-following time-varying formation control with cooperative online learning algorithm using GP regression.
	For high data efficiency with guaranteed control performance, two event-triggered mechanisms are proposed for online learning, namely the centralized and distributed version.
	Inspired from the centralized event-triggered learning design, the distributed trigger condition is evaluated only based on the local and neighboring information on each agent.
	Moreover, we show the exclusion of the Zeno behavior on each agent for both centralized and distributed event-trigger.
	Finally, the effectiveness of both proposed event-triggered cooperative online learning strategies is demonstrated in the simulations, which show that each agent collects fewer data and triggers less GP model update compared with other approaches while achieving a guaranteed control performance.

	{\appendix[]	 	
		\begin{IEEEproof} [Proof of \cref{proposition_centralized_ET}]
			The proof follows \cite{dai2023can}, where the sign of $\dot{V}$ for $\| \bm{z} \| > \chi^{-1} \bar{\vartheta}_c$ is investigated such that \eqref{eqn_centralized_trigger} is reformulated as $\bar{\rho} = \| \bm{z} \|$.
			Then, two cases divided by \eqref{eqn_centralized_trigger} is considered.
			In the case with $\rho(t) < \bar{\rho}(t)$ indicating no model update, the negativity of $\dot{V}$ is shown as
			\begin{align}
				\dot{V} < - \underline{\lambda}(\bm{Q}_z) \| \bm{z} \| ( \xi \rho - \xi \| \bm{\iota} + \hat{\bm{\eta}}_{\delta}(\bm{x}) \| ) \le 0 \nonumber
			\end{align}
			by considering \eqref{eqn_dotV_trigger_prepare} with \eqref{eqn_bound_Upsilon}.
			If $\rho(t) \ge \bar{\rho}(t)$, then the model update is activated, and with $\| \bm{z} \| > \chi^{-1} \bar{\vartheta}_c$ it has
			\begin{align}
				\dot{V} &< - \underline{\lambda}(\bm{Q}_z) \| \bm{z} \| ( \chi^{-1} \bar{\vartheta}_c - \xi \| \bm{\iota} + \underline{\tilde{\bm{\eta}}}_{\delta}(\bm{x}) \| ) \nonumber \\
				&\le - \underline{\lambda}(\bm{Q}_z) \| \bm{z} \| \xi ( \| \bm{\iota} + \underline{\hat{\bm{\eta}}}_{\delta} + \bm{\epsilon} \| - \| \bm{\iota} + \underline{\tilde{\bm{\eta}}}_{\delta}(\bm{x}) + \bm{\epsilon} \| ) \le 0, \nonumber
			\end{align}
			where the second inequality is derived by using the definition of $\bar{\vartheta}_c$ and the result from \eqref{eqn_centralized_ET_update_agent_selection}.
			Until here, the negativity of $\dot{V}$ when $\| \bm{z} \| > \chi^{-1} \bar{\vartheta}_c$ is proven, which concludes the proof by letting $\bar{z} = \chi^{-1} \bar{\vartheta}_c$ and using the result in \cref{lemma_boundness_z_vartheta}.
		\end{IEEEproof}
		
		\begin{IEEEproof} [Proof of \cref{lemma_zi}]
			The proof considers the contradiction, where $\| \bm{z}_i \| \le \bar{z} / \sqrt{N}$ for all $i \in \mathcal{V}$.
			Then, $\| \bm{z} \|$ is bounded by
			\begin{align}
				\| \bm{z} \|^2 = \sum\nolimits_{i \in \mathcal{V}} \| \bm{z}_i \|^2 \le \sum\nolimits_{i \in \mathcal{V}} \bar{z}^2 / N = \bar{z}^2, \nonumber
			\end{align}
			which is contradict to $\| \bm{z} \| > \bar{z}$ in the lemma.
		\end{IEEEproof}
	}
	
	\bibliographystyle{IEEEtran}
	\bibliography{refs}
	
	\vspace{-11pt}
	\begin{IEEEbiography}[{\includegraphics[width=1in,height=1.25in,clip,keepaspectratio]{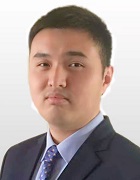}}]{Xiaobing Dai}
		received the B.Sc. mechanical engineering from the Tongji University, Shanghai, China, in 2018 with direction in mechatronics, building environment and civil engineering. He received double M.Sc degrees in Mechanical Engineering, Mechatronics and Robotics from the Technical University of Munich, Munich, Germany, in 2021. Since February 2022, he is a PhD student at the Chair of Information-oriented Control, TUM School of Computation, Information and Technology at the Technical University of Munich, Munich, Germany. His current research interests include efficient machine learning, networked control systems, safe learning-based control.
	\end{IEEEbiography}
	\vspace{-11pt}
	
	\begin{IEEEbiography}[{\includegraphics[width=1in,height=1.25in,clip,keepaspectratio]{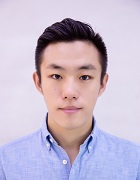}}]{Zewen Yang}
		received the M.S. degree in control engineering from Northeast Forest University, in 2017. He pursued a Ph.D. in control science and engineering at College of Intelligent Systems Science and Engineering, Harbin Engineering University, Harbin, China, from 2017 to 2019 and joined the Chair of Information-oriented Control, School of Computation, Information and Technology, the Technical University of Munich, Munich, Germany, in machine learning and data-driven control until 2023. His current research interests include multi-agent systems, cooperative learning, control theory, and general robotics.
	\end{IEEEbiography}
	\vspace{-11pt}
	
	\begin{IEEEbiography}[{\includegraphics[width=1in,height=1.25in,clip,keepaspectratio]{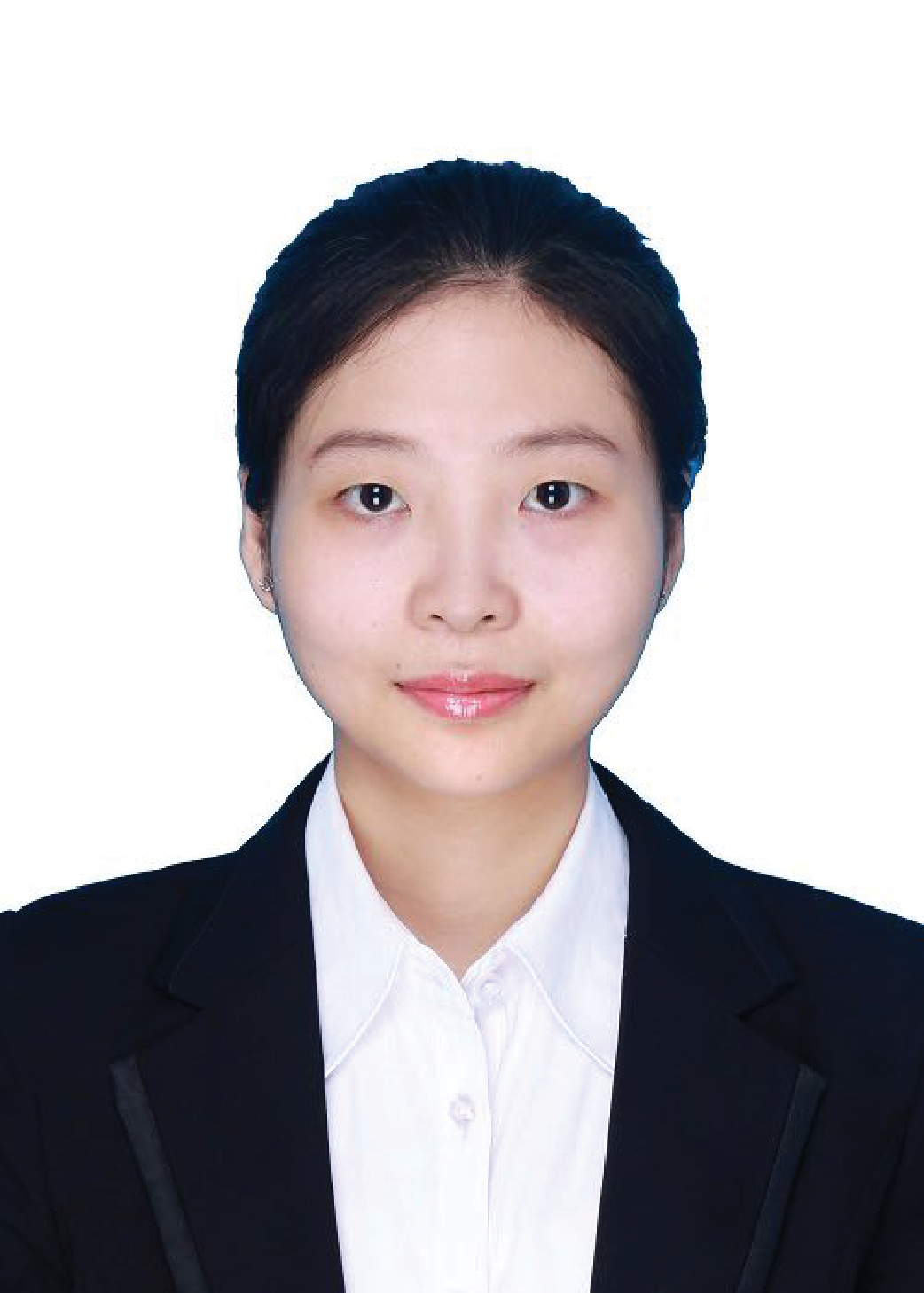}}]{Sihua Zhang}
		received the M.Eng. degree in control engineering from the Beijing Institute of Technology, Beijing, China, in 2021. She is currently pursuing the doctor's degree in control engineering with the School of Automation, Beijing Institute of Technology, Beijing, China. From 2023 to 2025, she is additionally working as research assistant at the Chair of Information-oriented Control, TUM School of Computation, Information and Technology at the Technical University of Munich, Germany. Her current research interests include safety-critical robotic control, control barrier functions, machine learning and optimal control.
	\end{IEEEbiography}
	\vspace{-11pt}
	
	\begin{IEEEbiography}[{\includegraphics[width=1in,height=1.25in,clip,keepaspectratio]{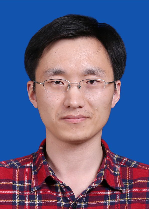}}]{Di-Hua Zhai}
		received the B.Eng. degree in automation from Anhui University, Hefei, China, in 2010, the M.Eng. degree in control science and engineering from the University of Science and Technology of China, Hefei, in 2013, and the Dr.Eng. degree in control science and engineering from the Beijing Institute of Technology, Beijing, China, in 2017. Since 2017, he has been with the School of Automation, Beijing Institute of Technology, where he is currently an Associate Professor. His research interests include teleoperation, intelligent robot, human robot collaboration, switched control, optimal control, constrained control, and networked control.
	\end{IEEEbiography}
	\vspace{-11pt}
	
	\begin{IEEEbiography}[{\includegraphics[width=1in,height=1.25in,clip,keepaspectratio]{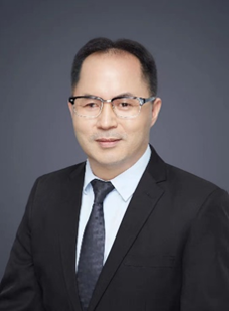}}]{Yuanqing Xia}
		(Fellow, IEEE) received his M.S. degree in Fundamental Mathematics from Anhui University, China, in 1998 and his Ph.D. degree in Control Theory and Control Engineering from Beijing University of Aeronautics and Astronautics, Beijing, China, in 2001. During January 2002-November 2003, he was a Postdoctoral Research Associate with the Institute of Systems Science, Academy of Mathematics and System Sciences, Chinese Academy of Sciences, Beijing, China. From November 2003 to February 2004, he was with the National University of Singapore as a Research Fellow, where he worked on variable structure control. From February 2004 to February 2006, he was with the University of Glamorgan, Pontypridd, U.K., as a Research Fellow. From February 2007 to June 2008, he was a Guest Professor with Innsbruck Medical University, Innsbruck, Austria. Since 2004, he has been with the Department of Automatic Control, Beijing Institute of Technology, Beijing, first as an Associate Professor, then, since 2008, as a Professor. His current research interests are in the fields of cloud control systems, networked control systems, robust control and signal processing, active disturbance rejection control, unmanned system control, and flight control.
	\end{IEEEbiography}
	\vspace{-11pt}
	
	\begin{IEEEbiography}[{\includegraphics[width=1in,height=1.25in,clip,keepaspectratio]{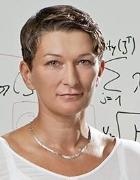}}]{Sandra Hirche}
		(M'03--SM'11--F'20) 
		received the Dipl.-Ing degree in aeronautical engineering from the Technical University of Berlin, Berlin, Germany, in 2002, and the Dr. Ing. degree in electrical engineering from the Technical University of Munich, Munich, Germany, in 2005. From 2005 to 2007, she was awarded a Post-doctoral scholarship from the Japanese Society for the Promotion of Science at the Fujita Laboratory, Tokyo Institute of Technology, Tokyo, Japan. From 2008 to 2012, she was an Associate Professor with the Technical University of Munich. Since 2013, she has served as Technical University of Munich Liesel Beckmann Distinguished Professor and has been with the Chair of Information-Oriented Control, Department of Electrical and Computer Engineering, Technical University of Munich. She has authored or coauthored more than 150 papers in international journals, books, and refereed conferences. Her main research interests include cooperative, distributed, and networked control with applications in human--machine interaction, multirobot systems, and general robotics. 
		
		Dr. Hirche has served on the editorial boards of the IEEE Transactions on Control of Network Systems, the IEEE Transactions on Control Systems Technology, and the IEEE Transactions on Haptics. She has received multiple awards such as the Rohde \& Schwarz Award for her Ph.D. thesis, the IFAC World Congress Best Poster Award in 2005, and -- together with students -- the 2018 Outstanding Student Paper Award of the IEEE Conference on Decision and Control as well as Best Paper Awards from IEEE Worldhaptics and the IFAC Conference of Manoeuvring and Control of Marine Craft in 2009.
	\end{IEEEbiography}
	\vspace{-11pt}
	
	\vfill
	
\end{document}